\documentclass[pre,twocolumn,preprintnumbers,amsmath,amssymb,%
nofootinbib,floatfix]{revtex4}

\usepackage{graphicx,bm,epsfig}

\def\spose#1{\hbox to 0pt{#1\hss}}
\def\lesssim{\mathrel{\spose{\lower 3pt\hbox{$\mathchar"218$}}
 \raise 2.0pt\hbox{$\mathchar"13C$}}}
\def\gtrsim{\mathrel{\spose{\lower 3pt\hbox{$\mathchar"218$}}
 \raise 2.0pt\hbox{$\mathchar"13E$}}}

\def\<{\langle}
\def\>{\rangle}

\begin{document}

\title{Depletion effects in colloid-polymer solutions}

\author{Giuseppe D'Adamo$^a$
\thanks{$^\dagger$Email: Giuseppe.Dadamo@aquila.infn.it 
       \vspace{0pt}}, 
Andrea Pelissetto$^{a,b}$
\thanks{$^\ddagger$Email: Andrea.Pelissetto@roma1.infn.it
       \vspace{0pt}}
 and Carlo Pierleoni$^c$
\thanks{$^\S$Email: Carlo.Pierleoni@aquila.infn.it
       \vspace{6pt}} 
\\\vspace{6pt}
$^a$Dipartimento di Fisica, Sapienza Universit\`a di Roma,
P.le Aldo Moro 2, I-00185 Roma, Italy
\\
$^b$INFN, Sezione di Roma I, P.le Aldo Moro 2, I-00185 Roma, Italy \\
$^c$ Dipartimento di Scienze Fisiche e Chimiche, Universit\`a dell'Aquila and
CNISM, UdR dell'Aquila, V. Vetoio 10, Loc. Coppito, I-67100  L'Aquila, Italy
 }

\begin{abstract}
The surface tension, the adsorption, and the depletion thickness of polymers 
close to a single nonadsorbing colloidal sphere are computed by means of Monte Carlo
simulations. We consider polymers under good-solvent conditions and in the thermal
crossover region between good-solvent and $\theta$ behavior. In the dilute regime we 
consider a wide range of values of $q$, from $q = 0$ (planar surface) up to $q\approx 30$-50,
while in the semidilute regime, for $\rho_p/\rho_p^*\le 4$ ($\rho_p$ is the 
polymer concentration and $\rho_p^*$ is its value at overlap), we  
only consider
$q = 0,0.5,1$ and 2. 
The results are compared with the available theoretical predictions,
verifying the existing scaling arguments. Field-theoretical results, 
both in the dilute and in the semidilute regime, are in good agreement with the 
numerical estimates for polymers under good-solvent conditions.

\end{abstract}

\maketitle


\section{Introduction}

The study of the fluid phases in mixtures of colloids and
nonadsorbing neutral polymers has become increasingly
important in recent years; see
Refs.~\cite{Poon-02,FS-02,TRK-03,MvDE-07,FT-08,ME-09} for recent reviews.
These systems show a very interesting phenomenology, which only depends
to a large extent on the nature of the polymer-solvent system and
on the ratio $q \equiv \hat{R}_g/R_c$, where $\hat{R}_g$ is the 
zero-density radius of
gyration of the polymer and $R_c$ is the radius of the colloid.
Experiments and numerical simulations indicate that polymer-colloid mixtures
have a fluid-solid coexistence line and, for $q > q^*$, where
\cite{Poon-02} $q^* \approx 0.3$-0.4, also a fluid-fluid coexistence line 
between  a
colloid-rich, polymer-poor phase (colloid liquid) and a
colloid-poor, polymer-rich phase (colloid gas).
On the theoretical side, research has mostly concentrated on mixtures of
neutral spherical colloids and polymers in solutions under 
good-solvent or $\theta$ conditions. In the former case,
predictions for the colloid-polymer interactions 
have been obtained by using full-monomer representations of polymers
(for instance, the self-avoiding walk model was used in Refs.~\cite{LBMH-02-a,LBMH-02-b}),
field-theoretical methods \cite{EHD-96,HED-99,MEB-01}, or fluid integral equations
\cite{FS-01}.
Moreover, some general properties have been derived by using general scaling arguments
\cite{deGennes-CR79,JLdG-79,Odijk-96}. 
At the $\theta$ point, the analysis is simpler, since polymers behave approximately 
as ideal chains. These theoretical results have then been used as 
starting points to develop 
a variety of coarse-grained models and approximate methods, see 
Refs.~\cite{LBHM-00,LBMH-02-a,FT-08} and references therein, 
which have been employed to predict colloid-polymer phase diagrams. 

In this paper, we consider polymer solutions that either show
good-solvent behavior or are in the thermal crossover region between
good-solvent 
and $\theta$ conditions. We study the solvation of a single colloid
in the solution, assuming that the monomer-colloid potential is
purely repulsive. We determine the distribution of the polymer chains 
around a single colloidal particle, which is the simplest property that 
characterizes polymer-colloid interactions. 
We investigate numerically, by means of 
Monte Carlo simulations, how it depends on the quality of the solution,
which is parametrized \cite{Nickel-91,DPP-thermal}
in terms of the second-virial combination 
$A_{2,pp} = B_{2,pp}/\hat{R}_g^3$, where $B_{2,pp}$ is the second virial coefficient 
and $\hat{R}_g$ is the zero-density radius of gyration. This adimensional combination
varies between 5.50 in the good-solvent 
case \cite{CMP-06} and zero ($\theta$ point). 
Beside good-solvent solutions, we consider two intermediate 
cases: solutions such that 
$A_{2,pp}$ is approximately one half of the good-solvent value, 
which show intermediate properties
betweeen good-solvent and $\theta$ behavior,
and solutions such that $A_{2,pp}$ is 20\% of the good-solvent  value, 
which are close to the 
$\theta$ point. In each case we compute the polymer density profile around the colloid. 
These results are used to 
determine thermodynamic properties, like the surface tension and the 
adsorption \cite{Widom}, which are then compared with the available 
theoretical predictions. Note that an analysis of the polymer depletion around a
colloid in the thermal crossover region was already performed in 
Refs.~\cite{ALH-04,PH-06}, but without a proper identification of the universal crossover 
limit \cite{Sokal-94}.  Here, we wish to perform a much more careful
analysis of the crossover behavior, following Refs.~\cite{CMP-08,DPP-thermal}. 
We focus on the dilute and semidilute regimes, in which the monomer density is small
and a universal behavior, i.e., independent of chemical details, is obtained in 
the limit of large degree of polymerization. In the dilute regime, in which polymer-polymer
overlaps are rare, solvation properties are determined for a wide range of values of 
$q$, from 0 up to 30-50. In the semidilute regime, simulations of systems with large $q$
require considering a large number of colloids, which makes 
Monte Carlo simulations very expensive. Hence, we only present results for $q=0,0.5,1$ and 2.

The paper is organized as follows. In Sec.~\ref{sec2} we define the basic quantities
we wish to determine. First, in Sec.~\ref{sec2.1} and \ref{sec2.2} we introduce 
the surface tension, the adsorption, and the depletion thickness, 
and discuss their relation to the density profile of the polymers around the 
colloid. In Sec.~\ref{sec2.3} we discuss the low-density behavior and the 
relation between solvation properties and colloid-polymer virial coefficients
that parametrize  the (osmotic) pressure of a polymer-colloid binary solution
in the low-density limit. 
Sec.~\ref{sec2.4} discusses the different behavior that are expected as 
a function of $q$ and density and gives an overview of 
theoretical and numerical predictions. Sec.~\ref{sec3} summarizes our polymer model 
and gives a brief discussion on how one can parametrize in a universal fashion 
the crossover between the good-solvent and the $\theta$ behavior 
(for more details, 
see Ref.~\cite{DPP-thermal}). In Sec.~\ref{sec4} we present our results for the dilute 
regime, while finite-density results are presented in Sec.~\ref{sec5}. Sec.~\ref{sec6}
discusses a simple coarse-grained model in which each polymer is represented by a
monoatomic molecule, which represents a more rigorous version of the well-known
Asakura-Oosawa-Vrij model. In Sec.~\ref{sec7} we present our conclusions. 
Two appendices are included, one explaining how to compute the virial coefficients of 
a binary mixture of flexible molecules, and one discussing the 
small-$q$ behavior of the virial coefficients. 
Tables of results are reported in the supplementary material.

\section{Adsorption and depletion thickness} \label{sec2}

\subsection{Definitions} \label{sec2.1}

Let us  consider a solution of nonadsorbing polymers
in the grand canonical ensemble at fixed volume $V$
and chemical potential $\mu_p$. Temperature is also present, but, since 
it does not play any role in our discussion, we will omit writing it 
explicitly in the following. Let us indicate with 
$\Omega(\mu_p,V)$ the corresponding grand potential. 
Let us now add a spherical colloidal particle of radius $R_c$ to the solution
and let $\Omega_{c}(\mu_p,V)$ be the corresponding grand potential.
The insertion free energy can be written as the sum of two
terms, one proportional to the volume $V_c = {4\over3} \pi R_c^3$ 
of the colloid and one proportional to its surface area $A_c = 4 \pi R_c^2$
\cite{Widom}:
\begin{equation}
\Omega_c(\mu_p,V)-\Omega(\mu_p,V) = P V_c + \gamma A_c,
\label{gamma-def}
\end{equation}
where $P$ is the bulk pressure and $\gamma$ is the surface tension.
The latter quantity can be related to the adsorption
$\Gamma(\mu_p)$ defined in terms of the change in the mean number
of polymers due to the presence of the colloid:
\begin{equation}
\langle N_{p} \rangle^{(c)} -\langle N_{p}\rangle = 
   - \rho_p V_c + \Gamma A_c,
\end{equation}
where $N_{p}$ indicates 
the numbers of polymers present in the solution, 
$\langle \cdot \rangle^{(c)}$ and $\langle \cdot \rangle$ are averages
in the presence and in the absence of the colloidal particle,
respectively.
Differentiating Eq.~(\ref{gamma-def}) with respect to $\mu_p$, we obtain 
\begin{equation}
\Gamma(\mu_p)=-\left( \frac{\partial \gamma}{\partial \mu_p} \right)_{T,V}.
\end{equation}
We also define the average bulk polymer density $\rho_p$ as 
\begin{equation}
\rho_p(\mu_p) = - {1\over V} 
  \left({\partial \Omega \over \partial \mu_p}\right)_{T,V}.
\end{equation}
The surface tension can also be defined in the canonical ensemble, 
as a function of the bulk polymer density $\rho_p$. 
Then, we have 
\begin{equation}
\Gamma(\rho_p) = - \left(\frac{\partial \beta\gamma}{\partial \rho_p}
     \right)_{T,V}
     {\rho_p\over K_p(\rho_p)},
\label{Gamma-gamma}
\end{equation}
where 
\begin{equation}
K_p = \left({\partial \beta P\over \partial \rho_p}\right)_{T,V},
\end{equation}
for the bulk system ($\beta = 1/k_B T$ as usual).

The adsorption coefficient can be easily related to  
the polymer density profile around the colloid.  Assume that each polymer 
consists of $L$ monomers and define the average bulk monomer density 
$\rho_{{\rm mon}} = L \rho_p$. Then, we write 
\begin{eqnarray}
&& \langle N_{p} \rangle^{(c)} -\langle N_{p}\rangle
\nonumber \\ 
&& =\, 
 \int d^3{\bf r}\, 
 \left[{1\over L} \left \langle \sum_{\alpha i} \delta ({\bf R}_c - {\bf
r}_{\alpha}^{(i)} -
       {\bf r})\right\rangle^{(c)} - \rho_p(\mu_p)\right],
\end{eqnarray}
where the average is performed at chemical potential $\mu_p$ and volume $V$,
${\bf R}_c$ is the colloid position, and ${\bf r}_{\alpha}^{(i)}$, 
$\alpha = 1,\ldots,L$, $i=1,\ldots N_p$, are the monomer positions.
If we now define the 
monomer-colloid pair correlation function
\begin{equation}
g_{{\rm mon},cp}(r;\mu_p) = 
   {1\over \rho_{{\rm mon}}} 
   \left \langle \sum_{\alpha i} \delta ({\bf R}_c - {\bf r}_{\alpha}^{(i)} -
       {\bf r})\right\rangle^{(c)},
\label{gmon-def}
\end{equation}
and the integral
\begin{equation} 
G_{{\rm mon},cp}(\mu_p) = \int d{\bf r}\, [g_{{\rm mon},cp}(r;\mu_p) - 1],
\label{Gmoncp}
\end{equation}
we obtain 
\begin{equation}
\Gamma(\mu_p) = {\rho_p(\mu_p)\over A_c} \left[ G_{{\rm mon},cp}(\mu_p) + V_c\right].
\end{equation}
Since $g_{{\rm mon},cp}(r;\mu_p) = 0$ for $r\le R_c$, a more transparent
relation is obtained by defining 
\begin{equation} 
\hat{G}_{{\rm mon},cp}(\mu_p) = 4\pi \int_{R_c}^\infty r^2 d{r}\, 
[g_{{\rm mon},cp}(r;\mu_p) - 1],
\label{Gmoncp2}
\end{equation}
in which one only integrates the density profile outside the colloid. 
Since ${G}_{{\rm mon},cp}(\mu_p) = \hat{G}_{{\rm mon},cp}(\mu_p) - V_c$,
 we have
\begin{equation}
\Gamma(\mu_p) = {\rho_p(\mu_p)\over A_c} \hat{G}_{{\rm mon},cp}(\mu_p).
\label{Gamma-Gmoncp}
\end{equation}
In the previous discussion we have considered the monomer-colloid correlation
function, but it is obvious that any other polymer-colloid distribution
function could be used.
In order to compare our results with those 
obtained in coarse-grained models (we will discuss them in Sec.~\ref{sec6}), 
we will also use the pair distribution function between the colloid and the 
polymer centers of mass. If ${\bf r}_{\alpha}^{(i)}$,
$\alpha=1,\ldots,L$, are the positions of the monomers belonging to polymer $i$,
we first define the polymer center of mass
\begin{equation}
  {\bf r}^{(i)}_{CM} = {1\over L} \sum_\alpha {\bf r}_\alpha^{(i)}.
\end{equation}
Then, the pair distribution function between a colloid and a polymer 
center of mass is defined by
\begin{equation}
g_{CM,cp}(r;\mu_p) = 
   {1\over \rho_{p}} 
   \left \langle \sum_{i} \delta ({\bf R}_c - {\bf r}_{CM}^{(i)} -
       {\bf r})\right\rangle^{(c)},
\label{gCM-def}
\end{equation}
where the average is taken at a given value $\mu_p$.
In terms of this quantity
\begin{eqnarray}
\Gamma(\mu_p) &=& {\rho_p(\mu_p)\over A_c} [V_c + G_{CM,cp}(\mu_p)], 
\label{Gamma-GCMcp}
\\
G_{CM,cp}(\mu_p) &=& 4\pi \int_0^\infty r^2 dr\, [g_{{CM},cp}(r;\mu_p) -1].
\nonumber 
\end{eqnarray}
If we define\footnote{Note that,
since $g_{{CM},cp}(r;\mu_p)\not = 0$ for $r \le R_c$, $\hat{G}_{CM,cp}(\mu_p)$
cannot be obtained directly by performing the integration
from $r=R_c$ to $\infty$.}
$\hat{G}_{CM,cp}(\mu_p) = V_c + G_{CM,cp}(\mu_p)$,
we can write a relation analogous to Eq.~(\ref{Gamma-Gmoncp}).
Comparison of Eqs.~(\ref{Gamma-Gmoncp}) and (\ref{Gamma-GCMcp}) implies
$G_{CM,cp} = G_{{\rm mon},cp}$ and 
$\hat{G}_{CM,cp} = \hat{G}_{{\rm mon},cp}$, 
hence in the following we will simply refer 
to these quantities as $G_{cp}$ and $\hat{G}_{cp}$.

It is interesting to relate the pair correlation functions $g_{cp}(r;\mu_p)$ 
to the analogous correlation functions
$\hat{g}_{cp}(r;\mu_p,\mu_c)$ that are appropriate for a binary 
system consisting of polymers and colloids at polymer and colloid chemical
potentials $\mu_p$ and $\mu_c$, respectively. Indeed, one can show 
that, in the limit $\mu_c\to -\infty$, i.e., when the colloid density goes to 
zero, one has
\begin{equation}
g_{cp}(r;\mu_p) = \lim_{\mu_c\to -\infty}
    \hat{g}_{cp}(r;\mu_p,\mu_c).
\label{equality-g}
\end{equation}
Eq.~(\ref{equality-g}) allows us to relate $G_{cp}$ to thermodynamic 
properties of the binary mixture in the limit of vanishing colloid density.
For this purpose we use the Kirkwood-Buff relations
between structural and thermodynamic properties of fluid mixtures
\cite{KB-51,BenNaim}. 
The integral $G_{cp}$, which is relevant to determine 
adsorption properties, corresponds to one of the 
Kirkwood-Buff integrals \cite{KB-51,BenNaim} defined as 
\begin{equation}
G_{\alpha\beta} = \int d{\bf r}\, (g_{\alpha\beta}(r) -1),
\end{equation}
where $\alpha$ and $\beta$ label the different species of the mixture.
The integrals 
$G_{\alpha\beta}$ can be related to derivatives of the pressure with 
respect to the polymer and colloid densities. 
For $\rho_c = 0$ we have \cite{KB-51,BenNaim}
\begin{eqnarray}
 K_c &=& \left({\partial \beta P\over \partial \rho_c}\right)_{\rho_p} = 
   1 - {\rho_p G_{cp}\over 1 + \rho_p G_{pp}} ,
\\
 K_p &=& \left({\partial \beta P\over \partial \rho_p}\right)_{\rho_c=0} = 
    {1 \over 1 + \rho_p G_{pp}},
\end{eqnarray}
which imply
\begin{equation}
G_{cp} = {1-K_c\over \rho_p K_p}.
\end{equation}
Eqs.~(\ref{Gamma-Gmoncp}) and (\ref{Gamma-gamma}) can then be rewritten as
\begin{eqnarray}
\Gamma &=& {1\over A_c} \left[{1 - K_c\over K_p} + \rho_p V_c\right], 
\label{Gamma-KB}
\\
\beta \gamma &=& 
{1\over A_c} \int_0^{\rho_p} {K_c - 1\over \rho_p'} 
    d\rho_p' - \beta P {V_c\over A_c} .
\label{gamma-KB}
\end{eqnarray}

\subsection{Depletion thickness} \label{sec2.2}

Depletion effects can be equivalently parametrized by 
introducing the depletion thickness
$\delta_s$ \cite{FT-07,FST-07,FT-08,LT-11}, which is an
average width of the depleted layer around the colloid. 
It is defined in terms of the integral $G_{cp}$ as 
\begin{eqnarray}
\frac{4\pi}{3}\left(R_c+\delta_s\right)^3 = - G_{cp} = V_c - \hat{G}_{cp},
\end{eqnarray}
so that 
\begin{eqnarray}
{\delta_s\over R_c} = \left( 1 - {\hat{G}_{cp}\over V_c}\right)^{1/3} - 1. 
\label{dsoverRcvsGcp}
\end{eqnarray}
Since $\delta_s$ is only determined by $G_{cp}$, knowledge of $\delta_s$
is completely equivalent to that of the adsorption. The two quantites 
are related by
\begin{equation}
\Gamma = - {\rho_p V_c\over A_c} \left[
\left(1 + \delta_s/R_c\right )^3 - 1 \right].
\label{Gamma-ds}
\end{equation}
As we shall discuss below, $\delta_s/R_c\to 0$ 
for large polymer densities, hence in this limit 
\begin{equation}
\Gamma = - \rho_p \delta_s.
\label{Gamma-deltas-largePhi}
\end{equation}
It is interesting to discuss the limit $q\to 0$, in which the colloid
degenerates into an impenetrable plane. Setting $r = R_c + z$ in
Eq.~(\ref{Gmoncp2}), 
we obtain 
\begin{equation}
\hat{G}_{cp} =  4 \pi
     \int_0^\infty dz\, (R_c + z)^2 [g_{{\rm mon},cp}(R_c + z)-1].
\end{equation}
For $R_c\to \infty$, we have $g_{{\rm mon},cp}(R_c + z) \approx 
g_{\rm mon,pl}(z)$, where $g_{\rm mon,pl}(z)$ is the pair distribution
function between an impenetrable plane at $z=0$ and a polymer. Then,
we obtain for $R_c\to \infty$
\begin{eqnarray}
&& \hat{G}_{cp} = 4 \pi R_c^2 G_{pl},
\end{eqnarray}
with 
\begin{eqnarray}
&& G_{pl} = \int_0^\infty dz\, [g_{\rm mon,pl}(z)-1].
\end{eqnarray}
Taking the limit $R_c\to \infty$ 
in Eqs.~(\ref{dsoverRcvsGcp}) and (\ref{Gamma-ds}),
we obtain 
\begin{equation}
\delta_s = - {\Gamma\over \rho_p} = - G_{pl}.
\label{deltas-piano}
\end{equation}

\subsection{Low-density expansions} \label{sec2.3}

For $\rho_p\to 0$ the depletion thickness $\delta_s$ and the surface quantities
$\Gamma$ and $\gamma$ 
can be related to the virial coefficients that parametrize the 
expansion of the pressure of a binary colloid-polymer system in powers
of the concentrations.  These relations have already 
been discussed in the literature \cite{Bellemans-63,SS-80,MQR-87,YSEK-13}. They
can be easily
derived by using Eqs.~(\ref{Gamma-KB}) and (\ref{gamma-KB}).
We start by expanding the pressure as
\begin{eqnarray}
&& \beta P = \rho_c + \rho_p + B_{2,cc} \rho_c^2 + B_{2,pp} \rho_p^2 + 
        B_{2,cp} \rho_c \rho_p 
\nonumber \\
        && \quad + B_{3,ccc} \rho_c^3 + B_{3,ppp} \rho_p^3 + 
        B_{3,ccp} \rho_c^2 \rho_p + B_{3,cpp} \rho_c \rho_p^2 + 
        \ldots,
\label{virial-expansion}
\end{eqnarray}
where $\rho_c$ and $\rho_p$ are the colloid and polymer concentrations and 
we have neglected fourth-order terms. 
Then, Eqs.~(\ref{Gamma-KB}) and (\ref{gamma-KB}) give
\begin{eqnarray}
\Gamma &=& {\rho_p\over A_c} \left[
   V_c - B_{2,cp} - (B_{3,cpp} - 2 B_{2,pp} B_{2,cp}) \rho_p + \ldots\right],
\nonumber \\
\\
\beta\gamma &=& 
   - {\rho_p\over A_c} \left[V_c - B_{2,cp} + {1\over2} 
      (2 B_{2,pp} V_c - B_{3,cpp})\rho_p + \ldots \right].
\nonumber \\
\label{Gamma-gamma-smallrhop}
\end{eqnarray}
In the limit $R_c\to\infty$ one should recover the results for an 
infinite impenetrable plane. This requires the coefficients appearing 
in the previous two expressions to be of order $A_c$ as $R_c\to \infty$. 
This is explicitly checked in App.~\ref{AppB} and allows us to write 
\begin{eqnarray}
\Gamma &=& - \rho_p P_{1,p} - \rho_p^2 P_{2,pp} + \ldots, \\
\beta\gamma &=& 
   P_{1,p} \rho_p + {1\over 2} (2 B_{2,pp} P_{1,p} + P_{2,pp}) \rho_p^2 + 
    \ldots
\end{eqnarray}
Explicit expressions for $P_{1,p}$ and $P_{2,pp}$ are 
reported in Appendix~\ref{AppB}.

For the depletion thickness we obtain 
\begin{equation}
{\delta_s\over R_c} = - 1 + \left({3 q^3 A_{2,cp} \over 4\pi}\right)^{1/3}
   \left[1 + {\Phi\over 4 \pi}\left( {A_{3,cpp}\over A_{2,cp}} - 2 A_{2,pp}
      \right)\right] + \ldots,
\label{deltas-smallrhop}
\end{equation}
where we have defined the polymer volume fraction
\begin{equation}
\Phi = {4\pi \hat{R}_g^3\over 3} \rho_p
\end{equation}
and the adimensional combinations $A_{2,\#} = B_{2,\#} \hat{R}_g^{-3}$ and
$A_{3,\#} = B_{3,\#} \hat{R}_g^{-6}$, where $\hat{R}_g$ is the zero-density
polymer radius of gyration.
In the limit $R_c\to \infty$, we should obtain the density expansion of the 
depletion thickness for an impenetrable plane. Using Eq.~(\ref{deltas-piano})
we obtain
\begin{equation}
\delta_s = P_{1,p} + \rho_p P_{2,pp} + O(\rho_p^2).
\end{equation}

\subsection{Theoretical predictions and scaling arguments} \label{sec2.4}

Depletion properties have been extensively studied in the past. 
Here we present scaling arguments and literature results, that will be 
checked in the following sections by using our 
accurate Monte Carlo estimates.

For an ideal (noninteracting) 
polymer solution the insertion free energy is exactly known
\cite{EHD-96}:
\begin{equation}
\beta \gamma = {2\over \sqrt{\pi}} \rho_p \hat{R}_g 
    \left(1 + {\sqrt{\pi}\over 2} q\right) = 1.128 \rho_p \hat{R}_g
    (1 + 0.886 q),
\end{equation}
where $\hat{R}_g$ is the zero-density radius of gyration. 
The depletion thickness follows immediately \cite{FT-08,LT-11}:
\begin{equation}
{\delta_s \over R_c} = 
  \left(1 + {6 q\over \sqrt{\pi}} + 3 q^2\right)^{1/3} - 1.
\end{equation}
For good-solvent polymers there are several predictions 
obtained by using the field-theoretical renormalization group.
In the dilute limit $\Phi\to 0$, the surface tension has been 
determined \cite{HED-99} 
both in the colloid limit in which $q\to 0$ and in the so-called
protein limit $q\to\infty$. Setting $R_x^2 = 2 \hat{R}_g^2$ and 
$\epsilon = 1$ in the results of Ref.~\cite{HED-99}, we obtain 
for $q\to 0$ and $\Phi\to 0$
\begin{equation}
\beta \gamma \approx 1.071 \rho_p \hat{R}_g (1 + 0.811 q - 0.037 q^2).
\label{FT-Phi0}
\end{equation}
Note that the dilute behavior in the colloidal regime $q\lesssim 1$ 
is similar to that observed
in the ideal case. The coefficients corresponding to the planar term and to 
the leading curvature correction are close, while the second-curvature 
correction is absent in the ideal case and quite small for good-solvent 
chains.

In the opposite limit $q\to \infty$
general arguments predict \cite{deGennes-CR79,HED-99}
\begin{equation}
\beta\gamma \approx A_{\gamma,\infty} \rho_p R_c q^{1/\nu}.
\label{gamma-largeq}
\end{equation}
The constant $A_{\gamma,\infty}$ has been estimated by Hanke {\em et al}.
\cite{HED-99}:
\begin{equation}
A_{\gamma,\infty} = 1.41 \pm 0.04.
\end{equation}
Eq.~(\ref{Gamma-gamma}) gives then 
$\Gamma = A_{\gamma,\infty} \rho_p R_c q^{1/\nu}$.
For the depletion thickness we obtain $\delta_s/R_c\sim q^{1/(3\nu)}$.

Finite-density corrections have been computed by Maassen {\em et al.}
\cite{MEB-01} in the renormalized tree approximation. For 
$\Phi\to 0$ they obtain 
\begin{equation}
\beta\gamma = 1.129 \rho_p \hat{R}_g 
   [1 + 0.698 \Phi + 0.886 q (1 - 0.094 \Phi)].
\label{FT-tree-Phi0}
\end{equation}
In this approximation one does not recover the correct large-$q$ behavior
(\ref{gamma-largeq}),
hence we  expect it to be valid only in the colloid regime. 
The zero-density behavior can be compared with that given in 
Eq.~(\ref{FT-Phi0}), which includes the leading (one-loop) 
$\epsilon$ correction. Differences are small, of order 5\%. We expect
an error of the same order for the coefficients of the density correction.

The behavior of $\gamma$ in the semidilute regime is expected to have 
universal features. If the polymer volume fraction $\Phi$ is large,
we expect, on general grounds, the behavior \cite{JLdG-79}
\begin{equation}
\beta \gamma = \rho_p \hat{R}_g A(q) \Phi^\alpha,
\label{gamma-largePhi}
\end{equation}
where $\alpha$ is an exponent to be determined and $A(q)$ is a 
coefficient, which {\em a priori} can depend on $q$. 
However, deep in the semidilute regime, the coil radius of gyration 
is no longer the relevant length scale. One should rather consider
the density-dependent correlation length $\xi$ \cite{deGennes-79}, 
which measures the polymer mesh size. 
The scaling behavior (\ref{gamma-largePhi}) should be valid 
for $\hat{R}_g,R_c\gg \xi$, and in this regime $\hat{R}_g$
plays no role. Therefore, $q$ is not the relevant parameter and
$A(q)$ is independent of $q$. To determine the 
exponent $\alpha$ we can use the same argument which allows one 
to determine the scaling behavior of the osmotic pressure in the 
semidilute regime. For large $\Phi$ we expect thermodynamic properties
to depend only on the monomer density $\rho_{\rm mon} = \rho_p L$
and not on the number $L$ of monomers per chain. This requirement
gives \cite{JLdG-79}
\begin{equation}
 \alpha = {1 - \nu\over 3\nu - 1} \approx 0.541,
\label{gamma-largePhi-alpha}
\end{equation}
where \cite{Clisby-10} $\nu = 0.587597(7)$.
Predictions (\ref{gamma-largePhi}) and (\ref{gamma-largePhi-alpha}) can also
be obtained \cite{JLdG-79} by noting that $\beta\gamma$ can only depend
on the correlation length $\xi$ deep in the semidilute regime, i.e.,
when $\xi \ll R_c,\hat{R}_g$. Then, dimensional analysis gives
\begin{equation}
\beta\gamma \sim \xi^{-2}.
\end{equation}
Using $\xi\sim \hat{R}_g \Phi^{-\nu/(3\nu-1)}$
\cite{deGennes-79,dCJ-book,Schaefer-99}, 
we obtain again Eq.~(\ref{gamma-largePhi}) with 
$\alpha$ given by Eq.~(\ref{gamma-largePhi-alpha}). 
Eq.~(\ref{gamma-largePhi}) allows us 
to obtain the large-$\Phi$ behavior of the adsorption and of the depletion 
thickness. Using Eq.~(\ref{Gamma-gamma}) and the general scaling of the osmotic 
pressure \cite{deGennes-79,dCJ-book,Schaefer-99} 
$\beta P/\rho_p \sim \Phi^{1/(3\nu-1)}$,
we obtain 
\begin{equation}
\Gamma \sim \rho_p \hat{R}_g \Phi^{-\nu/(3\nu-1)} = 
\rho_p \hat{R}_g \Phi^{-0.770}.
\label{Gamma-largePhi}
\end{equation}
Equivalently, one could have observed that 
$\delta_s \sim\xi$, since $\xi$ is the only relevant length scale. 
Using $\xi\sim \hat{R}_g \Phi^{-\nu/(3\nu-1)}$, we obtain 
$\delta_s\sim \hat{R}_g \Phi^{-\nu/(3\nu-1)} \sim 
\hat{R}_g \Phi^{-0.770}$. Eq.~(\ref{Gamma-deltas-largePhi}) implies then
Eq.~(\ref{Gamma-largePhi}).

The large-$\Phi$ behavior was determined in the renormalized 
tree-level approximation obtaining \cite{MEB-01}
\begin{equation}
\beta\gamma = 1.563 \rho_p \hat{R}_g \Phi^{(1-\nu)/(3\nu-1)} 
    [1 + 0.650 q \Phi^{-\nu/(3\nu-1)}] .
\label{gamma-largePhi-MEB}
\end{equation}
This result is fully consistent with Eq.~(\ref{gamma-largePhi}), since 
the $q$ correction appearing in Eq.~(\ref{gamma-largePhi-MEB})
vanishes for $\Phi\to \infty$. The exponent of the $q$-dependent correction
in Eq.~(\ref{gamma-largePhi-MEB}) can be easily interpreted. Consider the ratio 
$\gamma(q,\Phi)/\gamma(0,\Phi)$. This quantity is adimensional, hence it is 
a universal function of adimensional ratios of the 
relevant length scales in the system. Deep in the semidilute regime the 
relevant polymer scale is the correlation length $\xi$, hence we expect
\begin{equation}
{\gamma(q,\Phi)\over \gamma(0,\Phi)} = f(\xi/R_c).
\end{equation}
Now we take $\Phi$ large so that $\xi/R_c\ll 1$. Then, we can expand
\begin{equation}
{\gamma(q,\Phi)\over \gamma(0,\Phi)} = 1 + 
   a_1 {\xi\over R_c} + a_2 \left( {\xi\over R_c} \right)^2 + \ldots
\label{Helfrich-xi}
\end{equation}
Since $\xi\sim \hat{R}_g \Phi^{-\nu/(3\nu-1)}$, we obtain 
\begin{equation}
{\gamma(q,\Phi)\over \gamma(0,\Phi)} = 1 + 
   b_1 q \Phi^{-\nu/(3\nu-1)} + b_2 q^2 \Phi^{- 2 \nu/(3\nu-1)} + \ldots
\end{equation}
which reproduces the behavior (\ref{gamma-largePhi-MEB}).
Eq.~(\ref{Helfrich-xi}) is the semidilute analogue of the Helfrich
expansion in powers of $q$ that holds for $\Phi\to 0$. The only difference is 
the expansion variable: in the semidilute region, polymer size is 
characterized by $\xi$, hence one should consider $\xi/R_c$ instead of 
$q = \hat{R}_g/R_c$. 

Quantitative predictions for the large-$\Phi$
behavior of $\Gamma$ and $\delta_s$ can be derived from
Eq.~(\ref{gamma-largePhi-MEB}), by using Eq.~(\ref{Gamma-gamma}) 
and the large-$\Phi$ behavior of $K_p(\rho_p)$.
The latter can be derived from the results of Ref.~\cite{Pelissetto-08},
which give $K_p \simeq 3.71 \Phi^{1.311}$ for $\Phi\to \infty$.
Thus, we obtain 
\begin{eqnarray}
{\delta_s\over \hat{R}_g} \approx -{\Gamma\over \rho_p \hat{R}_g} &\approx& 
    0.649 \Phi^{-0.770}. 
\label{deltas-largePhi}
\end{eqnarray} 
In the protein limit, in which 
$q$ is large, beside the regime $R_c\gg \xi$ in which 
Eqs.~(\ref{gamma-largePhi}), (\ref{Helfrich-xi}) 
and (\ref{deltas-largePhi}) hold, there is a second interesting 
regime in which one has both $R_c\ll \hat{R}_g$ and $R_c\ll \xi$.
For $q$ large, these conditions are satisfied both in the dilute limit
and in the semidilute region, as long as $\Phi$ is not too large.
Under these conditions,
Eq.~(\ref{gamma-largeq}) holds irrespective of the 
polymer density.  Therefore, Eq.~(\ref{gamma-KB}) can be rewritten as 
\begin{equation}
{1\over \rho_p} \int_0^{\rho_p} {K_c-1\over \rho_p'} d\rho_p' = 
    V_c\left[{\beta P\over \rho_p} + 3 A_{\gamma,\infty} q^{1/\nu}\right].
\end{equation}
For $q\to \infty$, the pressure term can be neglected
compared with the term proportional to $q^{1/\nu}$, hence the right-hand
side is density independent. This implies that the integrand that appears in 
left-hand side is also density independent in the density region 
where $R_c\ll \xi$ and is equal to 
$3 V_c A_{\gamma,\infty} q^{1/\nu}$. 
For $\Phi\to 0$, using the virial expansion (\ref{virial-expansion})
we can write
\begin{equation}
{K_c-1\over \rho_p} = B_{2,cp} \left[1 + {B_{3,cpp}\over B_{2,cp}} \rho_p + 
  \ldots\right].
\end{equation}
Therefore, we can identify $B_{2,cp} = 3 V_c A_{\gamma,\infty} q^{1/\nu}$. 
Moreover, $B_{3,cpp}/B_{2,cp}$ vanishes for $q\to\infty$ 
(a similar result holds for the higher-order virial coefficients).
By using Eq.~(\ref{Gamma-gamma}) and 
Eq.~(\ref{gamma-largeq}) we also predict for the adsorption
\begin{eqnarray}
\Gamma &=& - A_{\gamma,\infty} q^{1/\nu} R_c {\rho_p\over K_p(\rho_p)},
\\
{\delta_s\over R_c} &=& 
   \left( {3 A_{\gamma,\infty}\over K_p(\rho_p)}\right)^{1/3} q^{1/(3\nu)}.
\label{deltas-largePhi-largeq}
\end{eqnarray}
Since $K_p(\rho_p) \sim \Phi^{1.311}$ for large $\Phi$ \cite{deGennes-79}, this 
relation predicts $\delta_s \sim \Phi^{-0.437}$. 
Note that Eq.~(\ref{deltas-largePhi-largeq}) holds only for 
$R_c\ll \xi \ll \hat{R}_g$.
As $\Phi$ further increases, $\xi$ decreases and 
one finds eventually $R_c\sim \xi$. 
Then, Eq.~(\ref{deltas-largePhi-largeq}) no longer
holds and a crossover occurs. For $R_c\gg \xi$ the asymptotic behavior 
$\delta_s\sim \Gamma\sim \Phi^{-0.770}$ sets in.
Eq.~(\ref{deltas-largePhi-largeq}) can be written in a more suggestive 
form, by noting that $K_p(\rho_p) \sim (\xi/\hat{R}_g)^{-1/\nu}$ 
\cite{deGennes-79}. Hence 
\begin{equation}
   {\delta_s\over R_c} \sim \left({\xi\over R_c}\right)^{1/3\nu}.
\end{equation}
We recover the same scaling that occurs in the dilute regime, with 
$\xi$ replacing $\hat{R}_g$ as relevant polymer scale.

\begin{figure}[t]
\begin{center}
\begin{tabular}{c}
\epsfig{file=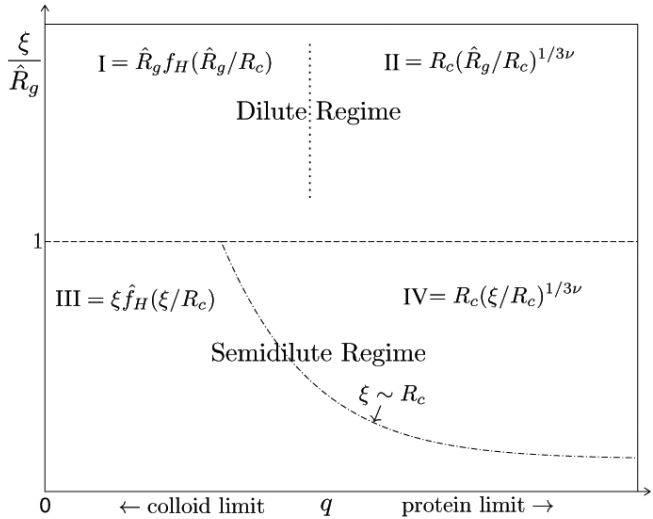,angle=0,width=9truecm} \hspace{0.5truecm} \\
\end{tabular}
\end{center}
\caption{Different regimes for the depletion thickness in terms of 
$\xi/\hat{R}_g$ and $q = \hat{R}_g/R_c$. The functions $f_H(\hat{R}_g/R_c)$
and $\hat{f}_H(\xi/R_c)$ have a regular expansion in powers of their argument.
}
\label{phase}
\end{figure}

To conclude, let us summarize the different types of behavior of the 
depletion thickness in the $\xi$-$q$ diagram for the good-solvent case. 
They depend on the 
relative size of the three different scales that appear in the problem:
the radius of gyration of the polymer, the radius of the colloid and the 
correlation length $\xi$. In the colloid regime in which $q < 1$, i.e.
$R_c > \hat{R}_g$, depletion shows two different behaviors, depending on the 
ratio $\xi/\hat{R}_g$. In the dilute regime in which the relevant scale is 
the radius of gyration (domain I in Fig.~\ref{phase}), $\delta_s$ is of 
order $\hat{R}_g$ with a proportionality constant that can be expanded 
in powers of $q$ (Helfrich expansion). If instead $\xi\ll \hat{R}_g$
(semidilute regime, domain III in Fig.~\ref{phase}), the relevant scale
is the correlation length $\xi$. The depletion thickness is 
proportional to $\xi\sim\hat{R}_g \Phi^{-0.770}$ with a proportionality
constant that admits an expansion in powers of $\xi/R_c$. Since $\xi\to 0$
for $\Phi\to \infty$, the limiting behavior is independent of the 
colloid radius. In the protein regime in which $q > 1$, i.e.,
$R_c < \hat{R}_g$, depletion shows three different behaviors. 
In the dilute regime (domain II in Fig.~\ref{phase}), 
$\delta_s \sim R_c q^{1/(3\nu)}\sim R_c^{1-1/(3\nu)} \hat{R}_g^{1/(3\nu)}$, 
i.e., $\delta_s$ is much larger than
the colloid radius but much smaller than $\hat{R}_g$. 
In the semidilute regime, two different behaviors occur. If 
$R_c\ll \xi\ll \hat{R}_g$ (domain IV), 
the role of the radius of gyration is now assumed 
by the correlation length and we have 
$\delta_s \sim R_c^{1-1/(3\nu)} \xi^{1/(3\nu)}$. Finally, as $\Phi$ increases
further, one finally finds $\xi\ll R_c$ and one observes again 
$\delta_s\sim \xi$ (domain III).

The surface tension $\gamma$ was also computed in the PRISM 
approach \cite{FS-01}, obtaining
\begin{equation}
\beta \gamma = 1.279 \rho_p \hat{R}_g 
  [1 + 1.06 \Phi + 0.634 q].
\label{PRISM}
\end{equation}
Such an expression does not have the correct behavior for 
$q\to\infty$ or $\Phi\to \infty$. 
In the dilute regime and for small $q$, 
comparison with the field-theoretical results
(we shall show that they are quite accurate)
shows that it only provides a very rough approximation, differences being of 
order 20-30\%.

The adsorption $\Gamma$ was computed numerically for the planar 
case ($q=0$) in Ref.~\cite{LBMH-02-a}, obtaining
\begin{equation}
\Gamma = - 1.074 \rho_p \hat{R}_g 
   \left(1 + 7.63 \Phi + 14.56 \Phi^3\right)^{-0.2565}.
\label{Gamma-planar}
\end{equation}
This expression allows us to compute $\gamma$ for $q=0$ using 
the expression of the compressibility factor given in 
Ref.~\cite{Pelissetto-08}. In the small-density limit we obtain 
\begin{equation}
\beta \gamma = 1.074 \rho_p \hat{R}_g \left[1 + 0.334 \Phi + O(\Phi^2)
   \right],
\label{LBMH-smallPhi}
\end{equation}
while for $\Phi\to \infty$ we obtain 
\begin{equation}
\beta \gamma = 1.30 \rho_p \hat{R}_g \Phi^{0.54}.
\label{LBMH-largePhi}
\end{equation}
We can compare these expressions with the field-theory results. The leading 
density correction in Eq.~(\ref{LBMH-smallPhi}) is approximately 
one half of that predicted by field theory, see Eq.~(\ref{FT-tree-Phi0}), 
while the large-$\Phi$ expression (\ref{LBMH-largePhi}) predicts 
a surface tension that is 17\% smaller than Eq.~(\ref{gamma-largePhi-MEB}).

Finally, we mention the phenomenological expression for the depletion thickness
of Fleer {\em et al.}
\cite{FST-07,FT-08}
\begin{equation}
{\delta_s\over R_c} = 0.865 q^{0.88} 
   \left(1 + 3.95 \Phi^{1.54} \right)^{-0.44},
\label{FT-deltas}
\end{equation}
which should be only valid in an intermediate range of values of $q$
\cite{FST-07,LT-11}, since it does not have the correct behavior in the limits
$q\to 0$ and $q\to\infty$.

There are no predictions for polymers in the thermal crossover region. 
In this case, a new scale comes in, the dimension $R_T$ of the so-called
thermal blob \cite{deGennes-79}. On scales $r\ll R_T$, 
the polymer behaves as an ideal chain, hence for $R_c\ll R_T$ 
the surface tension should coincide with that appropriate for 
an ideal chain. This implies that for any finite value of $z$ 
we should recover the ideal result for the surface tension, 
provided that $q$ is large enough. In particular, we predict 
\begin{equation} 
   \beta\gamma = \rho_p R_c q^2 
\end{equation} 
for all finite values of $z$ and $q\to \infty$. In practice, 
Eq.~(\ref{gamma-largeq}) holds also for finite $z$, with the values appropriate
for the ideal chain, $\nu = 1/2$ and 
$A_{\gamma,\infty} = 1$. If instead
$R_c \gtrsim R_T$, we expect to observe a nontrivial
crossover behavior. Its determination is one of the purposes of the present
paper.

\section{Polymer model and crossover behavior} \label{sec3}

In order to determine full-monomer properties, 
we consider the three-dimensional lattice Domb-Joyce 
model \cite{DJ-72}. We consider $N_p$ chains of $L$ monomers each
on a finite cubic lattice of linear size $M$ with periodic boundary
conditions. Each polymer chain is modeled by a random walk
$\{{\mathbf r}_1^{(i)},\ldots,{\mathbf r}_L^{(i)}\}$ with
$|{\mathbf r}_\alpha^{(i)}-{\mathbf r}_{\alpha+1}^{(i)}|=1$ (we take the
lattice
spacing as unit of length) and
$1\le i \le N_p$. The Hamiltonian is given by
\begin{equation}
H = \sum_{i=1}^{N_p} \sum_{1\le \alpha < \beta \le L}
   \delta({\mathbf r}_\alpha^{(i)},{\mathbf r}_\beta^{(i)}) +
   \sum_{1\le i < j \le N_p} \sum_{\alpha=1}^L \sum_{\beta=1}^L
    \delta({\mathbf r}_\alpha^{(i)},{\mathbf r}_\beta^{(j)}),
\end{equation}
where $\delta({\mathbf r},{\mathbf s})$ is the Kronecker delta.
Each configuration is weighted by $e^{-w H}$, where $w > 0$ is a free
parameter that plays the role of inverse temperature.
This model is similar
to the standard lattice self-avoiding walk (SAW) model, which is
obtained in the limit  $w \to +\infty$. 
For any positive $w$, this model has the same scaling limit as the
SAW model \cite{DJ-72} and thus allows us to compute the
universal scaling functions that are relevant for polymer solutions
under good-solvent conditions. In the absence of colloids, there is a significant 
advantage in using Domb-Joyce chains instead of SAWs. For SAWs
scaling corrections that decay as $L^{-\Delta}$ ($\Delta = 0.528(12)$,
Ref.~\cite{Clisby-10}) are particularly strong, hence the 
universal, large--degree-of-polymerization limit is only observed for 
quite large values of $L$. Finite-density properties are those 
that are mostly affected by scaling corrections, and indeed it is 
very difficult to determine universal thermodynamic properties of polymer 
solutions for $\Phi\gtrsim 5$ by using lattice SAWs \cite{Pelissetto-08}.
These difficulties are overcome by using the Domb-Joyce model for a 
particular value of $w$ \cite{BN-97,CMP-08}, $w = 0.505838$. For this value 
of the repulsion parameter, the leading scaling corrections have a negligible
amplitude \cite{BN-97,CMP-08}, so that scaling corrections decay faster, 
approximately as $1/L$.  As a consequence, scaling results are obtained 
by using significantly shorter chains. Unfortunately, in the presence of 
a repulsive surface, new renormalization-group operators arise, which are 
associated with the surface \cite{DDE-83}. 
The leading one gives rise to corrections that scale as $L^{-\nu}$
\cite{DDE-83}, where $\nu$ is the Flory exponent (an explicit test
of this prediction is presented in the supplementary material), hence it 
spoils somewhat the nice scaling behavior observed in the absence of colloids.
Nonetheless, 
the Domb-Joyce model is still very convenient from a computational point
of view. Since interactions are soft, the Monte Carlo dynamics for 
Domb-Joyce chains 
is much faster than for SAWs. We shall use the algorithm 
described in Ref.~\cite{Pelissetto-08}, which allows one to obtain 
precise results for quite long chains ($L\gtrsim 1000$) deep in the 
semidilute regime.

The Domb-Joyce model can also be used to derive the crossover 
functions that parametrize the crossover between the good-solvent and 
$\theta$-point regimes, at least not too close to the $\theta$ point,
see Refs.~\cite{CMP-08,DPP-thermal}
for a discussion. Indeed, if one neglects tricritical effects,
which are only relevant close to the $\theta$ point \cite{Duplantier}, 
this crossover can be 
parametrized by using the two-parameter model 
\cite{dCJ-book,Schaefer-99,Sokal-94}. 
Two-parameter-model results are obtained \cite{BD-79} by taking the limit
$w\to 0$, $L\to \infty$ at fixed $x = w L^{1/2}$.
The variable $x$ interpolates between the ideal-chain limit ($x=0$) and
the good-solvent limit ($x=\infty$). Indeed, for $w = 0$ the Domb-Joyce model is
simply the random-walk model, while for any $w\not=0$ and $L\to \infty$
one always obtains the good-solvent scaling behavior. The variable
$x$ is proportional to the variable $z$ that is used
in the context of the two-parameter model. We normalize $z$ as in 
Refs.~\cite{CMP-08,DPP-thermal}, setting
\begin{equation}
  z \equiv \left({3\over 2\pi}\right)^{3/2} w L^{1/2}.
  \label{zdef-DJ}
\end{equation}
Note that the crossover can be equivalently parametrized 
\cite{Nickel-91,BN-97,PH-05,CMP-08,DPP-thermal} by using 
the second-virial combination
$A_{2,pp} = B_{2,pp} \hat{R}_g^{-3}$ ($\hat{R}_g$ is the 
zero-density radius of gyration), which varies between 
the good-solvent value \cite{CMP-06} $A_{2,pp} = 5.500(3)$ 
and $A_{2,pp} = 0$ at the $\theta$ point. 
With normalization (\ref{zdef-DJ})
we have $A_{2,pp}(z) \approx 4 \pi^{3/2} z$ for small $z$
\cite{BD-79,BN-97}. The correspondence between $A_{2,pp}$ and $z$ 
in the whole crossover region is given in Ref.~\cite{CMP-08}.

As discussed in Ref.~\cite{CMP-08}, the two-parameter-model 
results can be obtained
from Monte Carlo simulations of the Domb-Joyce model
by properly extrapolating the numerical results to $L\to\infty$.
For each $z$
we consider several chain lengths $L_i$. For each of them
we determine the interaction parameter $w_i$ by using Eq.~(\ref{zdef-DJ}),
that is we set
$w_i = (2\pi/3)^{3/2} z L_i^{-1/2}$. Simulations of chains of $L_i$
monomers are then performed setting
$w=w_i$. Simulation results are then extrapolated to $L\to \infty$,
taking into account that corrections are of order
$1/\sqrt{L}$ \cite{BD-79,BN-97}.

In this paper we have performed a detailed study of the depletion 
for two values of $z$: $z = z^{(1)} = 0.056215$ and 
$z = z^{(3)} = 0.321650$, which correspond to \cite{CMP-08}
$A_{2,pp}(z^{(1)}) = 0.9926(10)$ and
$A_{2,pp}(z^{(3)}) = 2.9621(27)$. They correspond to polymer solutions of 
intermediate quality. Since $A_{2,pp} \approx 5.50$ \cite{CMP-06} under
good-solvent conditions, we have $A_{2,pp}(z)/A_{2,pp}(z=\infty) = 0.18$
and 0.54 for $z = z^{(1)}$ and $z^{(3)}$, respectively. 
Hence, for $z = z^{(1)}$ we are quite close to the $\theta$ point, 
while $z = z^{(3)}$ is intermediate between the good-solvent and 
$\theta$ regimes. 

In this paper we discuss depletion effects close to neutral colloids, which
are modelled as hard spheres that can move everywhere in space: their centers
are not constrained to belong to a lattice point. This choice is particularly
convenient since it drastically reduces lattice oscillations in colloid-polymer
correlation functions. Such oscillations are instead present if colloids 
are required to sit on lattice points, as was done in Ref.~\cite{PH-06}. 
Colloids and monomers interact by means of a simple exclusion potential.
If ${\bf r}_c$ and ${\bf r}_m$ are the coordinates of a monomer and of a 
colloid, we take as interaction potential
\begin{eqnarray}
 U = +\infty && \qquad |{\bf r}_c - {\bf r}_m| \le R_c,  \\
 U = 0 && \qquad |{\bf r}_c - {\bf r}_m| > R_c  .
\end{eqnarray}

\section{Dilute behavior}  \label{sec4}

As we have seen in Sec.~\ref{sec2.3}, the
low-density behavior of the surface tension or, equivalently, of the 
depletion thickness can be obtained by computing
the virial coefficients $B_{2,cp}$ and $B_{3,cpp}$. 
We will thus report the computation of these two quantities 
and also of $B_{3,ccp}$, which would be relevant to characterize the 
effective interaction between two colloids in a dilute solution of 
polymers. Then, we shall
discuss the depletion thickness $\delta_s$ for $\Phi = 0$ and its first density 
correction.

\subsection{Virial coefficients} \label{sec4.1}

\begin{table}
\caption{Estimates of the universal surface combinations
$R_{1,p}$ and $R_{2,p}$. We report full-monomer (FM)
and single-blob (SB) results (see Sec.~\protect\ref{sec6}).}
\label{table:surface-int}
\begin{tabular}{ccccc}
\hline\hline
$z$ & $R_{1,p}$ (FM) &   $R_{2,p}$ (FM) & $R_{1,p}$ (SB) &
       $R_{2,p}$ (SB) \\
\hline
$\infty$ & 1.0605(3) & $-4.50(5)$  & 1.0514(2) & $-$4.44(1) \\
$z_3$    & 1.1066(1) & $-2.366(7)$  & 1.1065(3) & $-$2.387(8)\\
$z_1$    & 1.1221(4) & $-0.765(3)$  & 1.1222(3) & $-$0.771(4)\\
\hline\hline
\end{tabular}
\end{table}

To determine the virial coefficients under good-solvent conditions
we have simulated the Domb-Joyce model at $w = 0.505838$. We consider 
chains of length $L = 240, 600, 2400$ for $q\le 3$, and $L=6000$, 24000 
to derive the results corresponding to $4\le q\le 50$. Long chains are needed 
for large values of 
$q$ to ensure that the colloid radius is somewhat 
larger than the lattice spacing. 
Virial coefficients are determined as explained in App.~\ref{AppA}.
The universal extrapolations of the finite-$L$ results for the
adimensional combinations $A_{2,cp} = B_{2,cp} \hat{R}_g^{-3}$ 
and $A_{3,\#} = B_{3,\#} \hat{R}_g^{-6}$ are explicitly 
reported in the supplementary material.
In the case of the two-parameter model, we have
considered $L=120,240,600,1200,2400$ for $q\le 3$ (for both $z = z^{(1)}$ and 
$z = z^{(3)}$) and 
$L = 6000$, 30000 for $4\le q \le 30$ (only for $z = z^{(1)}$). The results at 
the same value of $z$ have then been extrapolated taking into account the 
$L^{-1/2}$ scaling corrections.
Results are reported in the supplementary material.
We also computed the 
adimensional combinations $R_{1,p} = P_{1,p} \hat{R}_g^{-1}$
and $R_{2,p} = P_{2,p} \hat{R}_g^{-4}$, 
which parametrize the depletion thickness in 
the presence of an impenetrable planar surface, 
see Table \ref{table:surface-int}.
The behavior of the adimensional combinations for $q\to 0$ is discussed
in detail in Appendix~\ref{AppB}. We have
\begin{eqnarray}
A_{2,cp} &\approx& {4\pi\over 3 q^3} + {4\pi\over q^2} R_{1,p}, 
\label{A2cp-largeRc} \\
A_{3,cpp} &\approx& {8\pi\over 3 q^3} A_{2,pp} + {4\pi\over q^2} 
\left( 2 A_{2,pp} R_{1,p} + R_{2,pp}\right),
\label{A3cpp-largeRc} \\
A_{3,ccp} &\approx& {16\pi^2\over 9 q^6}.
\label{A3ccp-largeRc}
\end{eqnarray}
\begin{figure}[t]
\begin{center}
\begin{tabular}{c}
\epsfig{file=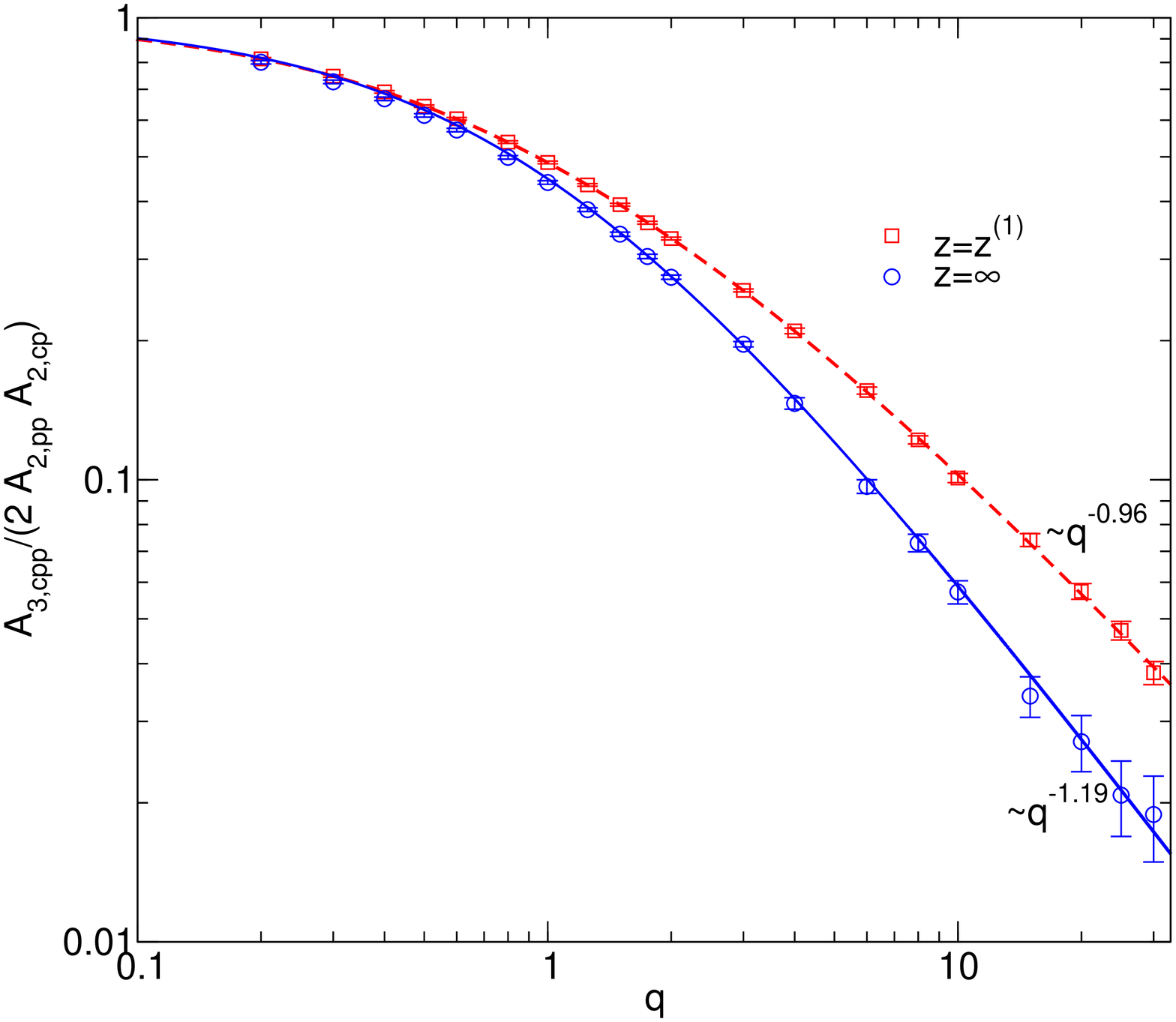,angle=0,width=9truecm} \hspace{0.5truecm} \\
\end{tabular}
\end{center}
\caption{Combination ${A_{3,cpp}/(2 A_{2,cp} A_{2,pp})}$ vs $q$.
For $q\to 0$, such a combination converges to 1 for all values of $z$.
We also report the approximate large-$q$ behavior.
}
\label{fig:ratioA3A2}
\end{figure}
For large values of $q$, we have $A_{2,cp} \sim q^{1/\nu-3}$, 
a behavior which can be derived by means of a blob argument \cite{PH-06} 
or from the large-$q$ behavior of $\gamma$, 
as discussed in Sec.~\ref{sec2.4}. More precisely, we predict
$A_{2,cp} = 4 \pi A_{\gamma,\infty} q^{1/\nu-3}$, 
where $A_{\gamma,\infty}$ is the constant parametrizing the large-$q$ 
behavior of $\gamma$, defined by Eq.~(\ref{gamma-largeq}).
Note that this relation holds both in the good-solvent regime with 
$\nu \approx 0.5876$ and in the crossover regime with $\nu = 1/2$.
As for the 
third virial coefficient $B_{3,cpp}$, we have already shown that 
$B_{3,cpp}/B_{2,cp}$ vanishes as $q\to \infty$. Hence, we expect
\begin{equation} 
{A_{3,cpp}\over A_{2,cp} } \sim q^{-\alpha},
\end{equation}
with $\alpha > 0$ for $q\to \infty$. 
We have been unable to predict the value of $\alpha$.
A numerical fit of the data indicates $\alpha \approx 1$, both in the 
good-solvent limit
and in the crossover region, see Fig.~\ref{fig:ratioA3A2}. As for
$A_{3,ccp}$, a blob argument implies 
$A_{3,ccp} \sim q^{1/\nu - 6}$. 

Since knowledge of the virial coefficients for all values of $q$ allows us to 
have a complete control of the depletion effects in the dilute regime, it 
is useful to determine interpolations of the data, with the correct 
limiting behavior for $q\to0$ and $q\to\infty$. 
We parametrize the data as 
\begin{eqnarray}
A_{2,cp}&=& \frac{4\pi }{3q^3}
  \left[ \frac{1+a_1 q + a_2 q^2+a_3 q^3}{1+a_4 q}
  \right]^{1/(2\nu)}, 
\label{A2cp-param}
\\
A_{3,cpp}&=&
 \frac{8\pi A_{2,pp}}{3q^3}\left[ \frac{1+a_1 q + a_2 q^2}{1+a_3 q}
\right]^\beta ,
\label{A3cpp-param}
\\
A_{3,ccp}& = & 
\left(\frac{4\pi}{3q^3}\right)^2(1+a_1q+a_2 q^2+\dots +a_n q^n)^{1/(n \nu)},
\end{eqnarray}
We enforce the asymptotic behaviors (\ref{A2cp-largeRc}), 
(\ref{A3cpp-largeRc}), and (\ref{A3ccp-largeRc}) for $q\to 0$. 
In the case of $A_{2,cp}$ and $A_{3,ccp}$  we have chosen the parametrization
so to obtain the correct large-$q$ behaviors 
$A_{2,cp} \sim q^{1/\nu-3}$ and $A_{3,ccp} \sim q^{1/\nu-6}$ 
($\nu = 0.5876$ for the good-solvent case and 
$\nu=1/2$ for $z=z^{(1)}$ and $z^{(3)}$).
In the case of $A_{3,cpp}$, 
$\beta$ is a free parameter. Fitting the data, we 
estimate the constants $a_i$. They are reported in Table
\ref{tab:VirialPar}.

\begin{table}[t!]
\caption{Coefficients parametrizing the universal virial combinations.
For $A_{3,ccp}$ we use $n=3$ for $z=\infty$ and $z^{(3)}$, and $n=4$ for 
$z = z^{(1)}$.
The interpolation is accurate in the range $0\le q \le q_{\rm max}$ in which 
we have data.}
\label{tab:VirialPar}
\begin{center}
\begin{tabular}{ccccccccc}
\hline\hline
& $z$ & $a_1$ & $a_2$ & $a_3$ & $a_4$ & $\beta$ & $q_{\rm max}$\\
\hline
\hline
$A_{2,cp}$ & $\infty$ & 4.1329 & 5.4906 &  2.12578 & 0.3942 &  & 50 \\
& $z_3$ & 3.4774 & 3.37453 & 0.39752 & 0.15763 &  & 3\\
& $z_1$ & 3.4378 & 3.18934 & 0.20253 & 0.071526 & & 30\\
$A_{3,ccp}$ & 
$\infty$ &  12.9575 & 39.2297  & 152.514& &  & $10$\\
& $z_3$ & 7.66551 & 55.4536 & 43.4239 & &  &  $3$\\
& $z_1$ & 23.1533 & 0.0000 & 421.593 & 100.977 &  & $6$\\
$A_{3,cpp}$ &
$\infty$ & 4.0850 & 5.0910  & 0.296425 & & 0.51574 & $20$ \\
& $z^{(3)}$ & 3.36452 & 3.0418 & 1.3236 & & 1.040 & $3$\\
& $z^{(1)}$ & 2.55348 & 1.2711 & 0.42515 & & 1.038 & $30$\\
\hline\hline
\end{tabular}
\end{center}
\end{table}

Using parametrization (\ref{A2cp-param}), we can compute the large-$q$
behavior of $A_{2,cp}$. In the good-solvent case we obtain 
$A_{2,cp} \approx 17.57 q^{1/\nu-3}$. 
Since $A_{2,cp} = 4 \pi A_{\gamma,\infty} q^{1/\nu-3}$, 
we can estimate the constant $A_{\gamma,\infty}$ which appears
in Eq.~(\ref{gamma-largeq}). We obtain $A_{\gamma,\infty} \approx 1.40$, 
which is in excellent agreement with the field-theoretical estimate
1.41 of Ref.~\cite{HED-99}. For $z = z^{(1)}$ we obtain instead 
$A_{2,cp} \approx 11.9/q$. Since
$A_{2,cp} = 4 \pi A_{\gamma,\infty}/q$ we obtain 
$A_{\gamma,\infty} = 0.94$, which is close to the prediction 
$A_{\gamma,\infty} = 1$ of Sec.~\ref{sec2.4}.

\subsection{Zero-density depletion thickness} \label{sec4.2}

\begin{figure}[t]
\begin{center}
\begin{tabular}{c}
\epsfig{file=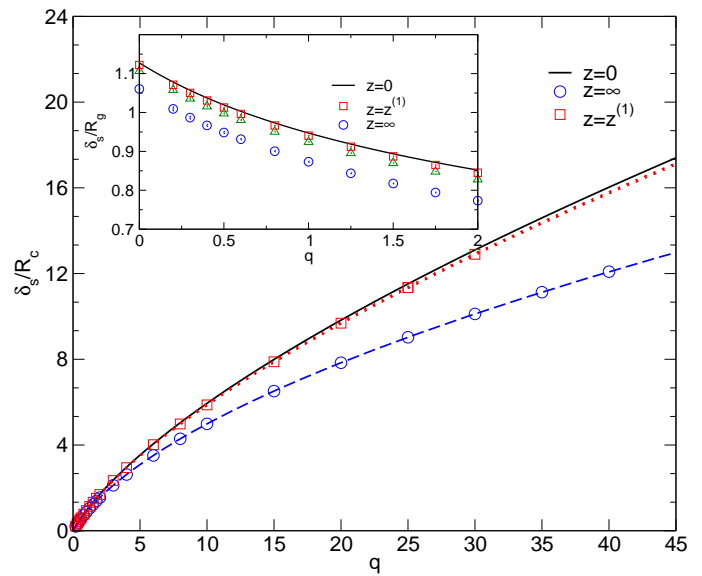,angle=0,width=9truecm} \hspace{0.5truecm} \\
\end{tabular}
\end{center}
\caption{Depletion thickness ratio $\delta_s/R_c$ vs $q$ for $q\le 40$
and $\Phi = 0$. In the 
inset we report $\delta_s/\hat{R}_g$ 
in the interval $0\le q\le 2$. The dotted and dashed lines that go through 
the points in the main panel are obtained by using interpolation
(\ref{A2cp-param}).
}
\label{fig:deltas-Phi0}
\end{figure}

Knowledge of $A_{2,cp}$ allows us to compute the depletion thickness in
the zero-density limit by using 
Eq.~(\ref{deltas-smallrhop}). In Fig.~\ref{fig:deltas-Phi0} we report 
our results. For $q\lesssim 2$, $\delta_s/\hat{R}_g$ has a tiny 
dependence on $z$: It slightly increases
as $z$ decreases, and for $z=z^{(3)}$ and $z=z^{(1)}$ it is very close to the 
ideal-case result. For the surface case, these small differences
can be appreciated by
looking at the results given in Table~\ref{table:surface-int}, since 
$\delta_s/\hat{R}_g = R_{1,p}$ ($\delta_s/\hat{R}_g = 1.128$ for $z=0$). 
The approximate independence of
$\delta_s/\hat{R}_g$ on $z$ implies
that the $z$-dependence of $\delta_s$ and of $\hat{R}_g$ are approximately 
the same: When $q$ is small, depletion effects are simply proportional 
to the typical size of the polymer and do not depend significantly
on the quality of the solution.
These considerations are valid only for $q$ not too large.
For large values of $q$, significant differences between the good-solvent 
and the finite-$z$ case are observed, since
the depletion thickness has a different asymptotic behavior for $q\to \infty$.
Indeed, while $\delta_s/R_c\sim q^{2/3}$ for any finite $z$ as discussed
in Sec.~\ref{sec2.4}, we have 
$\delta_s/R_c\sim q^{1/3\nu} \sim q^{0.567}$ in the good-solvent case.

To obtain a more quantitative comparison in the colloid regime, 
we can determine the small-$q$ behavior of $\delta_s(z)/\hat{R}_g$
by expanding parametrization (\ref{A2cp-param}) in powers of $q$.
We obtain 
\begin{eqnarray}
\frac{\delta_s(\infty)}{\hat{R}_g} &=&
1.0605 -0.281 q+0.140 q^2 + \dots  
\label{deltas-Phi0-RG} \\
\frac{\delta_s(z^{(3)})}{\hat{R}_g}&=&
     1.107 -0.274  q+ 0.138  q^2 + \dots \\
\frac{\delta_s(z^{(1)})}{\hat{R}_g}&=&
     1.122 -0.276  q+ 0.146  q^2 + \dots \\
\frac{\delta_s(0)}{\hat{R}_g}&=&
    1.128-0.273 q+0.138 q^2 +\dots 
\end{eqnarray}
Results for $z=z^{(1)}$ cannot be distinguished from the ideal ones.
Also the results for $z = z^{(3)}$ are very close to those corresponding to
$z = 0$.  Slightly larger differences are observed for the good-solvent case.

\begin{figure}[t]
\begin{center}
\begin{tabular}{c}
\epsfig{file=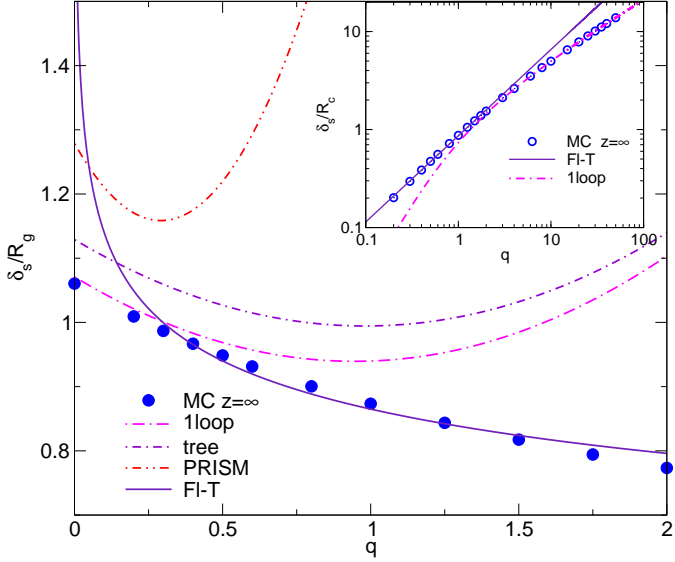,angle=0,width=9truecm} \hspace{0.5truecm} \\
\end{tabular}
\end{center}
\caption{Comparison of the variour predictions for the depletion thickness 
at zero density in the good-solvent regime.
We report the Monte Carlo estimates of $\delta_s/\hat{R}_g$ 
and (inset) $\delta_s/{R}_c$ versus $q$.
We also report: the PRISM prediction obtained by using 
Eq.~(\ref{PRISM}) (PRISM), Eq.~(\ref{FT-deltas}) (Fl-T), 
the field-theoretical Helfrich expansions at tree level (tree) and at one loop
(1loop), Eq.~(\ref{Helfrich-ds-FT}). 
In the inset we report (Fl-T) Eq.~(\ref{FT-deltas}) and the large-$q$
field-theory prediction $\delta_s/R_c = 1.62 q^{0.567}$ (1loop). 
}
\label{fig:deltas-comparison}
\end{figure}

In the good-solvent case,
we can compare our estimates of $\delta_s$ with the field-theoretical 
predictions \cite{HED-99,MEB-01}. For small 
values of $q$ we report the tree-level result, which can be
derived from Eq.~(\ref{FT-tree-Phi0}), and the one-loop result 
obtained from Eq.~(\ref{FT-Phi0}): 
\begin{eqnarray}
&& \left(\frac{\delta_s}{\hat{R}_g}\right)_{\rm tree}=1.13 -0.27 q + \dots 
\nonumber \\
&& \left(\frac{\delta_s}{\hat{R}_g}\right)_{\rm 1loop}=
1.07 q -0.28 q^2+0.18 q^3 + \dots 
\label{Helfrich-ds-FT}
\end{eqnarray}
Comparison with the Monte Carlo prediction (\ref{deltas-Phi0-RG}) 
shows that differences are tiny. Moreover, it is very reassuring that
loop corrections correctly change the values of the Helfrich coefficients
towards the numerically determined values.
For large values of $q$, we have 
$\delta_s/R_c \approx (3 A_{\gamma,\infty} q^{1/\nu})^{1/3}$. 
The Monte Carlo results imply $\delta_s/R_c \approx 1.61 q^{1/(3 \nu)}$,
while field theory predicts
$\delta_s/R_c \approx 1.62 q^{1/(3 \nu)}$. Again, field theory appears
to work very nicely. Other predictions are compared in 
Fig.~\ref{fig:deltas-comparison}. As already discussed, 
Eq.~(\ref{PRISM}) gives only a very rough approximation that fails 
completely for $q\gtrsim 0.7$. The phenomenological expression
(\ref{FT-deltas}), instead, provides a quite good approximation
in a quite large intermediate range, from $q\approx 0.2$ 
up to $q \approx 4$. The approximation
fails in the planar limit---it predicts $\delta_s = \infty$ for 
$q\to 0$---and for large values of $q$, as it predicts $\delta_s \sim q^{0.88}$,
while the correct behavior is $\delta_s\sim q^{0.567}$.

\subsection{Density correction to the depletion thickness} \label{sec4.3}

\begin{figure}[t]
\begin{center}
\begin{tabular}{c}
\epsfig{file=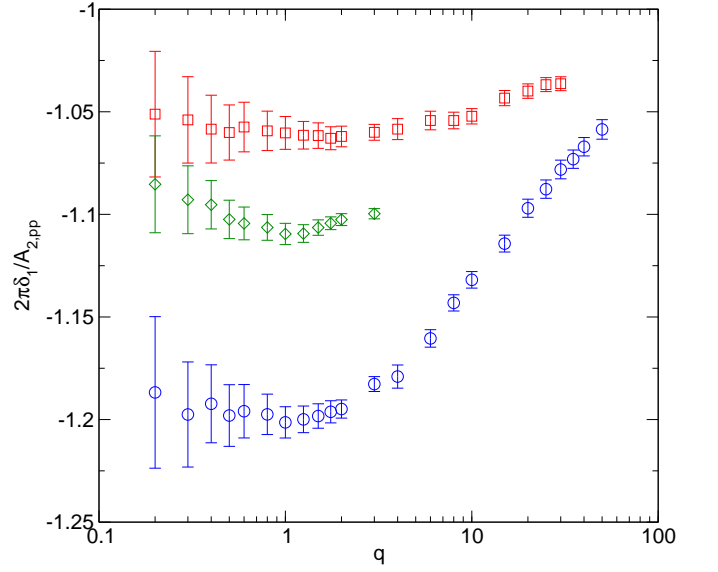,angle=0,width=9truecm} \hspace{0.5truecm} \\
\end{tabular}
\end{center}
\caption{Density-correction combination $2\pi \delta_1(q)/A_{2,pp}$ 
versus $q$ for $z = z^{(1)}$, $z^{(3)}$ and the good-solvent case $z = \infty$.
For $q=0$, $2\pi \delta_1(q)/A_{2,pp} = -1.030(4)$, $-$1.083(3),
$-$1.16(2) for the same values of $z$.
}
\label{fig:deltas-corr}
\end{figure}

Knowledge of the third virial coefficient $A_{3,cpp}$ allows one to 
determine the first density correction to $\delta_s(q,\Phi)$, see
Eq.~(\ref{deltas-smallrhop}). We define
\begin{equation}
{\delta_s(q,\Phi)\over \delta_s(q,0)} = 1 
     + \delta_1(q)\Phi + O(\Phi^2).
\label{delta1-def}
\end{equation}
For $q\to\infty$, since $A_{3,cpp}/A_{2,cp}\to 0$ in this limit,
Eq.~(\ref{deltas-smallrhop}) implies 
$\delta_1(q) \to - A_{2,pp}/(2 \pi)$. In Fig.~\ref{fig:deltas-corr} 
we report the combination $2\pi \delta_1(q)/A_{2,pp}$, which converges to 
$-1$ for $q\to\infty$ and any value of $z$. It is evident that there are two
different regimes. For $q$ not too large---$q\lesssim 2$ and $q\lesssim 10$ for 
the good-solvent case and for $z = z^{(1)}$, respectively---the density 
correction is mostly independent of $q$. In the opposite limit ($q$ large), 
a more pronounced $q$ dependence is observed, related to the fact that 
$2 \pi\delta_1(q)/A_{2,pp}$ always converges to $-1$ as $q\to\infty$.
The $z$ dependence of the combination $2\pi \delta_1(q)/A_{2,pp}$
is not large: it changes by at most 5\% as $z$ increases from $z^{(1)}$ 
to $z^{(3)}$ and by 10\% at most from $z^{(3)}$ to $\infty$ (the good-solvent
case). Hence, a rough approximation for $\delta_1(q)$ is simply
$\delta_1(q) = - A_{2,pp}/2\pi$, which relates directly solution quality
to depletion effects. The quality of this approximation improves as $z$ 
decreases.

We can compare our good-solvent results with several predictions that hold 
for small values of $q$. Field theory \cite{MEB-01}, see
Eq.~(\ref{FT-tree-Phi0}), gives
\begin{equation}
 \delta_1(q) = -1.23 - 0.015 q + \ldots
\end{equation}
The value for $q=0$ is not far from the numerical estimate 
$\delta_1(0) = 3 R_{2,pp}/(4\pi R_{1,p}) = -1.013(14)$, 
indicating that the renormalized tree-level approximation reasonably
predicts the low-density behavior in the colloid regime. Moreover, the 
leading $q$ 
correction is negative and very small, in agreement with our results: the $q$ 
dependence of $\delta_1(q)$ is tiny for $q\to 0$. 

For $q= 0$, 
we can also use the PRISM prediction (\ref{PRISM}) and the numerical 
expression (\ref{Gamma-planar}) for $\Gamma$ of Ref.~\cite{LBMH-02-a}.
We obtain in the two cases $\delta_1(0) = -0.51$
and $-1.96$, respectively. None of the two expressions 
appears to provide an accurate estimate of $\delta_1(q)$ for $q=0$.

\section{Finite-density results} \label{sec5}

\subsection{Numerical determination of $G_{cp}$} \label{sec5.1}

Let us now determine the depletion behavior 
at finite polymer density. 
For this purpose we perform finite-density simulations 
of the Domb-Joyce model in a cubic box 
in the presence of a single colloid and compute 
the density profile 
$\rho_{\rm mon}(r)$, which gives the density of monomers 
at distance $r$ from the colloid, and 
the analogous density $\rho_{CM}(r)$, 
which gives the density of polymer centers of mass. 
To compute $g_{{\rm mon},cp}(r)$ and 
$g_{CM,cp}(r)$ we should determine first the bulk polymer (or monomer) density. 
We proceed as follows. If the cubic 
box of volume $V=M^3$ contains $N_p$ polymers of 
$L$ monomers each, for each distance $\Lambda < M/2$ we define an 
effective bulk monomer density
\begin{equation}
\rho_{{\rm mon},b}(\Lambda) = 
{1\over V - V_\Lambda} \left(L N_p - \int_{r\le \Lambda} 
        d{\bf r}\, \rho_{\rm mon}(r) \right),
\end{equation}
where $V_\Lambda = {4 \pi\over3} \Lambda^3$. The quantity 
$\rho_{{\rm mon},b}(\Lambda)$ gives the average 
monomer density outside a sphere of 
radius $\Lambda$ centered on the colloid.
As a function of $\Lambda$, $\rho_{{\rm mon},b}(\Lambda)$ first increases,
then shows an approximate plateau, and finally shows a systematic upward or 
downward drift with a large statistical error. 
We take the approximately constant 
value of $\rho_{\rm mon}(\Lambda)$ in the plateau as 
an estimate of the bulk monomer density. Then, we estimate
$g_{{\rm mon},cp}(r) = \rho_{\rm mon}(r)/\rho_{{\rm mon},b}(\Lambda)$ and
\begin{equation}
 G_{cp} = \int_{r\le \Lambda} d{\bf r}\, 
    \left( g_{{\rm mon},cp}(r) - 1\right).
\end{equation}
The same calculation, {\em mutatis mutandis}, has been performed for the 
colloid polymer--center-of-mass distribution function. 

As a check, we computed $G_{cp}$ by using a third method. If 
$\hat{g}_{{\rm mon},cp}({\bf k})$ is
the Fourier transform of the pair distribution function,
the integral $G_{cp}$ can be computed as
\begin{eqnarray}
   G_{cp} = \lim_{k\to 0} \hat{g}_{{\rm mon},cp}({\bf k}).
\end{eqnarray}
Such a definition is much less sensitive to the definition of the bulk monomer 
density, but requires an extrapolation in $k$. 
Since we are considering a cubic box,
it is natural to restrict the calculation to ${\bf k} = (k,0,0)$ (or to 
$(0,k,0)$ and $(0,0,k)$, which are equivalent by symmetry). 
For $k\not=0$ the function $\hat{g}_{{\rm mon},cp}({\bf k})$ admits 
an expansion in powers of $k^2$, i.e.
\begin{equation}
  \hat{g}_{{\rm mon},cp}({\bf k}) = 
  G_{cp} + a_1 k^2 + a_2 k^4 + a_3 k^6 + \ldots
\label{expan-hatgk}
\end{equation}
To estimate $G_{cp}$,
we consider the smallest momenta available for a finite box of 
volume $V = M^3$, i.e.,
$k_1 = 2\pi/M$, $k_2 = 2 k_1$, $k_3 = 3 k_1$, $k_4 = 4 k_1$,
and the approximants
\begin{eqnarray}
G_{cp}^{(1)} &=& {4\over 3}\hat{g}_{{\rm mon},cp}({\bf k}_1) - 
                 {1\over 3}\hat{g}_{{\rm mon},cp}({\bf k}_2),
\label{Gncp} \\
G_{cp}^{(2)} &=& {3\over 2}\hat{g}_{{\rm mon},cp}({\bf k}_1) - 
                 {3\over 5}\hat{g}_{{\rm mon},cp}({\bf k}_2) + 
                 {1\over 10}\hat{g}_{{\rm mon},cp}({\bf k}_3),
\nonumber \\
G_{cp}^{(3)} &=& {8\over 5}\hat{g}_{{\rm mon},cp}({\bf k}_1) - 
                 {4\over 5}\hat{g}_{{\rm mon},cp}({\bf k}_2) 
\nonumber \\
   && + {8\over 35}\hat{g}_{{\rm mon},cp}({\bf k}_3) - 
                 {1\over 35}\hat{g}_{{\rm mon},cp}({\bf k}_4).
\nonumber 
\end{eqnarray}
Using Eq.~(\ref{expan-hatgk}), it is easy to show that 
$G_{cp}^{(n)} = G_{cp} + O(M^{-2n-2})$.
Note that we do not consider the volume corrections (of order $1/V = M^{-3}$
see, e.g., Ref.~\cite{LP-61}), which affect
$\hat{g}_{\rm mon}({\bf k})$ at fixed $k$. For the typical volumes we consider, 
such corrections are negligible (see Ref.~\cite{DPP-thermal} 
for the analogous discussion
concerning the polymer-polymer distribution function). 
On the other hand, we observe
a systematic difference between $G^{(1)}_{cp}$ and $G^{(2)}_{cp}$, while 
$G^{(2)}_{cp}\approx G^{(3)}_{cp}$ in all cases. 
Clearly, the $M^{-4}$ corrections that are present when considering 
$G^{(1)}_{cp}$ are not negligible. Therefore, we take $G^{(2)}_{cp}$ 
as the estimate of $G_{cp}$. 

\subsection{Colloid-monomer pair distribution functions} \label{sec5.2}

\begin{figure}[t]
\begin{center}
\begin{tabular}{c}
\epsfig{file=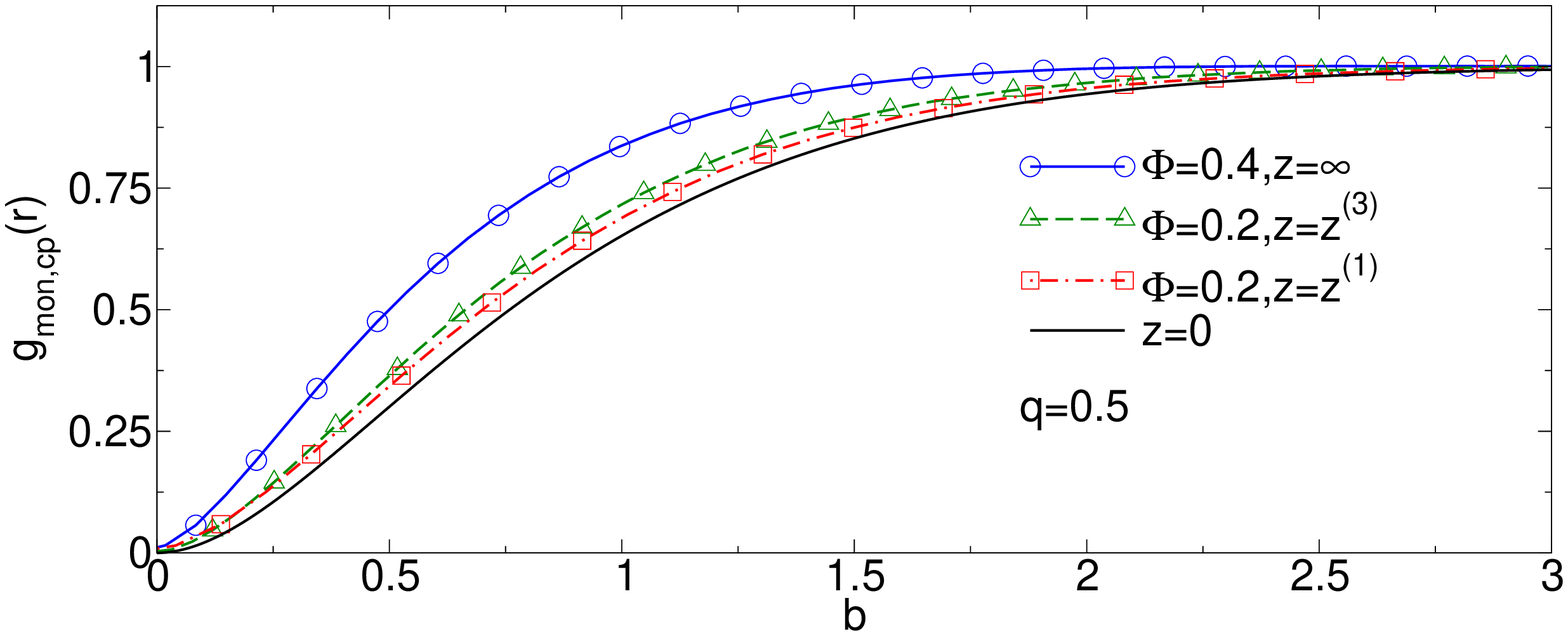,angle=0,width=9truecm} \hspace{0.5truecm} \\
\epsfig{file=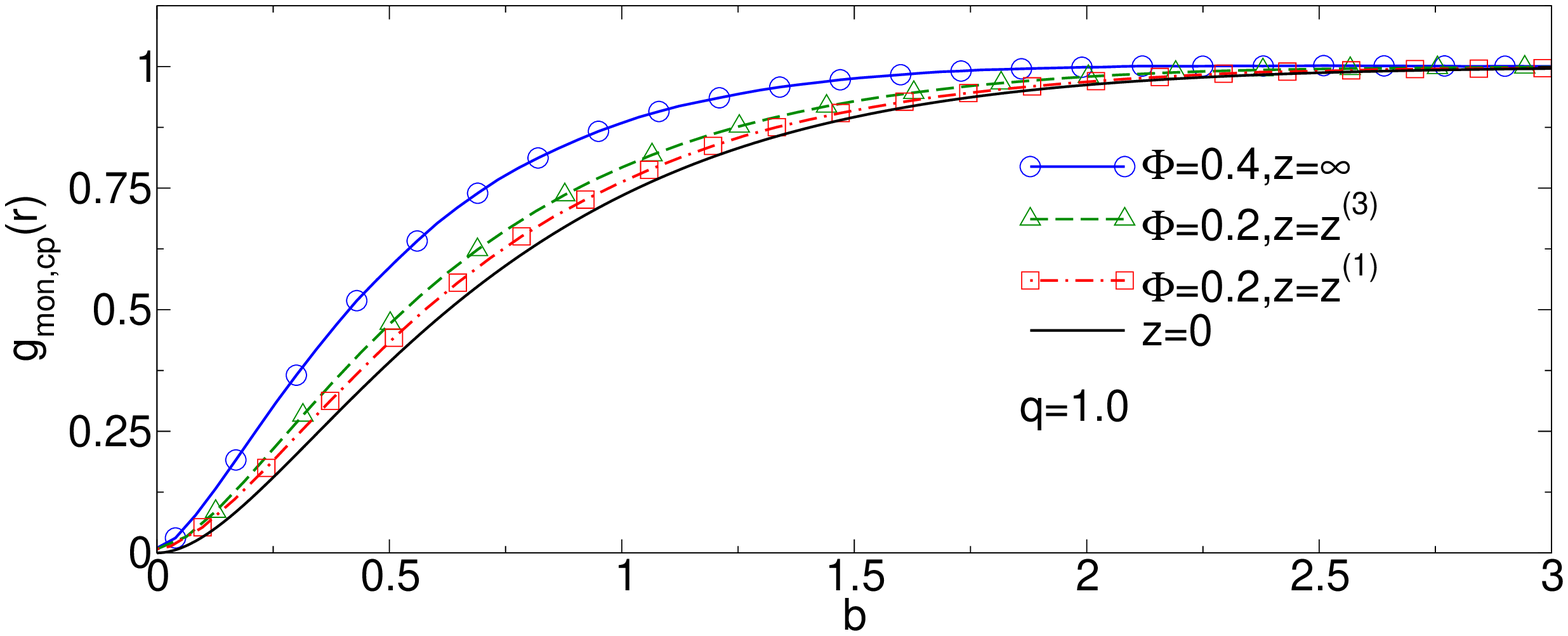,angle=0,width=9truecm} \hspace{0.5truecm} \\
\epsfig{file=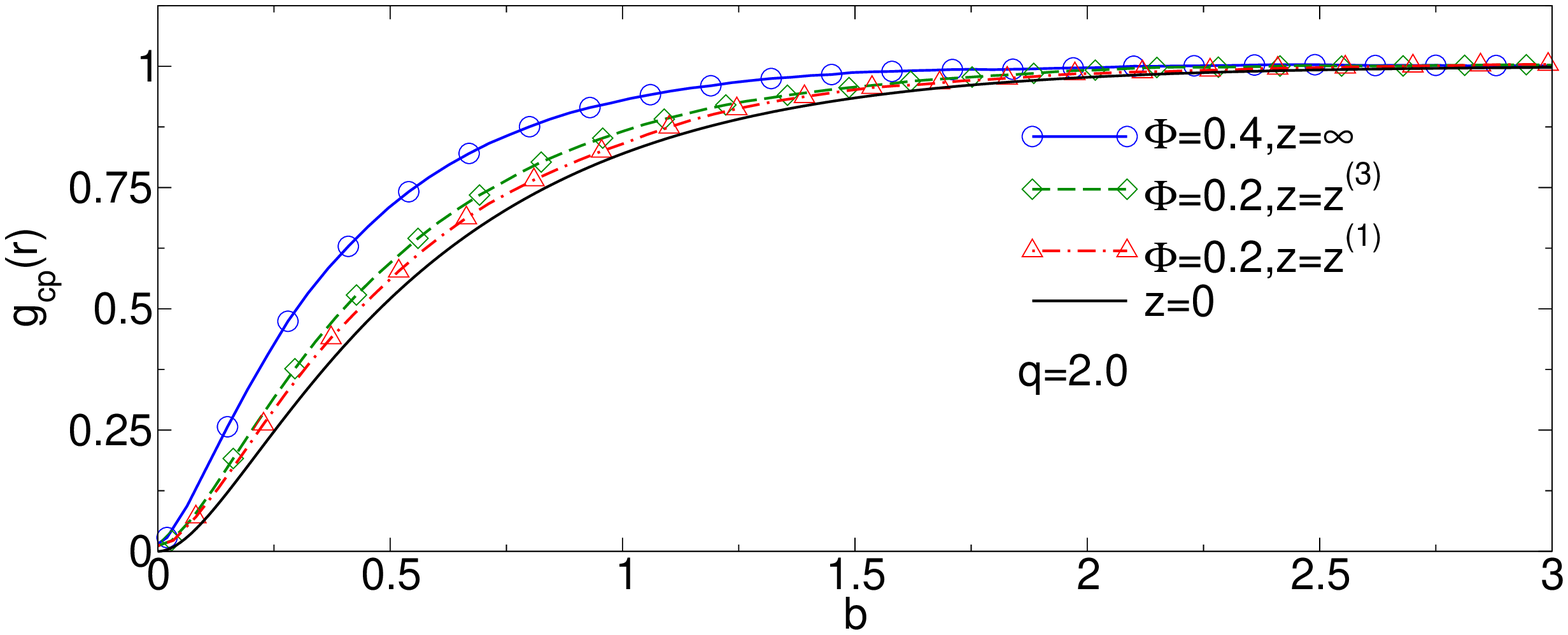,angle=0,width=9truecm} \hspace{0.5truecm} \\
\end{tabular}
\end{center}
\caption{Pair distribution function $g_{{\rm mon},cp}(r)$ 
as a function of $b = (r-R_c)/\hat{R}_g$ for different values of 
$q$, $z$, and $\Phi$ in the dilute regime. 
Eq.~(\ref{gmon-z0}) is used for $z = 0$.
}
\label{fig:gmon-smallPhi}
\end{figure}

\begin{figure}[t]
\begin{center}
\begin{tabular}{c}
\epsfig{file=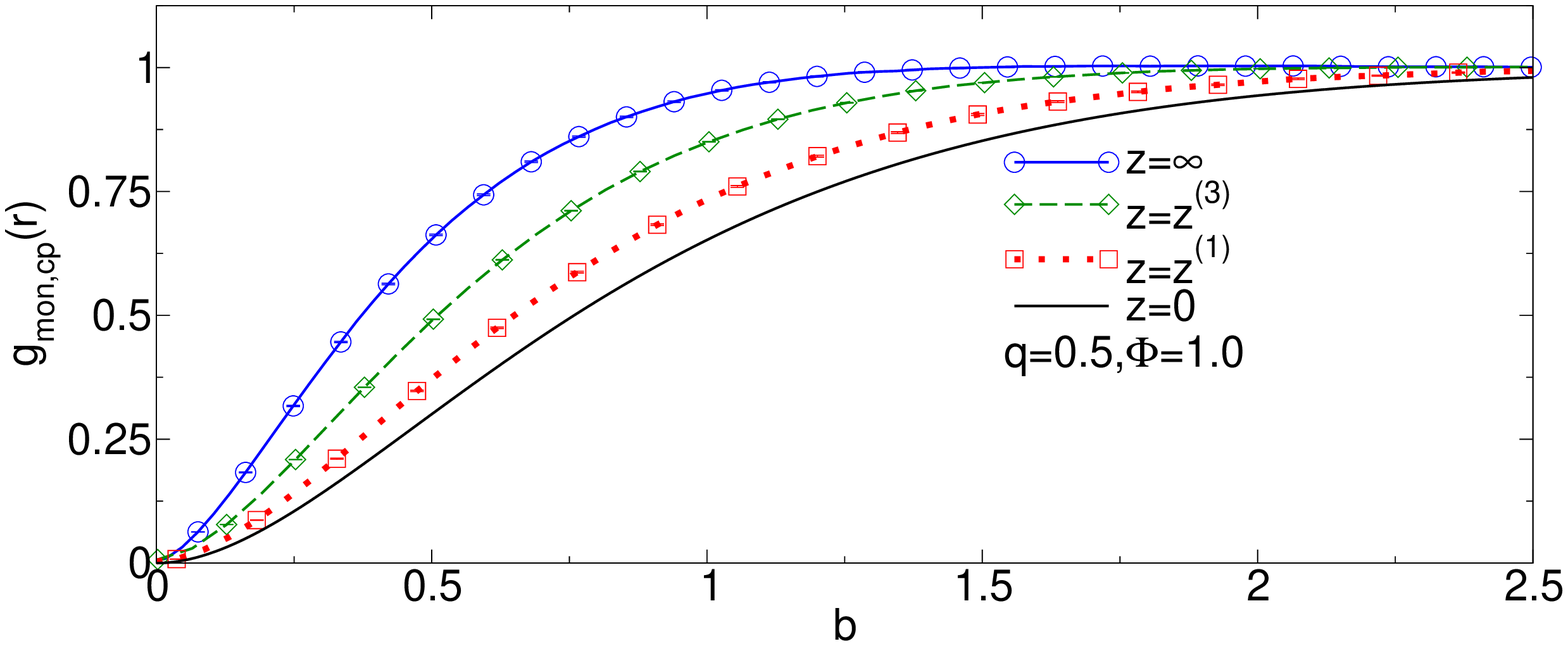,angle=0,width=9truecm} \hspace{0.5truecm} \\
\epsfig{file=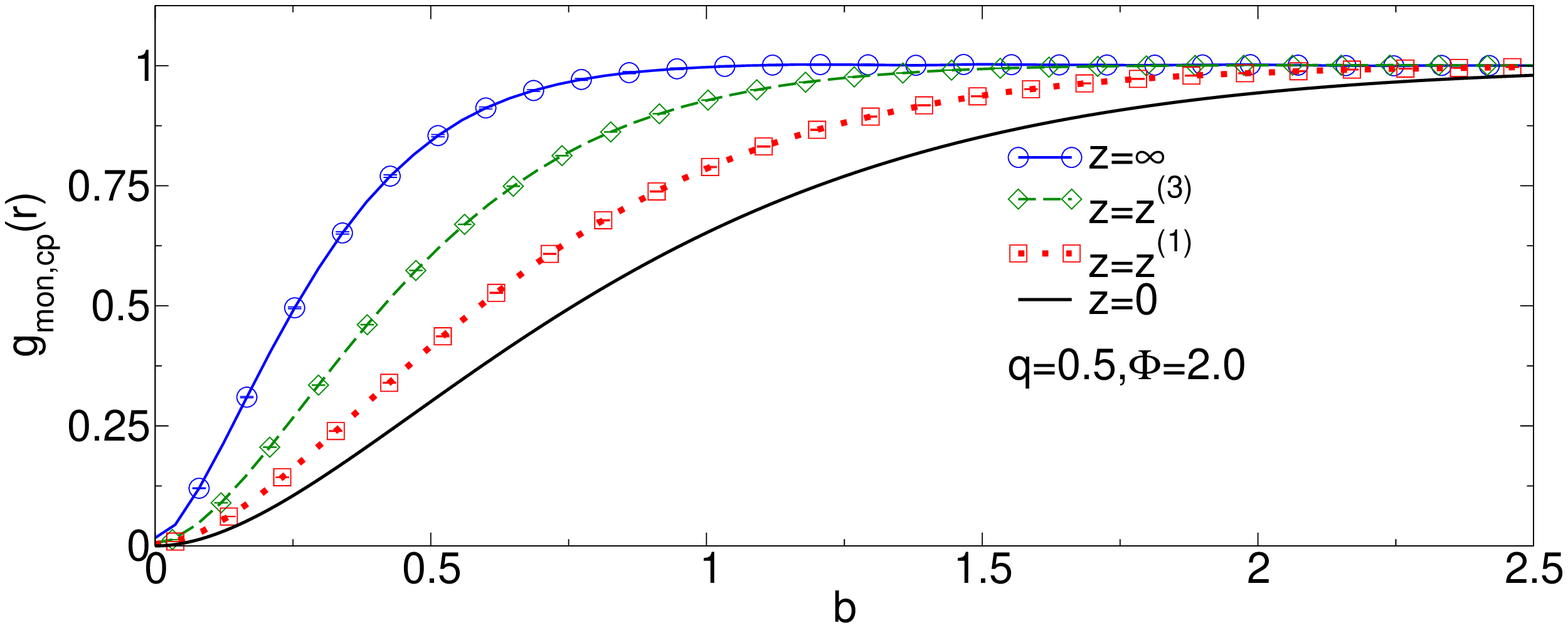,angle=0,width=9truecm} \hspace{0.5truecm} \\
\epsfig{file=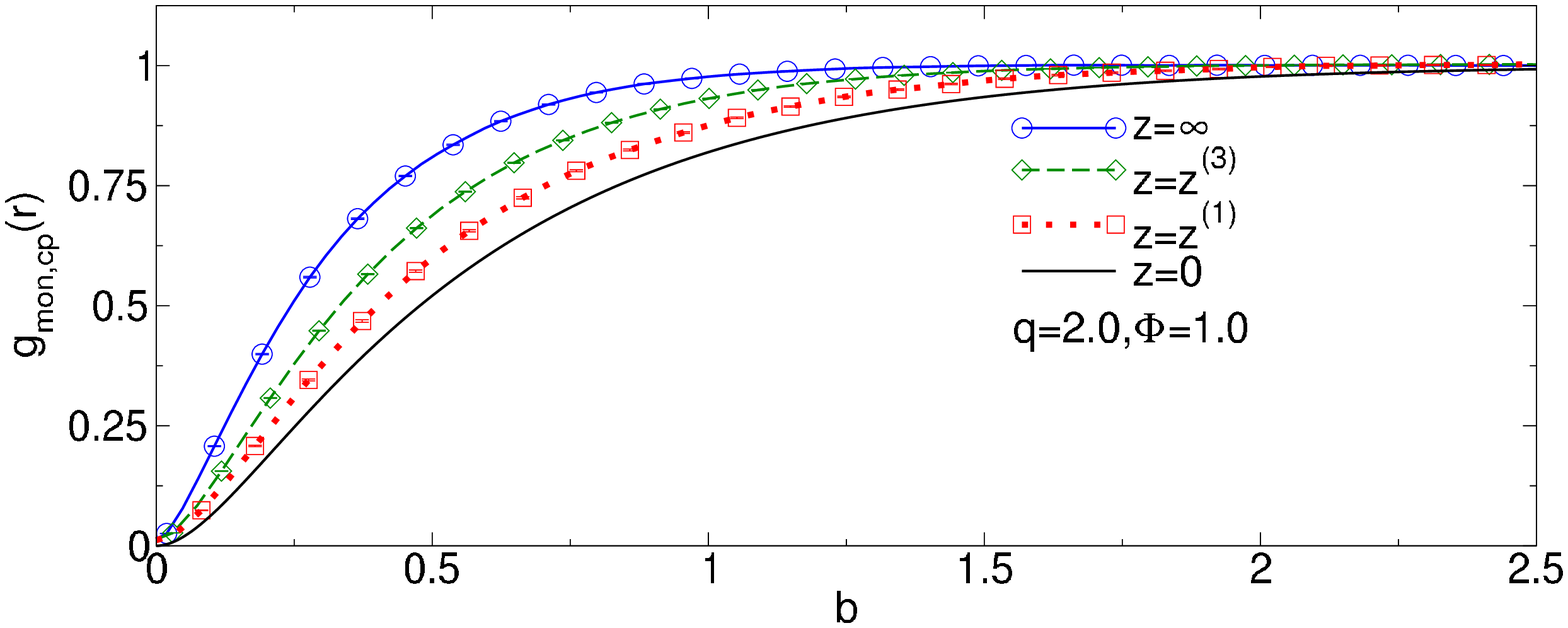,angle=0,width=9truecm} \hspace{0.5truecm} \\
\epsfig{file=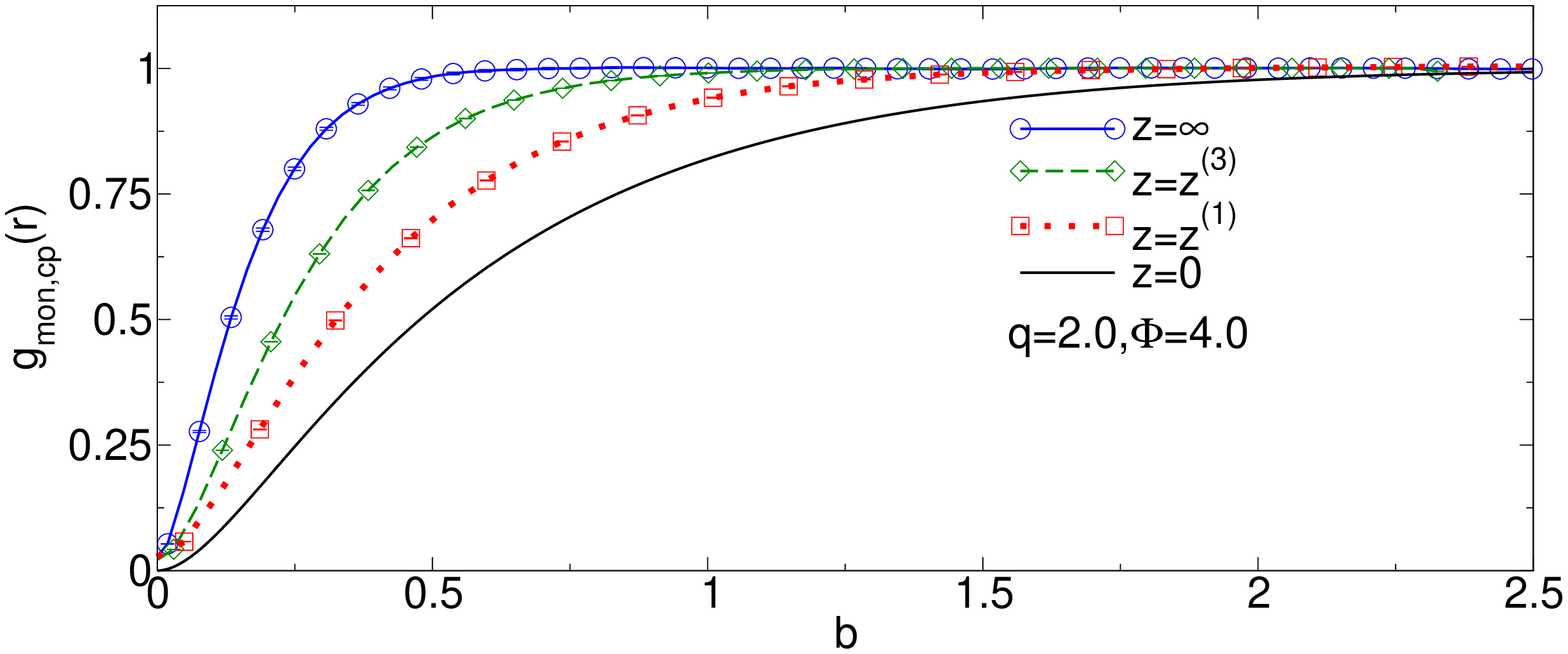,angle=0,width=9truecm} \hspace{0.5truecm} \\
\end{tabular}
\end{center}
\caption{Pair distribution function $g_{{\rm mon},cp}(r)$ 
as a function of $b = (r-R_c)/\hat{R}_g$ in the semidilute regime.
Eq.~(\ref{gmon-z0}) is used for $z = 0$.
}
\label{fig:gmon-Phi}
\end{figure}

We study the solvation properties of a single colloid 
in the semidilute regime 
for $q = 0.5$, 1 and 2, considering the good-solvent case and 
two values of $z$, $z = z ^{(1)}$ and $z = z^{(3)}$, in the 
thermal crossover region. 
In each case we compute numerically the pair correlation functions 
$g_{{\rm mon},cp}(r)$, $g_{CM,cp}(r)$, and the Fourier transform 
$\hat{g}_{{\rm mon},cp}(k)$ for the values of $k$ that are relevant for the 
computation of the approximants (\ref{Gncp}) for a few values of 
$\Phi$, up to $\Phi = 4$. We also present good-solvent results for the 
surface case ($q=0$) up to $\Phi = 8$. In this case, however, we have only
measured the monomer density profile close to the surface.

The function $g_{{\rm mon},cp}(r)$ is shown in Fig.~\ref{fig:gmon-smallPhi} 
as a function of $b = (r-R_c)/\hat{R}_g$ for the lowest values 
of $\Phi$ we have considered, together with expression
\cite{EHD-96}
\begin{eqnarray}
g_{{\rm mon},cp}(r) &= &
  \frac{ q^2 b^2+2bq \psi(b/2)+f(b/2)}{\left(bq+1\right)^2},
\label{gmon-z0} \\
 f(x)&=& 2\psi(x)-\psi(2x), \nonumber \\
\psi(x)&=&\hbox{erf}\,(x)+\frac{2x}{\sqrt\pi }e^{-x^2}-2x^2 \hbox{erfc}\, (x),
\nonumber 
\end{eqnarray}
which holds in the ideal case ($z=0$). Here $b = (r -
R_c)/\hat{R}_g$ is the distance from the colloid surface in units 
of $\hat{R}_g$, ${\rm erf}(x)$ is the error function and 
${\rm erfc}(x) = 1 - {\rm erf}(x)$.
For all values of $q$, the results for 
$z = z^{(1)}$ and, to a lesser extent, those for $z = z^{(3)}$ are very close 
to the ideal ones, indicating that in the dilute regime 
depletion effects for $q\le 2$ are 
little sensitive to solution quality at least up to $z = z^{(3)}$, as already 
discussed in the zero-density limit. 
In Fig.~\ref{fig:gmon-Phi} we show the same distribution
function for larger values of $\Phi$. Depletion effects are much more
dependent on solution quality and deviations from ideality are clearly visible,
even for $z = z^{(1)}$.

\subsection{Finite-density depletion thickness and adsorption} \label{sec5.3}

\begin{table}[ht!]
\caption{Full-monomer estimates of the depletion thickness ratio 
$\delta_s(q,\Phi)/R_c$ for $z = z^{(1)}$:
(a) is the estimate obtained by using extrapolation (\ref{Gncp}),
(b) is the direct estimate obtained by using $g_{{\rm mon},cp}(r)$,
and 
(c) is the direct estimate obtained by using $g_{CM,cp}(r)$.
The last column gives the final estimate.
}
\label{tab:FM-z1}
\begin{center}
\begin{tabular}{ccccccc}
\hline\hline
$q$ & $\Phi$ &  (a) & (b) & (c) & final\\
\hline
0.5& 0.2 & 0.48(2) & 0.486(5) & 0.490(3) & 0.48(2)\\
& 1.0 & 0.417(2) & 0.440(2) & 0.441(1) & 0.428(14) \\
& 2.0 & 0.383(2) & 0.392(2) & 0.390(2) & 0.387(6) \\
\hline
1 & 0.2 & 0.900(6) & 0.886(4) & 0.905(4) & 0.90(1)\\
& 1.0 & 0.778(3) & 0.81(1) & 0.820(4)  & 0.80(2) \\
& 4.0 & 0.641(3) & 0.615(4) & 0.610(5) & 0.62(2)  \\
\hline
2 & 0.2 & 1.7(1) & 1.63(6) & 1.66(2) & 1.7(1) \\
& 1.0 & 1.45(2)  & 1.44(1) & 1.45(1) & 1.45(2)\\
& 4.0 & 1.080(5) & 1.135(6) & 1.151(8) & 1.12(4) \\
\hline\hline
\end{tabular}
\end{center}
\end{table}

\begin{table}[t!]
\caption{Full-monomer estimates of the depletion thickness raio
$\delta_s(q,\Phi)/R_c$ for $z = z^{(3)}$:
(a) is the estimate obtained by using extrapolation (\ref{Gncp}),
(b) is the direct estimate obtained by using $g_{{\rm mon},cp}(r)$,
and 
(c) is the direct estimate obtained by using $g_{CM,cp}(r)$.
The last column gives the final estimate.
}
\label{tab:FM-z3}
\begin{center}
\begin{tabular}{ccccccc}
\hline\hline
$q$ & $\Phi$ &  (a) & (b) & (c) & final\\
\hline
0.5 & 0.2 & 0.45(1) & 0.47(1) & 0.466(5) & 0.46(2) \\
& 1.0 & 0.343(3) & 0.338(2) & 0.340(1) & 0.340(6) \\
& 2.0 & 0.2656(7) & 0.269(2) & 0.266(1) & 0.268(3) \\
\hline
1 & 0.2 & 0.826(9) & 0.842(15) & 0.837(5) & 0.84(2) \\
& 1.0 & 0.634(5) & 0.631(8) & 0.636(4) & 0.632(9)\\
& 2.0 & 0.498(2) & 0.47(2) & 0.507(3) & 0.503(7) \\
& 4.0 & 0.3699(6) &0.38(1) & 0.36(1) & 0.365(15) \\
\hline
2 & 0.2 & 1.50(4) & 1.51(5) & 1.56(2) & 1.52(6) \\
 & 1.0 & 1.15(1) & 1.13(1) & 1.134(8) & 1.14(2) \\
 & 2.0 &  0.92(1) & 0.96(2) & 0.96(1) & 0.94(3) \\
 & 4.0 & 0.498(2) & 0.47(2) & 0.507(3) & 0.503(7) \\
\hline\hline
\end{tabular}
\end{center}
\end{table}

\begin{table}[ht!]
\caption{Full-monomer estimates of the depletion thickness ratio
$\delta_s(q,\Phi)/R_c$ in the good-solvent case:
(a) is the estimate obtained by using extrapolation (\ref{Gncp}),
(b) is the direct estimate obtained by using $g_{{\rm mon},cp}(r)$,
and 
(c) is the direct estimate obtained by using $g_{CM,cp}(r)$.
The last column gives the final estimate. For $q=0.5$ and $\Phi = 4.0$, the box
was not large enough to allow us to estimate reliably $G_{cp}$ from
$g_{CM,cp}(r)$.
}
\label{tab:FM-GS}
\begin{center}
\begin{tabular}{ccccccc}
\hline\hline
$q$ & $\Phi$ &  (a) & (b) & (c) & final\\
\hline
0.5 & 0.4 & 0.335(25) & 0.340(5)  & 0.337(4) &  0.335(25) \\
& 1.0 & 0.239(6)    & 0.236(3) & 0.236(2) & 0.239(6) \\
& 2.0 &	 0.168(5) & 0.162(4) &  0.155(4) & 0.162(11) \\
& 4.0 & 0.110(2) & 0.096(4) & --- & 0.102(10)\\
\hline
1&0.4 & 0.625(15) & 0.615(3) & 0.612(5) & 0.624(17) \\
&1.0 & 0.439(8) & 0.435(3) & 0.427(3) & 0.436(11) \\
&2.0 & 0.35(3) & 0.313(8) & 0.30(1) & 0.335(45) \\
&4.0 & 0.195(6) & 0.192(3) & 0.168(7) & 0.18(2) \\
\hline
2 & 0.4 & 1.07(5) & --- & 1.175(8) & 1.10(8) \\
&1.0 & 0.79(3) & 0.79(1) & 0.78(1) & 0.795(25) \\
&2.0 & 0.67(4) & 0.58(1) & 0.59(2) & 0.65(8) \\
&4.0 & 0.39(2) &  0.37(1) & 0.34(3) &  0.36(5)\\
\hline\hline
 \end{tabular}
\end{center}
\end{table}

\begin{table}[tbp!]
\caption{Depletion thickness ratio $\delta_s/\hat{R}_g$ in the presence 
of a surface ($q=0$) in the good-solvent regime. Direct estimates obtained
by using the surface-monomer distribution function (monomer density 
profile).}
\label{tab:deltas-plane}
\begin{center}
\begin{tabular}{cc}
\hline
\hline
$\Phi$ & $\delta_s/\hat{R}_g$ \\
\hline
0.3	&	0.820(1)	\\
0.7	&	0.621(2)	\\
1.0	&	0.545(6)	\\
1.5	&	0.420(1)	\\
2.0	&	0.352(1)	\\
4.0	&	0.218(2)	\\	
6.0	&	0.162(3)	\\	
8.0	&	0.127(4)	\\
\hline 
\hline
\end{tabular}
\end{center}
\end{table}

By using the pair distribution function $g_{{\rm mon},cp}(r)$ 
we can compute the integral $G_{cp}$, as discussed in Sec.~\ref{sec5.1},
and the depletion thickness $\delta_s$.  The results for $q\not=0$
are reported in Tables~\ref{tab:FM-z1}, \ref{tab:FM-z3},
and \ref{tab:FM-GS} [estimates (b)], those for $q=0$ 
in Table~\ref{tab:deltas-plane}. Errors take only into 
account statistical fluctuations, hence they should not be taken
too seriously, as we shall discuss below. The same procedure can 
also be applied to $g_{CM,cp}(r)$. Although, this pair 
distribution function is quite different from $g_{{\rm mon},cp}(r)$
(it will be discussed in Sec.~\ref{sec6}), the estimates of 
$G_{cp}$ it provides are close to those obtained 
by using $g_{{\rm mon},cp}(r)$, see estimates (c) reported in 
Tables~\ref{tab:FM-z1}, \ref{tab:FM-z3},
and \ref{tab:FM-GS}. In most of the cases, estimates (b) and (c) 
are consistent within errors. In a few cases, however---mostly for the 
largest values of $\Phi$---differences are observed, indicating that 
systematic errors are larger than statistical ones. To obtain a better
control of systematic effects, it is important to have a 
different, conceptually independent method to estimate $G_{cp}$. For this 
purpose we compute $G_{cp}$ from the Fourier transform of the 
monomer distribution function. We use the method described in the 
previous section, and, in particular,
the approximant $G^{(2)}_{cp}$ defined in Eq.~(\ref{Gncp}). 
The corresponding results for $\delta_s$ are reported in 
Tables~\ref{tab:FM-z1}, \ref{tab:FM-z3},
and \ref{tab:FM-GS} [estimates (a)]. For small values of $\Phi$, estimates
(a) are consistent with the direct estimates (b) and (c). However, errors
are significantly larger than those on (b) and (c), hence we cannot 
exclude that the direct estimates show systematic deviations which 
are larger than their statistical errors. For $\Phi\ge 1$, all estimates
have comparable statistical errors, but results are sometimes not consistent.
In order to quote a reliable estimate with a correct error bar, 
we take a conservative attitude. 
We determine the largest interval that contains estimates (a), (b) and (c) with 
their errors.
The midpoint is the final estimate, while the half-width gives the error. 
The results of this procedure are reported (column ``final") in 
Tables~\ref{tab:FM-z1}, \ref{tab:FM-z3}, and \ref{tab:FM-GS}.

\begin{figure}[t]
\begin{center}
\begin{tabular}{c}
\epsfig{file=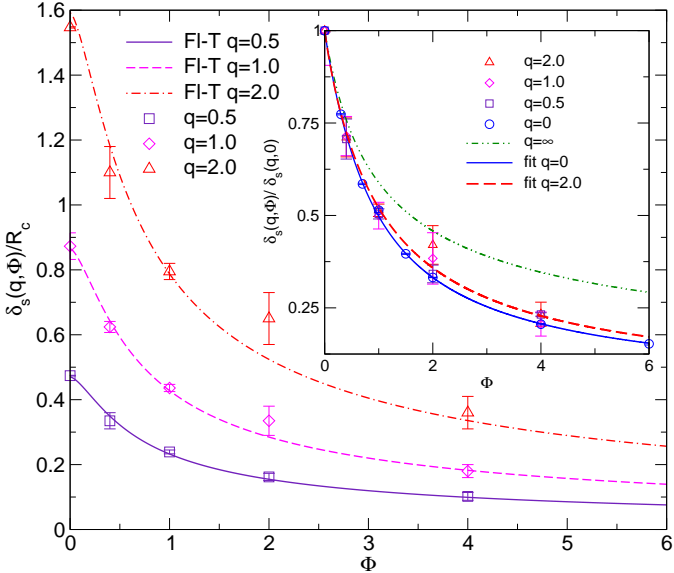,angle=0,width=9truecm} \hspace{0.5truecm} \\
\end{tabular}
\end{center}
\caption{Main panel: Depletion thickness ratio $\delta_s(q,\Phi)/\hat{R}_c$ 
as a function of $\Phi$ in the good-solvent  regime;
we report the Monte Carlo data (points) and the phenomenological prediction
(\ref{FT-deltas}) (lines, Fl-T). Inset: $\delta_s(q,\Phi)/\delta_s(q,0)$
as a function of $\Phi$; we report the data (points), the interpolations
(\ref{deltas-fit}) (fit), 
and the curve $K_p(\Phi)^{-1/3}$ ($q=\infty$), see text for a discussion.
}
\label{fig:deltas-Phi-GS}
\end{figure}

\begin{figure}[t]
\begin{center}
\begin{tabular}{c}
\epsfig{file=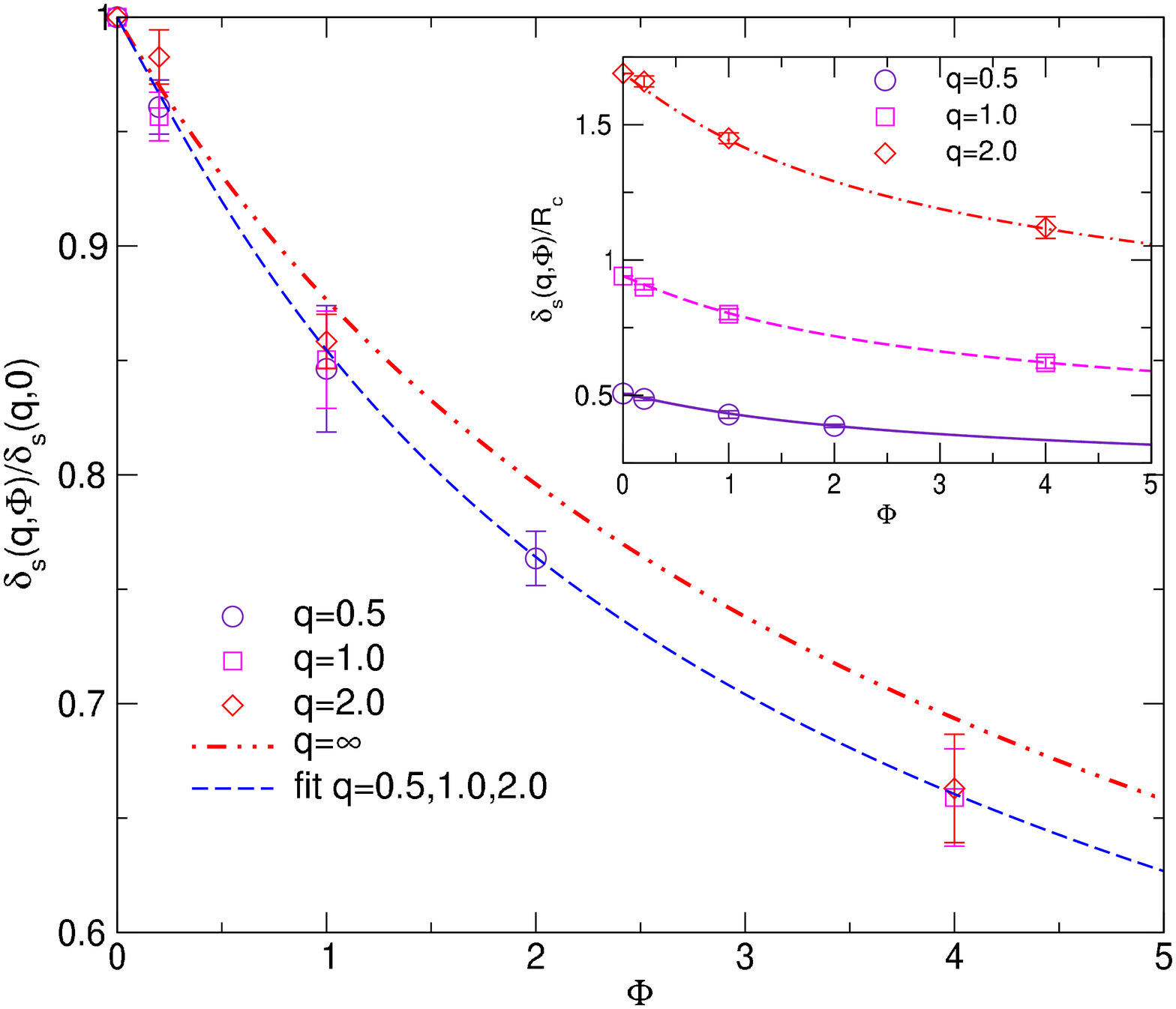,angle=0,width=7truecm} \hspace{0.5truecm} \\
\epsfig{file=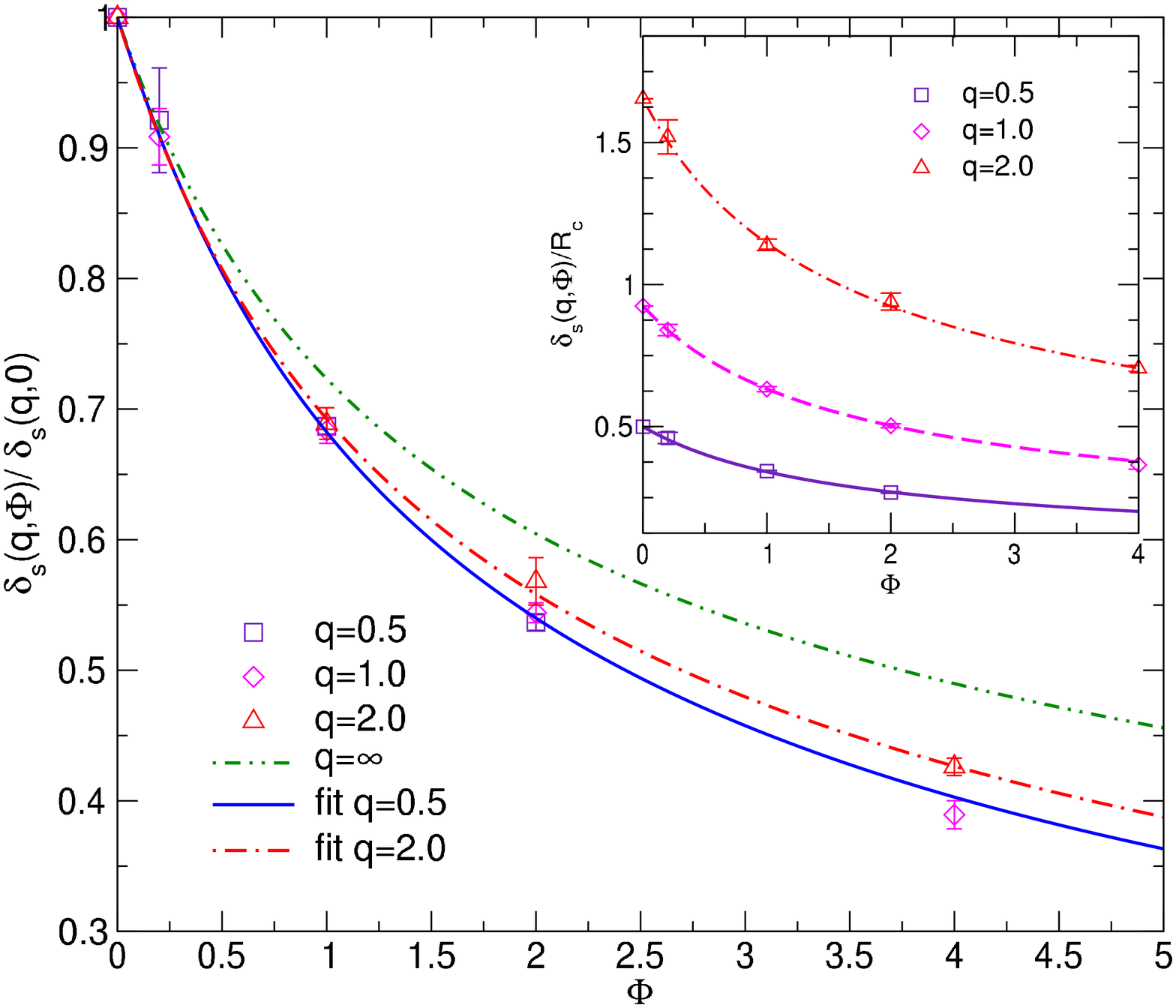,angle=0,width=7truecm} \hspace{0.5truecm} \\
\end{tabular}
\end{center}
\caption{Plot of $\delta_s(q,\Phi)/\delta_s(q,0)$ 
as a function of $\Phi$ for $q=0.5,1,2$ for $z = z^{(1)}$ (top)
and $z = z^{(3)}$ (bottom). Lines correspond to the interpolations
(\ref{deltas-fit}) (fit), while the curve $q=\infty$ corresponds to
$K_p(\Phi)^{-1/3}$, as discussed in the text. In the inset we report
$\delta_s(q,\Phi)/\hat{R}_c$ (points) and the corresponding 
interpolations (lines). The function $K_p(\Phi)$ is obtained by using 
the equation of state reported in Ref.~\cite{DPP-thermal}.
}
\label{fig:deltas-Phi-z}
\end{figure}

The good-solvent results are shown in Fig.~\ref{fig:deltas-Phi-GS}. 
The depletion thickness decreases very rapidly with $\Phi$. 
For instance, for $q=2$, we find $\delta_s/R_c = 1.5305(5)$ for $\Phi=0$
and $\delta_s/R_c = 1.10(6)$ for $\Phi = 0.4$. Even a small increase of the 
polymer density significantly reduces the width of the depleted layer around 
the colloid. An interesting feature of the results is that the $\Phi$
dependence for $0\le q\le 2$, the interval of $q$ we investigate, is
approximately independent of $q$. This is evident from the results 
reported in the inset of Fig.~\ref{fig:deltas-Phi-GS}, where we show 
the ratio $\delta_s(q,\Phi)/\delta_s(q,0)$ as a function of $\Phi$.
The $q$-dependence is practically absent. This result is far from obvious 
and is consistent with what we already observed in Sec.~\ref{sec4.3}, 
where we pointed out that the first density correction is approximately
$q$-independent for $q\lesssim 2$. 

The $q$ independence of the ratio is not expected to hold much beyond
$\Phi=4$, our largest density. Indeed, as we discussed in Sec.~\ref{sec2.4}, 
$\delta_s(q,\Phi)\sim \xi(\Phi)$ for $\Phi\to \infty$ and any $q$, so 
$\Delta(q,\Phi) = \delta_s(q,\Phi)/\delta_s(q,0) \sim 
\xi(\Phi)/\delta_s(q,0)$. Since $\delta_s(q,0)$ 
varies significantly with $q$, factorization breaks down deep in the 
semidilute regime (some differences are already observed for $\Phi=4$).
Analogously, such a property does not hold for large values of $q$. Indeed, 
as long as $R_c \ll \xi$, 
Eq.~(\ref{deltas-largePhi-largeq}) holds, which implies 
\begin{equation}
\Delta(q,\Phi) = {\delta_s(q,\Phi)\over \delta_s(q,0)} = K_p(\rho_p)^{-1/3}.
\label{Delta-K}
\end{equation}
Using the equation of state of Ref.~\cite{Pelissetto-08}, we can compute 
$K_p(\rho_p)$, hence $\Delta(q,\Phi)$ for $q\to\infty$. The corresponding 
curve is reported in Fig.~\ref{fig:deltas-Phi-GS} (line ``$q=\infty$" in the inset).
Differences with the Monte Carlo results are quite significant. For instance, 
for $\Phi= 4$, Eq.~(\ref{Delta-K}) predicts $\Delta(q,4) = 0.346$ for $q\to\infty$, 
to be compared with 
$\Delta(q,4) = 0.232(5)$ and 0.23(3) for $q=0.5$ and $q=2$, respectively.
Note that differences increase rapidly with $\Phi$. This is due to the fact 
that, 
for $q\le 2$, $\Delta$ already scales as  $\Phi^{-0.8}$ for $\Phi\gtrsim 2$, 
while $\Delta\sim \Phi^{-0.437}$ for large values of $q$.

In Fig.~\ref{fig:deltas-Phi-GS} we also report
the phenomenological approximation (\ref{FT-deltas}), which works 
quite well for $0.2\le q \le 4$ in the dilute limit.
Also the density dependence is well reproduced for $0.5\le q\le 2$:
Tiny differences are only observed in the dilute regime.

In Fig.~\ref{fig:deltas-Phi-z} we report the depletion thickness in the 
thermal crossover region, for $z=z^{(1)}$ and $z = z^{(3)}$. The qualitative
behavior is very similar to that observed in the good-solvent case. 
For all values of $q$ considered, the $\Phi$ dependence and the 
$q$ dependence appear to be factorized, i.e. $\delta_s(q,\Phi)/\delta_s(q,0)$
is essentially independent of $q$. Such a result is expected to hold 
in a $q$ interval that is larger than in the good-solvent case. 
First, we have already observed that the first  density correction
is essentially $q$ independent for $q\lesssim 10$ for $z = z^{(1)}$. 
Second, the difference between our data and the large-$q$ prediction
(\ref{Delta-K}), which should also hold in the thermal crossover region, 
decreases as $z$ decreases. Finally, it is interesting to observe
that in the crossover region the density dependence of $\delta_s$ 
is smaller than in the good-solvent case. For $\Phi = 4$, 
$\delta_s(q,4)/\delta_s(q,0) = 0.22$, 0.40, 0.66, for $z=\infty$, 
$z=z^{(3)}$, and $z=z^{(1)}$. This result is of course expected,
since $\delta_s$ becomes density independent for $z\to 0$. 

\begin{table}[tb!]
\caption{Coefficients parametrizing the depletion-thickness interpolations
(\ref{deltas-fit})
as function of density. The parametrizations should hold for 
$\Phi\le \Phi_{\rm max}$.}
\label{tab:depletion-fits}
\begin{center}
\begin{tabular}{ccccccccc}
\hline\hline
$z$ & $q$ & $n$ & $a_1$ & $a_2$ & $a_3$ & $\eta$ & $\Phi_{\rm max}$\\
\hline
$\infty$ & 0 & 3 & 3.9467 & 4.3305 & 5.8889 & 0.770 & $4$ \\
& 0.5 	&  3 & 4.0909   & 6.8272   & 2.6728 & 0.770 & $4$ \\
& 1.0 	&  3 & 4.0987   &  4.4818 & 4.92968 & 0.770 & $4$ \\
& 2.0   &  3 & 4.0753   & 6.12348   & 1.92624 & 0.770 & $4$ \\
\hline
$z^{(3)}$ & 0.5 & 2 & 1.8641 & 1.0753 & 0 &0.5579 & $2$ \\
& 1.0 & 2 & 1.8747  & 1.0279 & 0          &0.5579 & $4$ \\
& 2.0  & 2 & 2.0279 & 1.16915 & 0	  &0.5128 & $4$ \\
\hline
$z^{(1)}$ & 0.5/1.0/2.0 & 3 & 1.7682 & 1.8151 & 0.6591 & 0.2845 & $4$ \\
\hline\hline
\end{tabular}
\end{center}
\end{table}

To summarize our results in a simple way, we determine interpolations
of the Monte Carlo data for the depletion thickness. For this purpose 
we fit the results to 
\begin{equation}
{\delta_s(q,\Phi)\over \delta_s(q,0)} = 
    \left(1 + \sum_{k=1}^n a_k \Phi^k\right)^{-\eta/n}.
\label{deltas-fit}
\end{equation}
We set $n=2$ or $3$ and 
$a_1 = - n\delta_1/\eta$, where $\delta_1$ is the first 
density correction defined in Eq.~(\ref{delta1-def}), in such a way
to reproduce accurately the low-density behavior. In the good-solvent 
case, we have $\delta_s \sim \Phi^{-0.770}$ for $\Phi\to \infty$. 
We have enforced this condition in our interpolations, requiring 
$\eta = 0.770$. In the crossover region, we do not have predictions
for the large-$\Phi$ behavior, hence $\eta$ has been taken as 
a free parameter. For $z = z^{(3)}$ and $q = 0.5$, our data extends only 
up to $\Phi = 2$, hence fits are not very sensitive to $\eta$. 
To obtain stable fit results, we fix $\eta$ to be equal to the result
obtained for $q = 1$ and $z=z^{(3)}$. For the good-solvent case and 
for $z = z^{(3)}$ the results show a tiny $q$-dependence for $\Phi = 4$,
hence we determine an interpolation for each value of $q$. On the other 
hand, for $z = z^{(1)}$ results for different values of $q$ 
coincide within errors. Hence, we have performed a single fit,
considering simultaneously all value of $q$. The coefficients of the 
interpolations are reported in Table \ref{tab:depletion-fits}.
The interpolations are reported in Figs.~\ref{fig:deltas-Phi-GS}
and \ref{fig:deltas-Phi-z}. 

\begin{figure}[t]
\begin{center}
\begin{tabular}{c}
\epsfig{file=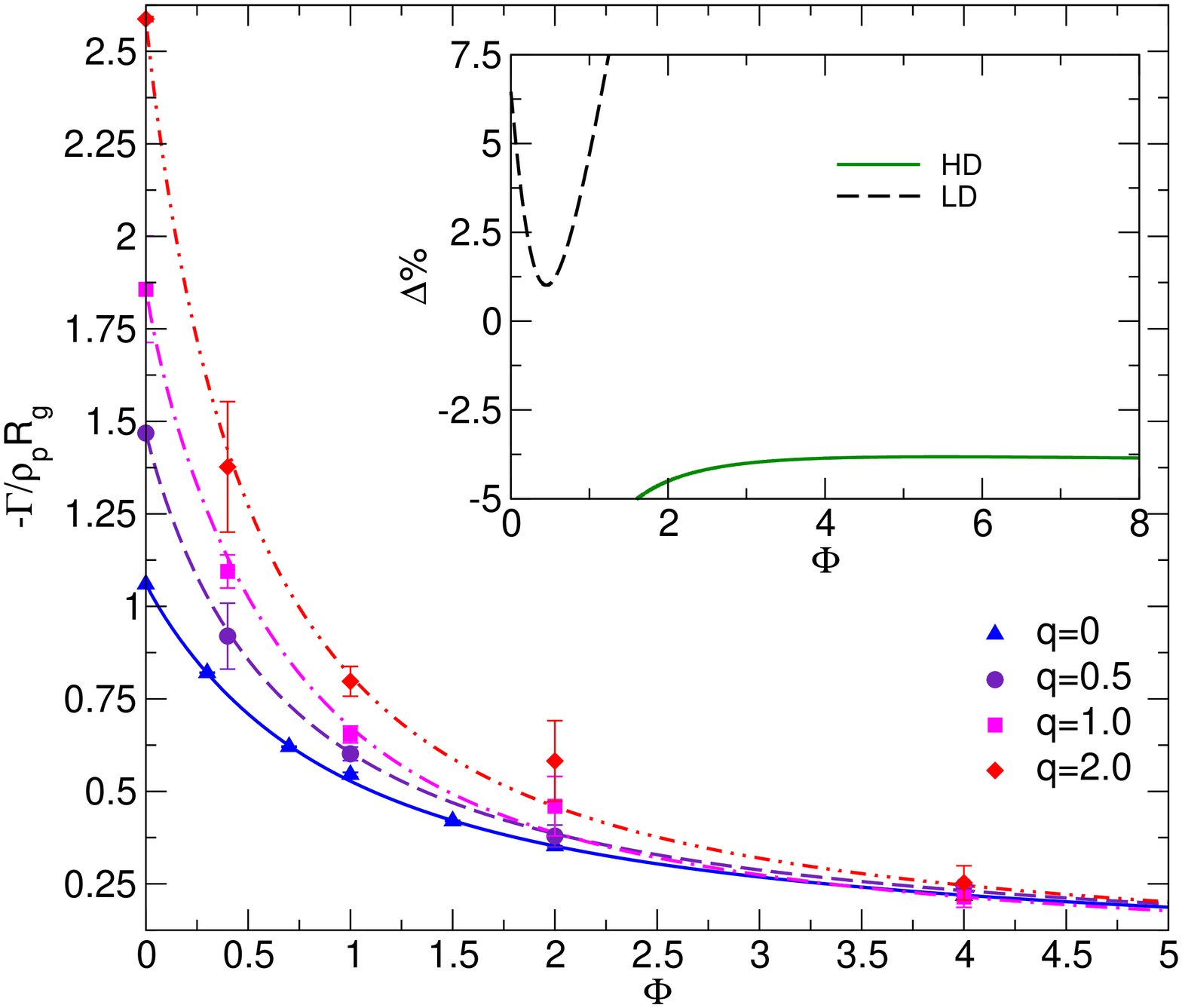,angle=0,width=9truecm} \hspace{0.5truecm} \\
\end{tabular}
\end{center}
\caption{Main panel: 
Rescaled adsorption $\hat{\Gamma} = -\Gamma/(\rho_p \hat{R}_g)$ 
as a function of $\Phi$ for $q=0,0.5,1.0,2.0$ as a function of $\Phi$
in the good-solvent regime; lines are obtained by using 
interpolations (\ref{deltas-fit}). In the inset we report
the relative deviations
$\Delta = 100 ({\Gamma}_{\rm pred}/{\Gamma}_{MC} - 1)$,
where ${\Gamma}_{\rm pred}$ is either Eq.~(\ref{deltas-largePhi}) 
(HD) or Eq.~(\ref{Gamma-smallPhi}) (LD), and $\Gamma_{MC}$ is obtained 
by using interpolations (\ref{deltas-fit}).
}
\label{fig:Gamma-Phi-GS}
\end{figure}

\begin{figure}[t]
\begin{center}
\begin{tabular}{c}
\epsfig{file=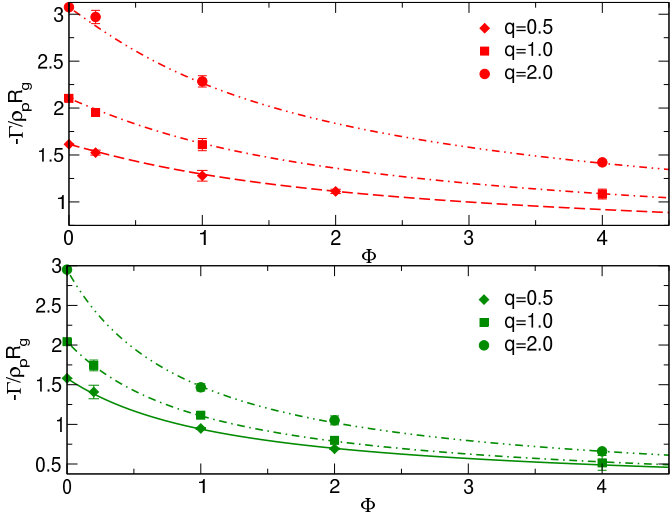,angle=0,width=9truecm} \hspace{0.5truecm} \\
\end{tabular}
\end{center}
\caption{Rescaled adsorption $-\Gamma/(\rho_p \hat{R}_g)$ 
as a function of $\Phi$ for $q=0.5,1,2$. We report 
data for $z = z^{(1)}$ (top) and $z = z ^{(3)}$ (bottom).
Lines are obtained by using 
interpolations (\ref{deltas-fit}). 
}
\label{fig:Gamma-Phi-z}
\end{figure}

Finally, we wish to compare our good-solvent data with the 
large-$\Phi$ field-theoretical predictions. 
In Fig.~\ref{fig:Gamma-Phi-GS} we show our results for the 
adsorption $\Gamma$. As predicted by theory, adsorption becomes 
independent of $q$ as $\Phi$ increases. On the scale of the figure,
all curves coincide for $\Phi\gtrsim 4$. This is consistent with 
the results of Ref.~\cite{LBMH-02-a}, where it was shown 
that $\gamma(q,\Phi)/\gamma(0,\Phi)$ converges to 1 for large $\Phi$
for all $q\le 1.68$. Note that this ratio becomes approximately 1
at densities which are significantly larger than $\Phi\approx  4$. 
This is due to the fact that $\beta \gamma(q,\Phi)$ is obtained by integrating 
$\Gamma(q,\Phi)$, see Eq.~(\ref{Gamma-gamma}), from 0 to $\Phi$, 
hence including the dilute region in which depletion effects are 
strongly $q$-dependent. 
Quantitatively, the field-theoretical prediction
(\ref{deltas-largePhi}), 
$\Gamma/(\rho_p \hat{R}_g) \approx - 0.649\Phi^{-0.770}$, 
holds quite precisely.  For the 
surface case ($q=0$) it is in good agreement with our data 
for $\Phi\gtrsim 2$ with deviations which are of order 4\% (see inset).  
For instance, interpolation (\ref{deltas-fit}) 
gives $\Gamma/(\rho_p \hat{R}_g) = -0.673 \Phi^{-0.770}$,
which is  compatible  with prediction (\ref{deltas-largePhi}). We can also
compare our results with interpolation (\ref{Gamma-planar}). 
For $\Phi\to \infty$
it predicts $\Gamma/(\rho_p \hat{R}_g) = -0.54 \Phi^{-0.770}$, 
which differs significantly from our result.
This is probably related to the fact that 
Eq.~(\ref{Gamma-planar}) is obtained by fitting SAW data. Indeed, such 
a model shows large finite-length corrections to scaling, 
especially in the semidilute regime \cite{Pelissetto-08}. Hence, even the results obtained 
from simulations of rather long walks ($L\approx 10^3$) do not probe the 
universal, infinite-length behavior.

We can also compare the results with low-density prediction
\begin{eqnarray}
-{\Gamma\over \rho_p \hat{R}_g} &=& 
    {1.129 (1 + 1.4\Phi)\over K_p(\rho_p)},
\label{Gamma-smallPhi}
\end{eqnarray}
where $K_p(\rho_p)$ is obtained from the equation of state of 
Ref.~\cite{Pelissetto-08}. Such an expression describes well the data up
to $\Phi\approx 0.5$, with deviations of less than 6\%.

\section{Comparison with single-blob results} \label{sec6}

\begin{figure}[t]
\begin{center}
\begin{tabular}{c}
\epsfig{file=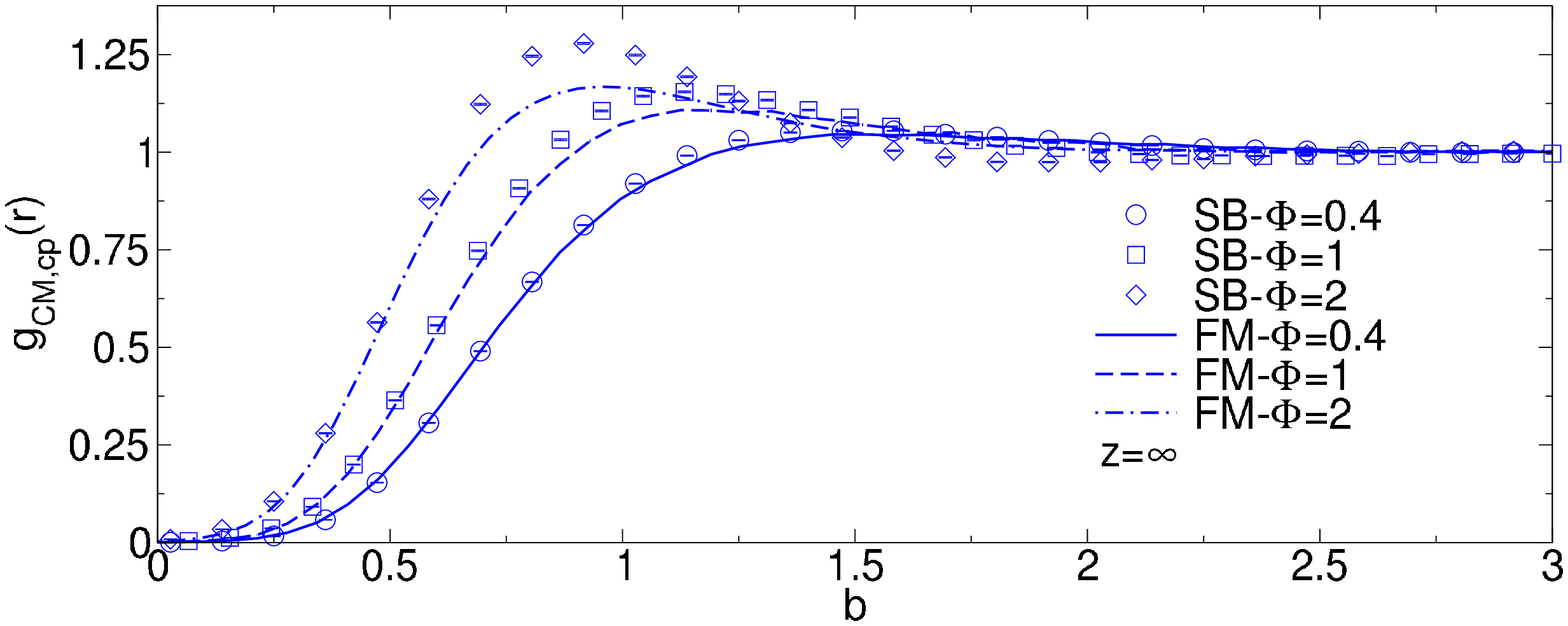,angle=0,width=9truecm} \hspace{0.5truecm} \\
\epsfig{file=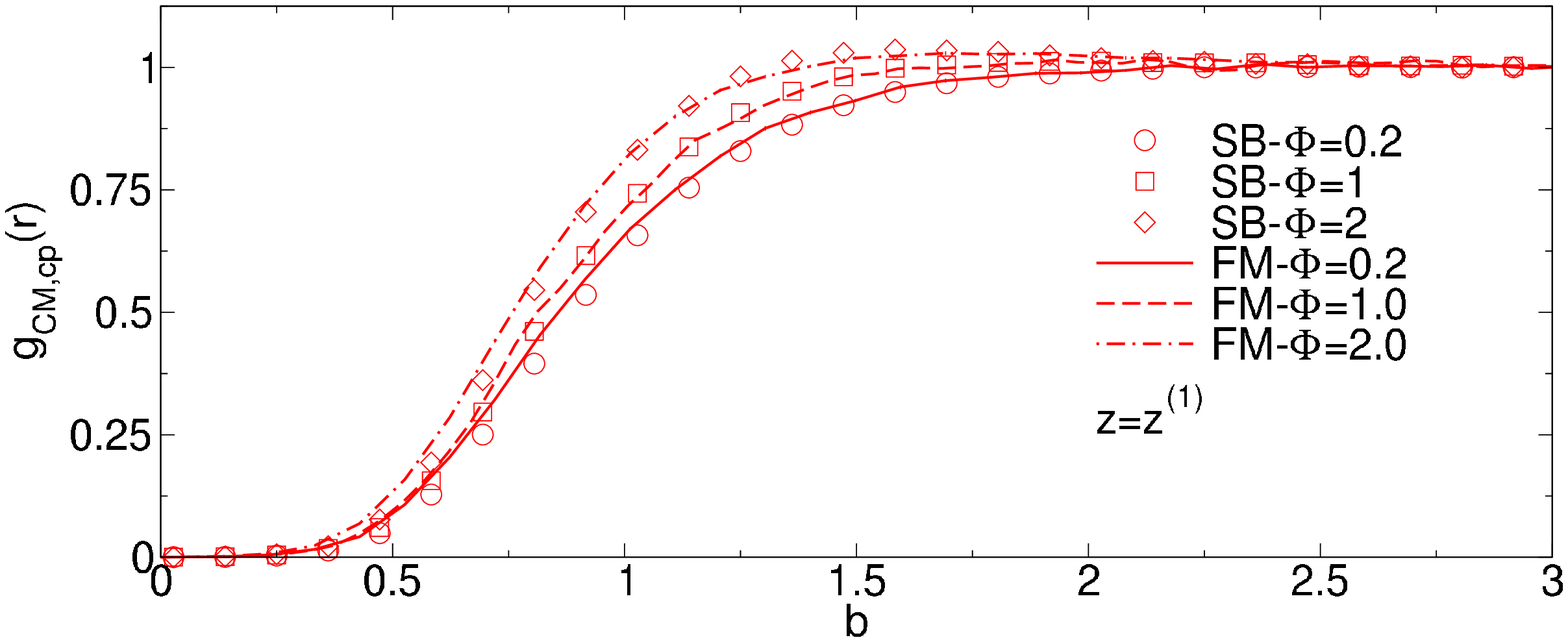,angle=0,width=9truecm} \hspace{0.5truecm} \\
\end{tabular}
\end{center}
\caption{Pair distribution function $g_{CM,cp}(r)$ for $q = 0.5$ 
as a function of $b = (r-R_c)/\hat{R}_g$ under good-solvent conditions (top), 
and $z = z^{(1)}$ (bottom). 
}
\label{fig:gCM-Phi-q05}
\end{figure}

\begin{figure}[t]
\begin{center}
\begin{tabular}{c}
\epsfig{file=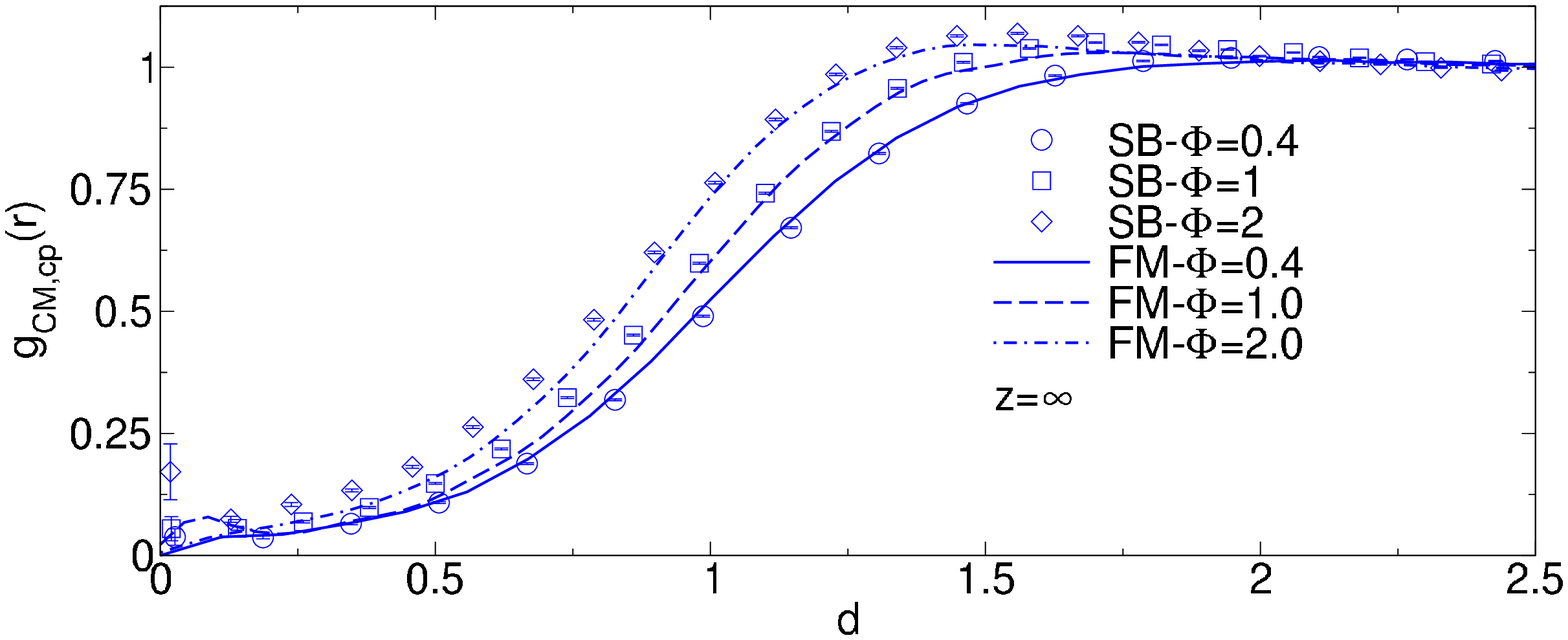,angle=0,width=9truecm} \hspace{0.5truecm} \\
\epsfig{file=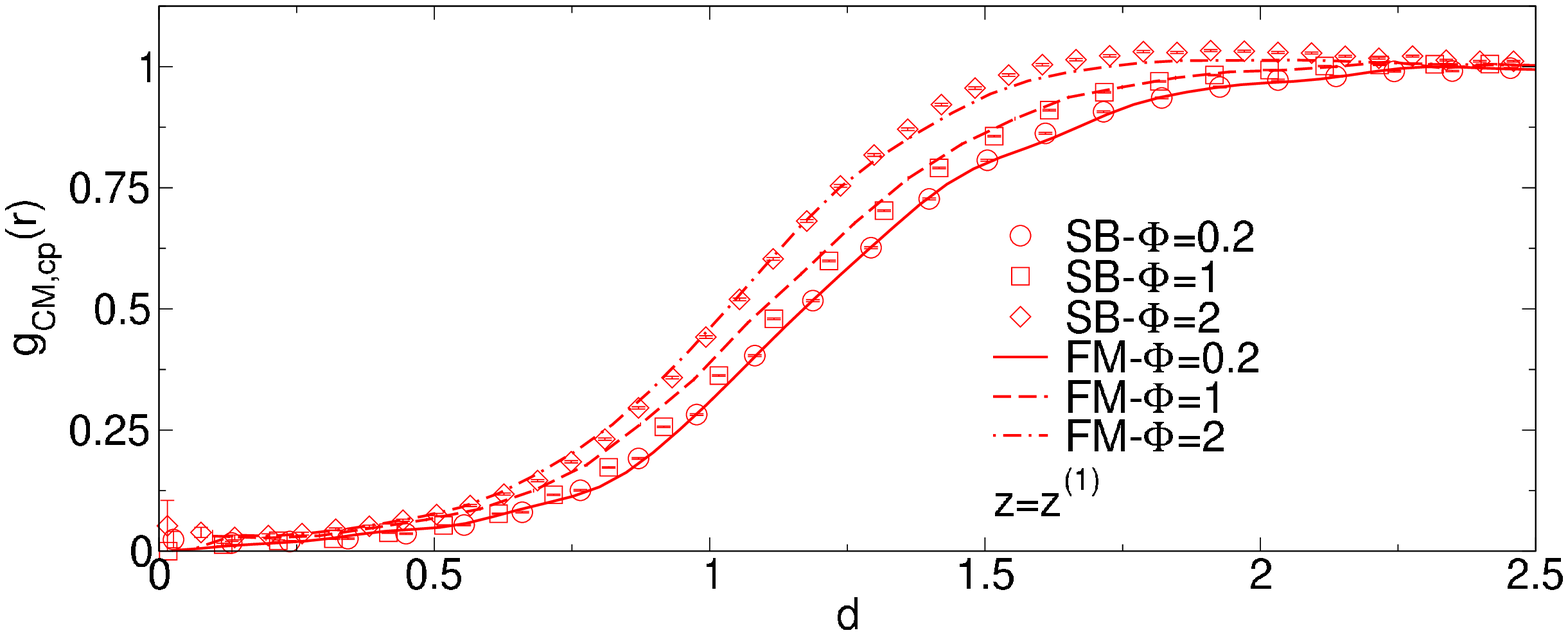,angle=0,width=9truecm} \hspace{0.5truecm} \\
\end{tabular}
\end{center}
\caption{Pair distribution function $g_{CM,cp}(r)$ for $q = 2$ 
as a function of $d = r/\hat{R}_g$ under good-solvent conditions (top), 
and $z = z^{(1)}$ (bottom). 
}
\label{fig:gCM-Phi-q2}
\end{figure}

Recently, there has been significant work dealing with coarse-grained 
models of polymer
solutions \cite{MullerPlathe-02,PK-09,Voth2009,PCCP2009,SM2009,FarDisc2010}.
The simplest model
\cite{Likos-01,LBHM-00,HL-02} is obtained by representing polymers with
monoatomic molecules (single-blob model) 
interacting via the polymer center-of-mass potential of
mean force.
By definition the model reproduces the dilute behavior of the solution, 
but fails to be accurate as soon as polymer-polymer overlaps become important, i.e. 
for $\Phi\gtrsim 1$.
This model can be extended to include colloids \cite{BLM-03}, 
taking the colloid-polymer potential of mean force as interaction
potential. In Ref.~\cite{PH-06}
the effective potential between a colloid and a coarse-grained molecule 
was computed in the whole crossover region, between the $\theta$ point and 
the good-solvent case, for several values of $q$. The calculation 
was performed using interacting SAWs, without identifying the value of 
$z$ associated with each potential. Here, we wish to perform a much more 
careful analysis, following Ref.~\cite{DPP-thermal}.  We repeat the calculation
using Domb-Joyce 
walks, determining the potential for the good-solvent case and for $z
= z^{(1)}$, $z = z^{(3)}$. We then use the coarse-grained single-blob 
model to determine
depletion properties for different values of $q$ and $\Phi$, which are then compared 
with the results of full-monomer simulations. 
The coarse-grained model is expected to be predictive 
as long as details of the polymer structure are not relevant. For the 
single-blob model,
 we thus expect to obtain accurate results only for $q \lesssim 1$, i.e., 
in the colloid regime. If one wishes to investigate larger values of $q$, 
multiblob models \cite{Pierleoni:2007p193,VBK-10,DPP-12-Soft,GDA-2012-transferability,%
DPP-thermal} should be considered, fixing the number of blobs in such a way
that the radius of gyration $\hat{r}_g$ of the blob satisfies 
$\hat{r}_g \lesssim R_c$.

By construction, the single-blob model reproduces the full-monomer 
second-virial combination
$A_{2,cp}$ or, for $q=0$, the surface quantity $R_{1,p}$. We have verified this condition
for all values of $q$ (see Table~\ref{table:surface-int} for $q=0$ and the supplementary
material for $q\ge 0.5$), confirming the accuracy of the effective potentials we use.
It is also interesting to compare full-monomer and single-blob 
results for the third-virial combination
$A_{3,cpp}$ (for $R_{2,pp}$ in the surface case), 
since this quantity gives us information on how well
the coarse-grained model reproduces the 
colloid-polymer-polymer three-body interactions. 
For $q=0$ the results are reported in Table~\ref{table:surface-int}. The 
single-blob model reproduces quite well the full-monomer 
results and the agreement improves as $z$ decreases.
On the other hand, for $q=2$, differences are significant, even for $z = z^{(1)}$. 
In the good-solvent regime we have $A_{3,cpp} = 19.16(2)$ and $25.49(6)$ 
for the single-blob and the full-monomer case, 
while, for $z = z^{(1)}$ the two representations give
$A_{3,cpp} = 5.299(6)$ and 6.73(2), respectively.  As 
expected, three-body forces are not well modelled by representing polymers with a 
single blob: since $R_c$ is small, the structure of the polymer plays an important role.

\begin{figure*}[t]
\begin{center}
\begin{tabular}{c}
\epsfig{file=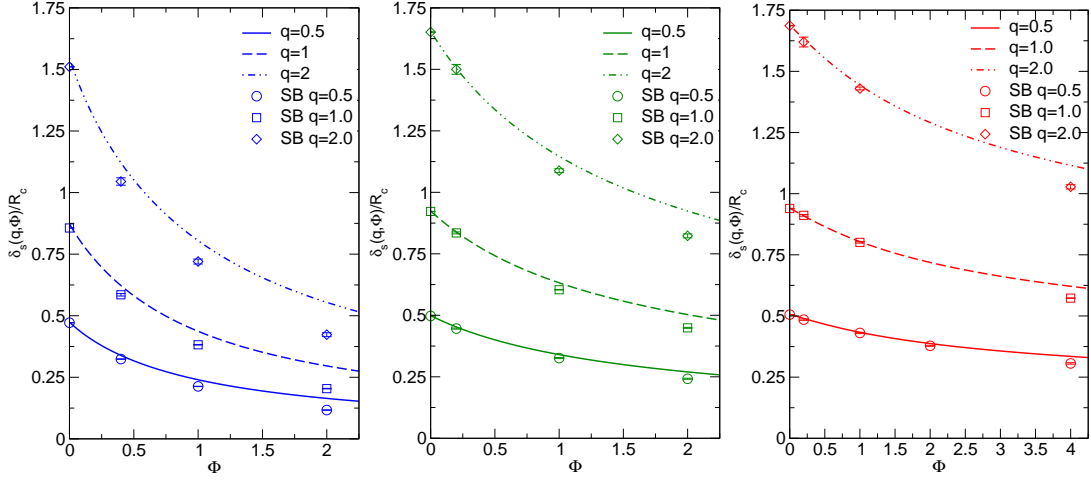,angle=0,width=15truecm} \hspace{0.5truecm} \\
\end{tabular}
\end{center}
\caption{Depletion ratio $\delta_s(q,\Phi)/R_c$ for $z=\infty$ (left),
$z=z^{(3)}$ (center), and $z=z^{(1)}$ (right). We report full-monomer
(lines) and single-blob (SB, points) data as a function of $\Phi$.
}
\label{fig:deltas-comparison-SB}
\end{figure*}

Let us now compare finite-density results.
In Fig.~\ref{fig:gCM-Phi-q05} we report the full-monomer and single-blob 
distribution function 
$g_{CM,cp}(r)$ for $q = 0.5$. The curves vanish on the surface of the colloid
($b=0$) and then show some oscillations that 
become stronger as $\Phi$ increases.
The coarse-grained model appears to reproduce well the full-monomer 
correlations for $z = z^{(1)}$, while 
deviations are observed already for $\Phi = 1$ in the good-solvent case.
Results for $q=2$ are reported in Fig.~\ref{fig:gCM-Phi-q2}. In this case 
correlations are non zero even for $r\le R_c$ ($d\le 1/2$): since $q > 1$, it is not 
unlikely that the polymer center of mass lies inside the colloid. Since 
the effective 
potential is soft, oscillations are tiny and, apparently, the coarse-grained 
model reasonably reproduces the full-monomer distribution function.
However, at a closer look one notices some systematic deviations on the tails
of the distributions, which are particularly relevant for the computation of 
$G_{cp}$, hence significantly affect the adsorption properties.

Finally, let us consider the depletion thickness. The single-blob 
results are compared with 
the full-monomer ones [we use interpolations (\ref{deltas-fit})] in 
Fig.~\ref{fig:deltas-comparison-SB}. In the good-solvent case, 
the single-blob model
provides reasonably accurate estimates up to $\Phi\approx 2$ for $q=0.5$. As $q$
increases, the agreement worsens, as expected. For $q = 2$, small deviations are 
already observed for $\Phi = 0.4$. The single-blob 
model appears to be more accurate 
in the crossover region. For $z = z^{(1)}$ and $q = 0.5$ good results are obtained 
up to $\Phi = 4$, while for $q = 2$, agreement is observed up to $\Phi\approx 1$.

\section{Conclusions} \label{sec7}

In this paper we perform a detailed study of the solvation properties of a single colloid
in a polymer solution. Beside the good-solvent case, which has already 
been extensively studied, 
see, e.g., Refs.~\cite{deGennes-CR79,JLdG-79,EHD-96,HED-99,MEB-01,FS-01,LBMH-02-a,FT-08}
and references therein, we also consider the crossover between the 
good-solvent and the $\theta$ regime. 
We perform a detailed study for two intermediate cases.
We consider $z = z^{(3)}$, which corresponds to 
$A_{2,pp}/A_{2,pp,GS} = \Psi/\Psi_{GS} = 0.54$ (here $\Psi$ is the interpenetratio ratio
often used in experimental work and $A_{2,pp,GS}$, $\Psi_{GS}$ are the 
good-solvent values), hence to solutions that have properties in between the 
good-solvent  
and the $\theta$ case.  Moreover, we consider $z = z^{(1)}$, 
corresponding to $\Psi/\Psi_{GS} = 0.18$, which 
corresponds to a solution close to the $\theta$ point. 

We perform a detailed study of the depletion thickness in the dilute regime. For this 
purpose we relate solvation properties to polymer-colloid virial coefficients. 
We compute the second and the third virial coefficients in a large $q$ interval 
($0\le q \le 50$ for the good-solvent case and 
$0\le q\le 30$ for $z = z^{(1)}$). The 
good-solvent results are compared with 
the existing field-theoretical predictions
\cite{EHD-96,HED-99,MEB-01}, finding a reasonable agreement in all cases. 
We also consider the PRISM prediction \cite{FS-01}, which appears to be of limited 
quantitative interest, and the phenomenological prediction of Ref.~\cite{FT-08}
[see Eq.~(\ref{FT-deltas})], which turns out to describe the numerical data quite 
accurately for $0.2\lesssim q \lesssim 4$. 

We also perform a careful study at finite density for $q=0,0.5,1,2$. 
For all these values of 
$q$ and both in the good-solvent and in 
the crossover regime, we find that the ratio
$\Delta(q,\Phi) = \delta_s(q,\Phi)/\delta_s(q,0)$ is approximately independent of $q$ 
for $\Phi\le 4$, 
so that the $\Phi$ dependence and the $q$ dependence are approximately factorized.
We do not have any
theoretical explanation of this phenomenon, but we can easily argue that it 
can only hold 
for $q$ and $\Phi$ not too large. First, we can compute exactly the ratio 
$\Delta(q,\Phi)$ for $q\to \infty$. Then, we find that the limiting $\Delta(\infty,\Phi)$
differs significantly from what we obtain for $q\le 2$, indicating that, as $q$ 
increases, $\Delta(q,\Phi)$ should gradually change from its value for $q\le 2$ to the 
infinite-$q$ limiting curve, hence it should be $q$-dependent. Second, a general 
argument predicts that $\delta(q,\Phi)$ becomes independent of $q$ for $\Phi\to \infty$. 
Hence, deep in the semidilute regime $\Delta(q,\Phi)$ should have the same 
$q$-dependence as $1/\delta(q,0)$, which is quite significant. 
We compare the 
numerical good-solvent results with field-theory predictions \cite{MEB-01}.
We find that 
the large-$\Phi$ prediction of Ref.~\cite{MEB-01} describes well the numerical data,
confirming the accuracy of the field-theory approach.

We also analyze depletion properties of the single-blob coarse-grained model. 
As expected, they are accurate 
as long as polymer-polymer overlaps are rare and colloids are large compared with the 
polymers, i.e. for $q\lesssim 1$. As already observed in Ref.~\cite{DPP-thermal}, we find that
the accuracy of the coarse-grained model increases as $z$ decreases. While in the
good-solvent case
some deviations are observed for $\Phi\gtrsim 1$, even for large colloids ($q=0.5$), 
for $z = z^{(1)}$ reasonably good results are obtained up to $\Phi\approx 4$, well
inside the semidilute regime.

\section*{Acknowledgments}

C.P. is supported by the Italian Institute of Technology (IIT) under the
SEED project grant number 259 SIMBEDD – Advanced Computational Methods for
Biophysics, Drug Design and Energy Research.

\appendix

\section{Virial expansion for multicomponent systems} \label{AppA}

In this Appendix we wish to study the virial expansion for a multicomponent
system of flexible molecules, computing explicitly the flexibility 
contribution due to the polyatomic nature of the molecules. We extend here 
the results presented in Refs.~\cite{CMP-06,CMP-08-star}.

We start by considering a multicomponent system in the grand canonical ensemble. 
The grand partition function is given by
\begin{eqnarray}
\Xi = \sum_{N_1,\ldots,N_k} {z_1^{N_1}\over N_1!} \ldots 
         {z_k^{N_k}\over N_k!} Q(N_1,\ldots,N_k),
\end{eqnarray}
where $k$ is the number of species present, $z_1$, $\ldots$, $z_k$ are the corresponding
fugacities, $Q(N_1,\ldots,N_k)$ is the canonical partition function of the system.
If $V$ is the volume of the box, we define reduced fugacities as
\begin{eqnarray}
&& \hat{z}_1 = z_1 Q(1,0\ldots,0)/V, \qquad \ldots  \nonumber \\
&& \hat{z}_k = z_k Q(0,\ldots,0,1)/V.
\end{eqnarray}
Then, at third order in the fugacities we obtain the expansion
\begin{eqnarray}
\beta P &=& {1\over V} \ln \Xi = \sum_\alpha \hat{z}_\alpha + 
      {1\over 2} \sum_{\alpha\beta} \hat{z}_\alpha \hat{z}_\beta I_2(\alpha,\beta) + 
\nonumber \\
     && {1\over 6} \sum_{\alpha\beta\gamma} \hat{z}_\alpha \hat{z}_\beta \hat{z}_\gamma
   \left[I_{3}(\alpha\beta\gamma) +  
         I_2(\alpha\beta) I_2(\alpha\gamma) + 
\right. \nonumber \\
&&     \qquad 
         I_2(\alpha\beta) I_2(\beta\gamma) + 
         I_2(\alpha\gamma) I_2(\beta\gamma) + 
\nonumber \\[2mm]
     && \qquad \left. 
         T_2(\alpha,\beta\gamma) +
         T_2(\beta,\alpha\gamma) + 
         T_2(\gamma,\alpha\beta) \right].
\label{exppress-z}
\end{eqnarray}
To define the integrals $I_2(\alpha\beta)$, $I_3(\alpha\beta\gamma)$ 
and $T_2(\alpha,\beta\gamma)$, we should associate to each molecule a specific point 
$X$. The choice of $X$ is irrelevant, as long as $X$ is a weighted average of the 
positions of the atoms belonging to the molecule. For instance one can take the 
center of mass of the molecule, but for a linear polymer an equally good choice 
corresponds to choosing the first or the central monomer. Then, we define 
the average $\langle \cdot \rangle_{\alpha,{\bf r}; \beta, {\bf s}}$ as 
the average over all pairs of isolated molecules of type $\alpha$ and $\beta$,
respectively, such that point $X$ of molecule $\alpha$ is fixed in ${\bf r}$
and point $X$ of molecule $\beta$ is fixed in ${\bf s}$. Analogously,
we define 
the average $\langle \cdot \rangle_{\alpha,{\bf r}; \beta, {\bf s}; \gamma, {\bf t}}$ 
over all triples of isolated molecules.
Then, the integral 
$I_2(\alpha\beta)$ is defined as 
\begin{eqnarray}
I_{2}(\alpha\beta) = \int d{\bf r} \langle 
    e^{-\beta U_{{\rm inter},\alpha\beta}} - 1 
  \rangle_{\alpha,{\bf 0}; \beta, {\bf r}}
\end{eqnarray}
where $U_{{\rm inter},\alpha\beta}$ is the intermolecular energy between
an $\alpha$ and a $\beta$ molecule. 
Analogously we define 
\begin{eqnarray}
&& I_3(\alpha\beta\gamma) = \int d{\bf r}d{\bf s} 
    \langle (e^{-\beta U_{\rm inter,\alpha\beta}} - 1) 
\nonumber \\
    && \qquad \times (e^{-\beta U_{\rm inter,\alpha\gamma}} - 1) 
            (e^{-\beta U_{\rm inter,\beta\gamma}} - 1) 
  \rangle_{\alpha,{\bf 0}; \beta, {\bf r}; \gamma, {\bf s}},
\\
&& T_2(\alpha,\beta\gamma) = \int d{\bf r}d{\bf s} 
    \langle (e^{-\beta U_{\rm inter,\alpha\beta}} - 1) 
\nonumber \\
    && \qquad \times (e^{-\beta U_{\rm inter,\alpha\gamma}} - 1)  
  \rangle_{\alpha,{\bf 0}; \beta, {\bf r}; \gamma, {\bf s}} 
     - I_{2}(\alpha\beta) I_{2}(\alpha\gamma).
\end{eqnarray}
The integral $T_2$ represents the flexibility contribution: 
if the $\alpha$ molecule is rigid, then $T_2(\alpha,\beta\gamma) = 0$.
It is easy to generalize the bounds for $T_2(\alpha,\alpha\alpha)$ found
in Ref.~\cite{CMP-08-star} obtaining
\begin{eqnarray}
&& T_2(\alpha,\beta\beta) \ge 0,
\\
&& |T_2(\alpha,\beta\gamma)| \le 
    {1\over2} \left[ T_2(\alpha,\beta\beta) + T_2(\alpha,\gamma\gamma)
    \right].
\end{eqnarray}
We wish now to express the pressure in terms of the concentrations 
$\rho_\alpha$:
\begin{equation}
\rho_\alpha = {\langle N_\alpha\rangle \over V} = 
    \hat{z}_\alpha {\partial \beta P\over \partial \hat{z}_\alpha}.
\end{equation}
Expressing the fugacities in terms of the concentrations, we obtain
\begin{eqnarray}
\hat{z}_\alpha &=& \rho_\alpha - 
\rho_\alpha \sum_\beta I_2(\alpha\beta) \rho_\beta 
 \\
&& + {1\over 2} \rho_\alpha \sum_{\beta\gamma} \rho_\beta\rho_\gamma
    \left[I_2(\alpha\beta) I_2(\alpha\gamma) - I_{3}(\alpha\beta\gamma)
       \right. \nonumber \\
     && \left.
         - T_2(\alpha,\beta\gamma) 
         - T_2(\beta,\alpha\gamma) 
         - T_2(\gamma,\alpha\beta) \right].
\nonumber 
\end{eqnarray}
Substituting this expression in Eq.~(\ref{exppress-z}), we obtain finally
\begin{eqnarray}
\beta P &=& \sum_\alpha \rho_\alpha - 
   {1\over 2} \sum_{\alpha\beta} \rho_\alpha \rho_\beta I_2(\alpha\beta) 
\nonumber \\
&& - {1\over 3} \sum_{\alpha\beta\gamma} \rho_\alpha \rho_\beta \rho_\gamma
\left[I_{3}(\alpha\beta\gamma) \right.
\nonumber \\
&& \qquad \left. + T_2(\alpha,\beta\gamma)
         + T_2(\beta,\alpha\gamma)
         + T_2(\gamma,\alpha\beta) \right].
\end{eqnarray}
We can now specialize this expression to a polymer-colloid mixture.
If the suffixes "c" and "p" refer to the colloids and polymers, 
respectively, we can expand the pressure as in Eq.~(\ref{virial-expansion}),
neglecting terms that are of fourth order in the concentrations. The 
virial coefficients are then given by
\begin{eqnarray}
B_{2,c} &=& -{1\over2} I_{2}(cc), \\ 
B_{2,p} &=& -{1\over2} I_{2}(pp), \\ 
B_{2,cp} &=& - I_{2}(cp), \\ 
B_{3,c} &=& -{1\over3} I_{3}(ccc), \\ 
B_{3,p} &=& -{1\over3} I_{3}(ppp) - T_2(p,pp), \\ 
B_{3,ccp} &=& - I_3(ccp) - T_2(p,cc), \\ 
B_{3,cpp} &=& - I_3(cpp) - 2 T_2(p,cp), 
\end{eqnarray}
where we used the fact that the colloid is rigid,
hence $T_2(c,\alpha\beta) = 0$ for any $\alpha$ and $\beta$.

For our lattice model, integrals over the polymer positions are replaced by 
lattice sums, which 
are evaluated by using the hit-or-miss procedure applied
in Refs.~\cite{LMS-95,CMP-06} to the computation of the polymer virial 
coefficients.
The flexibility contributions are usually quite 
small \cite{CMP-06,CMP-08-star,Kofke-11-12}.
For the mixed third virial coefficients, their relevance 
depends on $q$. The flexibility correction to $B_{3,ccp}$  and 
$B_{3,cpp}$ decreases as $q\to 0$. The ratio $T_2(p,cc)/B_{3,ccp}$ is equal to 
8\%, 3\%, 1\% for $q=3,1,0.4$, respectively, in the good-solvent case. 
The analogous ratio 
$2 T_2(p,cp)/B_{3,cpp}$ is slightly larger. It is 
equal to 9\%, 5\%, 2\% for  $q=3,1,0.4$, respectively.

The ratio $T_2/B_3$ gives us a hint on the role of the neglected three-body 
forces in single-blob coarse-grained models. Indeed, if the flexibility 
integral can be neglected, we can infer that 
the following factorization holds approximately:
\begin{eqnarray}
    && \langle (e^{-\beta U_{\rm inter,\alpha\beta}} - 1) 
            (e^{-\beta U_{\rm inter,\alpha\gamma}} - 1) 
  \rangle_{\alpha,{\bf 0}; \beta, {\bf r}; \gamma, {\bf s}} 
\\
  && = 
    \langle (e^{-\beta U_{\rm inter,\alpha\beta}} - 1) 
  \rangle_{\alpha,{\bf 0}; \beta, {\bf r}} 
    \langle (e^{-\beta U_{\rm inter,\alpha\gamma}} - 1) 
  \rangle_{\alpha,{\bf 0}; \gamma, {\bf s}} .
\nonumber 
\end{eqnarray}
By definition we have 
\begin{equation}
\langle (e^{-\beta U_{\rm inter,\alpha\beta}} - 1)
  \rangle_{\alpha,{\bf 0}; \beta, {\bf r}} = 
    e^{-\beta V_{SB,\alpha\beta}({\bf r})},
\end{equation}
where $V_{SB,\alpha\beta}({\bf r})$ is the pair potential in the 
single-blob model. Therefore, the previous factorization condition implies
\begin{eqnarray}
I_{3}(\alpha\beta\gamma) 
  &\approx& \int d{\bf r}d{\bf s}\, 
    \left( e^{-\beta V_{SB,\alpha\beta}({\bf r})} - 1\right)
\\
  &&\times  \left( e^{-\beta V_{SB,\alpha\gamma}({\bf s})} - 1\right)
    \left( e^{-\beta V_{SB,\beta\gamma}({\bf r}-{\bf s})} - 1\right).
\nonumber 
\end{eqnarray}
The right-hand side is the integral in the coarse-grained model. 
Hence, if $T_2$ is negligible, the third virial coefficient in the 
model is approximately equal to that in the coarse-grained  model,
indicating that the effective three-body forces are small (see
the appendix of
Ref.~\cite{DPP-12-softspheres} for the explicit expression of the
third virial coefficient in terms of three-body forces). 

\section{Asymptotic behavior of the virial coefficients for $q\to 0$}
\label{AppB}

To compute the limiting expression of the virial coefficients for $q\to 0$,
let us first define $U_{\rm int,cp}(r)$ as the intermolecular 
energy between a colloid in the origin and a polymer such that its
first monomer is in ${\bf r}$. The choice of the first monomer is arbitrary
and the same result would have been obtained by taking any other monomer.
Then, let 
\begin{equation}
f_{cp}(r) = e^{-\beta U_{\rm int,cp}(r)} - 1
\end{equation}
be the corresponding Mayer function, which satisfies $f_{cp}(r) = 1$ for 
$r < R_c$. Analogously we define the polymer-polymer Mayer function 
$f_{pp}(r)$, where $r$ is the distance between the first monomers of the 
two polymers. Using the fact that $f_{cp}(r) = - 1$ for $r \le R_c$ and
defining $z = r - R_c$, we obtain
\begin{eqnarray}
B_{2,cp} &=& {4\pi R_c^3 \over 3} - 4 \pi \int_{R_c}^\infty dr r^2 
   \langle f_{cp}(r) \rangle
\label{B2cp-exp} \\
\nonumber 
&\approx& {4\pi R_c^3 \over 3} - 4 \pi R^2_c \int_0^\infty dz\, 
   \langle \hat{f}_{cp}(z)\rangle 
\nonumber \\
 &=& {4\pi R_c^3 \over 3} + 4 \pi R^2_c P_{1,p},
\nonumber 
\end{eqnarray}
where $\hat{f}_{cp}(z) = f_{cp}(R_c+z)$ and 
\begin{equation}
P_{1,p} = - \int_0^\infty dz\, \langle \hat{f}_{cp}(z)\rangle.
\end{equation}
The integral $P_{1,p}$ corresponds to a polymer interacting with an 
impenetrable plane. In our model, in which we have a hard-interaction
between monomers and hard wall, we can further simplify $P_{1,p}$.
If $z_{\rm min}$ is the smallest value of the $z$ coordinate 
of the first point of the walk such that the walk does not intersect the wall,
we have $P_{1,p} = \langle z_{\rm min} \rangle$. This 
expression can be rewritten in a form which is independent of the coordinates
of the first monomer. Indeed, define
\begin{equation}
z_m = \min_k z_k, \qquad z_M = \max_k z_k,
\end{equation}
where $z_k$ is the $z$ coordinate of the $k$-th monomer. Then, we can rewrite
$z_{\rm min} = z_1 - z_m$. If we now consider the walk which is obtained by 
means of a specular reflection with respect to the plane $z = z_1$
we obtain $z_{\rm min} = z_M - z_1$. Hence, if we average the two contributions
we obtain 
\begin{equation}
P_{1,p} = {1\over 2} \langle z_M - z_m \rangle,
\end{equation}
in which there is no reference to the first monomer, which was arbitrarily
chosen to define the integrations.

Let us now consider the third virial coefficient $B_{3,cpp}$. 
We have
\begin{eqnarray}
B_{3,cpp} &=& - \int d{\bf r}\, d{\bf s}\,
     \left\{
   \langle f_{cp}(r) f_{cp}(s) f_{pp}(\rho) \rangle + \right.
\nonumber \\
&& \left.
   \langle f_{cp}(r) f_{pp}(\rho) \rangle +
   \langle f_{cp}(s) f_{pp}(\rho) \rangle \right.
\nonumber \\
&& \left. -
   \langle f_{cp}(r) \rangle \langle f_{pp}(\rho) \rangle -
   \langle f_{cp}(s) \rangle \langle f_{pp}(\rho) \rangle \right\},
\end{eqnarray}
where $\rho = |{\bf s} - {\bf r}|$.
We can further simplify this expression defining
\begin{equation}
A(r,s) = e^{-\beta U_{{\rm inter},cp}(r) - \beta U_{{\rm inter},cp}(s)} - 1,
\end{equation}
and the function $I(r)$ such that  $I(r)= 1$ for $r\le R_c$ and 
$I(r)= 0$ otherwise. Since $A(r,s) = - 1$ and $f_{cp}(r) = - 1$ for $r < R_c$, 
we can write 
\begin{eqnarray}
&& \int d{\bf r}\, d{\bf s}\,
     \left\{
   \langle f_{cp}(r) f_{cp}(s) f_{pp}(\rho) \rangle \right. + 
\nonumber \\
&& \qquad \left.
   \langle f_{cp}(r) f_{pp}(\rho) \rangle +
   \langle f_{cp}(s) f_{pp}(\rho) \rangle \right\} 
\nonumber \\
&& =  \int d{\bf r}\, d{\bf s}\, \langle A(r,s) f_{pp}(\rho) \rangle
\nonumber \\
&& =  \int d{\bf r}\, d{\bf s}\, \left\{
    - \langle f_{pp}(\rho) \rangle I(r) \right. 
\nonumber \\ 
&& \qquad \left. + 
    \langle A(r,s) f_{pp}(\rho) \rangle [1 - I(r)] \right\}
\nonumber \\
&&  =  \int d{\bf r}\, d{\bf s}\, \left\{
     \langle f_{cp}(r) \rangle \langle f_{pp}(\rho) \rangle I(r) \right.
\nonumber \\ 
&& \qquad +  \left. 
    \langle A(r,s) f_{pp}(\rho) \rangle [1 - I(r)] \right\}
\nonumber \\
&& =  \int d{\bf r}\, d{\bf s}\, \left\{
     \langle f_{cp}(r) \rangle \langle f_{pp}(\rho) \rangle  \right.
 \\ 
&& \qquad \left. + 
    \left[\langle A(r,s) f_{pp}(\rho) \rangle - 
     \langle f_{cp}(r) \rangle \langle f_{pp}(\rho) \rangle \right]
       [1 - I(r)] \right\}.
\nonumber 
\end{eqnarray}
Therefore, we can rewrite 
\begin{eqnarray}
&& B_{3,cpp} = - \int d{\bf r}\, d{\bf s}\,
\left\{ \vphantom{\hat{R}}
   \left[\langle A(r,s) f_{pp}(\rho) \rangle \right. \right.
\\
&& \qquad
   \left. \left. - \vphantom{\hat{R}}
     \langle f_{cp}(r) \rangle \langle f_{pp}(\rho) \rangle \right]
       [1 - I(r)] - 
    \langle f_{cp}(r) \rangle \langle f_{pp}(\rho) \rangle
   \right \}.
\nonumber 
\end{eqnarray}
To rewrite this term in a more transparent way, let us consider 
$B_{2,cp} B_{2,pp}$ which we rewrite as 
\begin{eqnarray}
2 B_{2,cp} B_{2,pp} &=& 
  \int d{\bf r}\, d{\bf s}\,
     \langle f_{cp}(r)\rangle 
    \langle f_{pp}(\rho)\rangle.
\end{eqnarray}
Using this expression we obtain finally
\begin{eqnarray}
&& B_{3,cpp} = 2 B_{2,cp} B_{2,pp}
\\
&& - \int d{\bf r}\, d{\bf s}\, [1 - I(r)]
\left\{\langle A(r,s) f_{pp}(\rho)\rangle  - 
        \langle f_{cp}(r) \rangle \langle f_{pp}(\rho) \rangle \right\}
\nonumber 
\end{eqnarray}
The remaining integral is a surface contribution. Indeed, for $r \ge R_c$ 
the function 
$A(r,s)$ is different from zero only if $r - R_c$ is of order of a few times
$\hat{R}_g$. Moreover, since the range 
of $f_{pp}(\rho)$ is also of order $\hat{R}_g$, the integral gets contributions
only if $|{\bf r} - {\bf s}|$ is of order $\hat{R}_g$. Hence, a nonvanishing 
contribution is obtained only if 
$|R-s|$ is of the order of a few times $\hat{R}_g$. To make this 
explicit, let us introduce bipolar coordinates so that 
\begin{eqnarray}
&& B_{3,cpp} = 2 B_{2,cp} B_{2,pp}
\nonumber \\
&& \qquad - 8 \pi^2 \int_{R_c}^\infty r dr\, \int_0^\infty s ds 
   \int_{|r-s|}^{r+s} \rho d\rho 
\nonumber \\
&& \qquad \left\{\langle A(r,s) f_{pp}(\rho)\rangle  - 
        \langle f_{cp}(r) \rangle \langle f_{pp}(\rho) \rangle \right\}
\end{eqnarray}
We now change variable, defining
\begin{equation}
z_1 = r - R_c, \quad z_2 = s - R_c \quad x = \sqrt{\rho^2 - (z_1-z_2)^2} .
\end{equation}
Taking the limit $R_c\to \infty$, we obtain 
\begin{eqnarray}
&& B_{3,cpp} = 2 B_{2,cp} B_{2,pp} 
\\
&& \qquad - 4 \pi R_c^2 \int_0^\infty (2 \pi x) d x
     \int_0^\infty dz_1
     \int_{-\infty}^\infty dz_2
\nonumber \\
&& \qquad \left\{\langle A(R_c+z_1,R_c+z_2) f_{pp}(\rho)\rangle  - 
        \langle \hat{f}_{cp}(z_1) \rangle \langle f_{pp}(\rho) \rangle \right\}.
\nonumber
\end{eqnarray}
We define $\hat{A}(z_1,z_2) = {A}(R_c+z_1,R_c+z_2)$, which takes the value
$-1$ whenever $z_1$ or $z_2$ is negative, and 
\begin{eqnarray}
&& P_{2,pp} = 2 P_{1,p} B_{2,pp} 
\\
&& \qquad - \int_0^\infty (2 \pi x) d x
     \int_0^\infty dz_1
     \int_{-\infty}^\infty dz_2
\langle \hat{A}(z_1,z_2) f_{pp}(\rho)\rangle .
\nonumber
\end{eqnarray}
This allows us to write
\begin{equation}
B_{3,cpp} = 2 B_{2,cp} B_{2,pp} + 4 \pi R_c^2 P_{2,pp}.
\label{B3cpp-exp}
\end{equation}
Using this expression we can compute the expansion of $\delta_s$ for $q\to 0$.
We obtain
\begin{equation}
\delta_s = P_{1,p} + \rho_p P_{2,pp} + O(\rho_p^2),
\end{equation}
which coincides with that valid for polymers in the presence 
of an impenetrable plane. From Eq.~(\ref{B3cpp-exp}) and (\ref{B2cp-exp})
we obtain finally Eq.~(\ref{A3cpp-largeRc}).

For our lattice model the integral $\hat{P}_{2,pp} = 
P_{2,pp} - 2 P_{1,p} B_{2,pp}$ can be given a simpler form, averaging again 
over the walks that are obtained by specular reflections with respect 
to the planes that go through the first monomer and are parallel the surface. 
Let $\omega_1$ and $\omega_2$ be two lattice chains and 
$z_k^{(1)}$ and $z_k^{(2)}$ be the $z$-coordinates of their 
$k$-th monomers.
Then, define $T(\omega_2,{\bf r})$ as the lattice walk that is obtained
by translating $\omega_2$ by the lattice vector ${\bf r}$ and the function 
$H(\omega,\overline{z})$ which takes the value $+1$ if the walk $\omega$ 
intersects the plane $z = \overline{z}$ and the value $0$ otherwise. 
If $z_m^{(i)} = \min z_k^{(i)}$, $z_M^{(i)} = \max z_k^{(i)}$,
$Z_m = z_m^{(1)} + z_m^{(2)} - z_M^{(2)}$, and 
$Z_M = z_M^{(1)} + z_M^{(2)} - z_m^{(2)}$
we have 
\begin{eqnarray}
&& \hat{P}_{2,pp} = - {1 \over2} \left\langle \sum_{z_1 = Z_m}^{Z_M} 
     \sum_{{\bf r}} \left\{ H[\omega_1,z_1] + H[T(\omega_2,{\bf r}),z_1]  \right. \right.
\nonumber \\ 
&& \qquad \left. \left. - 
     H[\omega_1,z_1] H[T(\omega_2,{\bf r}),z_1] \right\}
    \left(1 - e^{-w N_{\rm int}}\right)\right\rangle,
\end{eqnarray}
where the sum over ${\bf r}$ is over all lattice translations and 
$N_{\rm int}$ is the number of intersections between $\omega_1$ and the 
translated $T(\omega_2,{\bf r})$. The sums are evaluated by using the
obvious generalization of the hit-or-miss procedure applied
in Refs.~\cite{LMS-95,CMP-06} to the computation of the polymer virial 
coefficients.

Finally, we shall discuss the third-virial coefficient $B_{3,ccp}$.
Since this quantity is not relevant for the depletion we will 
only compute the leading term, which can be obtained by approximating
the polymer with a hard sphere of zero radius. Thus, we obtain for 
$q \to 0$
\begin{equation}
B_{3,ccp} \hat{R}_g^{-6} = {16 \pi^2\over 9 q^6}.
\end{equation}

\section{Supplementary material}

\subsection{Low-density virial coefficients and depletion thickness}

In this supplementary material we report the numerical estimates 
of the virial coefficients and of the depletion thickness in the 
low-density limit. In the good-solvent case, for $q \le 3$ 
we have results for $L=240$, 600, and 2400: in this case,
the universal large-$L$
limit has been obtained by performing an extrapolation of the results
with $a + b/L^\nu + c/L$. For $q \ge 4$, we have results for $L = 6000$ and
24000, which have been extrapolated to $a + b/L^\nu$. Here $\nu$ is the 
usual Flory exponent, $\nu = 0.587597$.
In the crossover region, for $q \le 3$ we have results for 
$L=120$, 240, 600, 1200, and 2400: they have been fitted
to $a + b/\sqrt{L} + c/L$. For $q \ge 4$ we only have results for 
$L = 6000$ and 30000: they have been extrapolated using 
$a + b/\sqrt{L}$. The results of the extrapolations are reported in 
Tables~\ref{table:viriali1}, \ref{table:viriali2}, and \ref{table:depletion}.
In Table~\ref{tab:zero-density-SB} we report single-blob results. Of course,
here no extrapolation is needed.

\begin{table*}[h]
\caption{Estimates of the asymptotic, universal adimensional 
virial combinations $A_{2,cp} = B_{2,cp} \hat{R}_g^{-3}$, 
$A_{3,cpp} = B_{3,cpp} \hat{R}_g^{-6}$,
$A_{3,ccp} = B_{3,ccp} \hat{R}_g^{-6}$,
for $z=\infty$ (good-solvent case) and for 
$z = z^{(1)}$.}
\label{table:viriali1}
\begin{tabular}{ccccccc}
\hline\hline
& \multicolumn{3}{c}{$z=\infty$} & 
 \multicolumn{3}{c}{$z=z^{(1)}$}  \\
\hline
$q$ & $A_{2,cp}$  &   $A_{3,cpp}$  & $A_{3,ccp}$  
    & $A_{2,cp}$  &   $A_{3,cpp}$  & $A_{3,ccp}$   \\
\hline
50 & 0.1097(2) & 0.015(5) &   \\
40 & 0.1467(4) & 0.023(6) &  \\
35 &  0.1744(4) & 0.029(7) &  \\
30 & 0.2131(5) & 0.0442(9) &   & 0.416(1) & 0.032(2) &  \\
25 & 0.2702(7) & 0.062(10) &   & 0.504(1) & 0.047(2) &  \\
20 & 0.3615(9) & 0.108(15) &    & 0.639(1) & 0.073(3) & \\
15 & 0.527(1) & 0.20(2) &    & 0.870(1) & 0.128(5) &  \\
10 & 0.899(2) & 0.57(3) & 0.018(2)& 1.360(2) & 0.272(6) & 0.014(1) \\
8 & 1.210(3) & 0.97(4) & 0.040(3) & 1.750(3) & 0.42(1) & 0.044(8) \\
6 & 1.782(4) & 1.90(6) & 0.144(6) & 2.446(4) & 0.76(1) & 0.156(6) \\
4  & 3.114(7) & 5.0(1) & 0.82(1)  & 4.010(6) & 1.67(2) & 0.95(2) \\
3  &  4.71(1) & 10.2(1) & 2.82(3)& 5.807(8) & 2.96(2) & 3.35(2) \\
2 &  8.65(2) & 26.1(2) & 16.8(1)&10.20(1) & 6.73(4) & 19.55(8)  \\
1.75 & 10.67(2) & 35.7(3) & 30.1(2)& 12.41(2) & 8.87(5) & 35.0(1) \\
1.5 & 13.69(3) & 51.2(4) & 59.4(3) &15.71(2) & 12.29(7) & 68.7(2) \\
1.25 &  18.60(4) & 78.6(1) &132.9(7)  &21.02(3) & 18.1(1) & 152.7(5) \\
1.0 &  27.54(6) &133.3(9) & 360(2)&30.61(5) & 29.5(1) & 409(1) \\
0.8 & 41.7(1) &228.5(15) &984(5) &45.62(7) & 48.7(2) & 1107(4) \\
0.6 & 73.45(20) & 462(3) & 3680(20) &79.1(1) & 94.9(5) & 4080(15)\\
0.5 & 107.4(3) & 726(5) & 8630(45) &114.5(2) & 146.4(7) & 9500(30)\\
0.4 & 174.5(4) &1281(8) & 24900(150) &184.4(3) & 253(1) & 27100(90)\\
0.3 & 337.7(9)  &2700(20) & 101000(550) & 352.9(6) & 525(2) & 108700(400)\\
0.2 & 909(2)& 8000(50) & 783000(4000) &937(2) & 1520(7) & 827000(3000)\\
\hline\hline
\end{tabular}
\end{table*}

\begin{table}[h]
\caption{Estimates of the asymptotic, universal adimensional 
virial combinations $A_{2,cp} = B_{2,cp} \hat{R}_g^{-3}$, 
$A_{3,cpp} = B_{3,cpp} \hat{R}_g^{-6}$,
$A_{3,ccp} = B_{3,ccp} \hat{R}_g^{-6}$,
for $z = z^{(3)}$.}
\label{table:viriali2}
\begin{center}
\begin{tabular}{cccc}
\hline
\hline
$q$ & $A_{2,cp}$ & $A_{3,cpp}$ & $A_{3,ccp}$ \\
\hline
3 & 5.53(2) & 7.7(1) & 3.31(2) \\
2 & 9.81(1) & 18.2(1) & 19.19(7) \\
1.75 &11.97(2) & 24.1(1) & 34.3(1) \\
1.5 & 15.20(1) & 33.7(1) & 67.1(2) \\
1.25 & 20.42(3) & 50.1(2) & 149.0(5) \\
1.0 & 29.85(4) & 82.6(3) & 399(1) \\
0.8 & 44.63(6) & 138.1(5) & 1082(3)  \\
0.6 & 77.7(1) & 272(1) & 4000(15) \\
0.5 & 112.8(2) & 423.2(15) & 9300(30) \\
0.4 & 182.0(3) & 737(3) & 26625(85) \\
0.3 & 350(5) & 1533(6) & 107100(300) \\
0.2 & 932(2) & 4470(20) & 818000(3000) \\
\hline\hline
\end{tabular}
\end{center}
\end{table}

\begin{table}[h]
\caption{Depletion thickness $\delta_{s,0}/R_c = \delta_s(\Phi=0)/R_c$ 
at zero density and first density correction $\delta_1(q)$.}
\label{table:depletion}
\begin{center}
\begin{tabular}{ccccccc}
\hline
\hline
& \multicolumn{2}{c}{ $z=\infty$}
& \multicolumn{2}{c}{ $z=z^{(3)}$}
& \multicolumn{2}{c}{ $z=z^{(1)}$}
\\
\hline
$q$ & $\delta_{s,0}/R_c$ & $\delta_1$ & 
      $\delta_{s,0}/R_c$ & $\delta_1$ &
      $\delta_{s,0}/R_c$ & $\delta_1$  \\
\hline
\hline
50 & 13.85(9) & $-$0.927(4) \\
40 & 12.09(1) & $-$0.934(4) \\
35 & 11.13(1) & $-$0.939(4) \\
30 & 10.117(9) & $-$0.944(4) &&& 12.897(7) & $-$0.1637(5)\\
25 & 9.027(8) & $-$0.952(4) &&& 11.345(6) & $-$0.1638(5)\\
20 & 7.838(7) & $-$0.960(4) &&& 9.685(6) & $-$0.1643(8)\\
15 & 6.515(6) & $-$0.975(4) &&& 7.883(5) & $-$0.1655(6)\\
10 & 4.987(5) & $-$0.991(4)&&&5.872(4) & $-$0.1665(7) \\
8 & 4.289(4) & $-$1.001(3)&&& 4.980(3) & $-$0.1672(8)\\
6 & 3.512(3) & $-$1.016(4)  &&& 4.015(3) & $-$0.1675(6)\\
4 & 2.624(3) & $-$1.032(4)  &&& 2.942(2) & $-$0.1678(8)\\
3 & 2.119(2) & $-$1.035(3) & 2.292(1) & $-$0.519(2) & 2.343(6) & $-$0.168(1)\\
2 & 1.547(2) & $-$1.046(4) & 1.656(1) & $-$0.520(2) &1.6893(5) & $-$0.168(1)\\
1.75 & 1.390(2) & $-$1.047(5)& 1.484(1)& $-$0.521(1)&1.5132(5) & $-$0.168(1)\\
1.5 & 1.226(2) & $-$1.049(5) & 1.306(1)& $-$0.522(2)&1.3304(4) & $-$0.168(1)\\
1.25 & 1.054(2) & $-$1.050(6)& 1.1194(9)& $-$0.523(2)&1.1401(4) & $-$0.168(1)\\
1.0 & 0.873(41) & $-$1.052(7) &  0.9243(8)& $-$0.523(3)&0.9408(4) & $-$0.1675(15)\\
0.8 & 0.720(1) & $-$1.048(9) & 0.7605(8)& $-$0.522(4)&0.7737(4) & $-$0.167(2)\\
0.6 & 0.559(1) & $-$1.05(1) & 0.5884(7)& $-$0.521(4)&0.5982(3) & $-$0.167(2)\\
0.5 & 0.474(1) & $-$1.05(1) & 0.4988(7)& $-$0.520(4)&0.5069(3) & $-$0.167(2)\\
0.4 &  0.387(1) & $-$1.04(2)& 0.4063(7)& $-$0.516(6)&0.4128(3) & $-$0.167(3)\\
0.3 & 0.296(1) & $-$1.05(2) & 0.3107(7)& $-$0.515(8)&0.3155(3) & $-$0.166(3)\\
0.2 & 0.202(1) & $-$1.04(3) & 0.2116(6)& $-$0.51(1)&0.2146(3) & $-$0.166(5)\\
\hline\hline
\end{tabular}
\end{center}
\end{table}

\clearpage

\begin{table}[h]
\caption{Virial coefficients, 
depletion thickness $\delta_{s,0}/R_c = \delta_s(\Phi=0)/R_c$ 
at zero density and first density correction $\delta_1(q)$ 
in the single-blob model.  }
\label{tab:zero-density-SB}
\begin{center}
\begin{tabular}{cccccccc}
\hline
\hline
 $z$ & $q$ & $A_{2,cp}$ & $\delta_{s,0}/R_c$ & $\delta_1$ & $A_{3,cpp}$ &
$A_{3,ccp}$ \\
\hline
$z^{(1)}$ & 0.5& 114.376(5)  & 0.50563(2)  & $-$0.1775(1) & 141.32(7) &
     9518(2)   \\
& 1.0     & 30.555(3)  & 0.9394(5) & $-$0.1838(1) & 26.64(2) & 399.0(3)  \\
& 2.0     & 10.155(1)  & 1.68675(9) & $-$ 0.18620(7) & 5.299(6) & 16.94(3) \\
\hline
$z^{(3)}$ & 0.5 & 112.496(5) & 0.49734(2) & $-$0.5491(3) & 409.2(2) & 9285(2) \\
& 1.0     & 29.750(2)  & 0.92221(5) & $-$0.5656(3) &  74.94(6) &  385.2(2) \\
& 2.0     & 9.762(1) & 1.6516(9) & $-$ 0.5707(2) & 14.27(1) & 16.01(3) \\
\hline
$\infty$ & 0.5 & 106.787(6) & 0.47156(3) & $-$1.0967(9) & 701.2(4) & 8565(2) \\
& 1.0 & 26.796(2) &  0.85635(4) & $-$1.1443(4) & 116.55(6) & 331.8(2)   \\
& 2.0 & 8.2866(9) &   1.51068(9) & $-$1.1468(3) & 19.16(2) & 12.55(2) \\
\hline
\hline
\end{tabular}
\end{center}
\end{table}

\subsection{Scaling corrections in the presence of colloid-polymer interactions}

In the field-theoretical approach to critical phenomena,
the presence of an impenetrable boundary gives rise to additional 
irrelevant surface operators, which, in turn, give rise to new 
corrections to scaling. For the case of a nonadsorbing boundary, 
the question was analyzed by H. W. Diehl, S. Dietrich, and E. 
Eisenriegler [Phys. Rev. B {\bf 27}, 2937 (1983)]. They found that the 
leading surface correction is associated with an exponent $\omega = -\nu$.
We have performed a careful check of this prediction,
by considering the universal combinations $R_{1,p}$. 
Estimates for several values of $L$ are reported in
Table~\ref{table:plane-zinf} (good-solvent case) and 
in Table~\ref{table:plane-z1} ($z = z^{(1)}$).
The results have been fitted to
$R_{1,p}^* + a/L^\theta$, where $R_{1,p}^*, a$, and $\theta$ 
are taken as free parameters. The results of the fits of the good-solvent 
data are reported in Table \ref{table:fit-plane-zinf}. 
They are clearly consistent with $\theta = \nu\approx 0.588$. For $z = z^{(1)}$,
the results reported in 
Table \ref{table:fit-plane-z1} are consistent with $\theta = 1/2$, as expected.

\begingroup
\begin{table*}[h]
\caption{Estimates of the 
combinations $R_{1,p}$ and $R_{2,pp}$ 
for $z=\infty$ (good-solvent case), as a function of the 
length $L$ of the chains.} 
\label{table:plane-zinf}
\begin{tabular}{lcc}
\hline\hline
  $L$&  $R_{1,p}$ & $R_{2,pp}$ \\
\hline
  120& 0.98351(3) & $-$5.354(3) \\
  240& 1.00905(7) & $-$5.184(10) \\
  480& 1.02597(3) & $-$4.928(4) \\
  600& 1.03019(4) & $-$4.890(5) \\
  900& 1.03650(2) & $-$4.789(3) \\
 1200& 1.04018(3) & $-$4.752(3) \\
 1800& 1.04446(4) & $-$4.704(4) \\
 2400& 1.04683(11)& $-$4.683(12) \\
 3600& 1.04978(3) & $-$4.633(3) \\
 4800& 1.05142(4) & $-$4.614(3) \\
 6000& 1.05255(5) & $-$4.602(6) \\
 9000& 1.05418(8) & $-$4.576(8) \\
12000& 1.05514(9) & $-$4.566(9) \\
24000& 1.05696(8) & $-$4.541(8) \\
\hline\hline
\end{tabular}
\end{table*}
\endgroup

\begingroup
\begin{table*}[h]
\caption{Fits of $R_{1,p}$ to $R_{1,p}^* + a/L^\theta$, including 
only data satisfying $L\ge L_{\rm min}$. $\chi^2$ is the sum of the residuals 
and DOF is the number of degrees of freedom of the fit.
Results for the good-solvent ($z = \infty$) case.}
\label{table:fit-plane-zinf}
\begin{tabular}{lccc}
\hline\hline
$L_{\rm min}$ & $\chi^2$/DOF & $R_{1,p}^*$ &  $\theta$ \\
\hline
120 & 4.75/11& 1.06057(5) & 0.578(1) \\ 
240 & 3.32/10& 1.06063(7) & 0.575(2) \\
480 & 2.32/9 & 1.06057(9) & 0.578(3) \\ 
600 & 2.17/8 & 1.06060(12)& 0.576(5) \\ 
900 & 1.26/7 & 1.06052(15)& 0.581(6) \\ 
1200& 1.23/6 & 1.06050(18)& 0.582(9) \\ 
1800& 1.19/5 & 1.06053(23)& 0.580(14) \\ 
\hline\hline
\end{tabular}
\end{table*}
\endgroup

\begingroup
\begin{table*}[h]
\caption{Estimates of the 
combinations $R_{1,p}$ and $R_{2,pp}$ 
for $z=z^{(1)}$, as a function of the chain length $L$.}
\label{table:plane-z1}
\begin{tabular}{lcc}
\hline\hline
  $L$&  $R_{1,p}$ & $R_{2,pp}$ \\
\hline
120  & 1.01296(10) & $-$1.093(3) \\
240  & 1.04504(9)  & $-$0.997(2) \\
600  & 1.07343(11) & $-$0.910(2) \\
1200 & 1.08795(12) & $-$0.873(3) \\
2400 & 1.09810(9)  & $-$0.839(2) \\
6000 & 1.10701(9)  & $-$0.812(2) \\
12000& 1.11140(19) & $-$0.797(4) \\
30000& 1.11539(19) & $-$0.787(4) \\
\hline\hline
\end{tabular}
\end{table*}
\endgroup

\begingroup
\begin{table*}[h]
\caption{Fits of $R_{1,p}$ to $R_{1,p}^* + a/L^\theta$, including 
only data satisfying $L\ge L_{\rm min}$. $\chi^2$ is the sum of the residuals 
and DOF is the number of degrees of freedom of the fit.
Results for $z = z^{(1)}$.
}
\label{table:fit-plane-z1}
\begin{tabular}{lccc}
\hline\hline
$L_{\rm min}$ & $\chi^2$/DOF & $R_{1,p}^*$ &  $\theta$ \\
\hline
120 & 3.35/5 & 1.1224(2) & 0.501(2) \\ 
240 & 2.80/4 & 1.1223(2) & 0.503(3) \\ 
600 & 0.14/3 & 1.1218(4) & 0.514(7) \\ 
1200& 0.11/2 & 1.1219(5) & 0.512(14) \\
\hline\hline
\end{tabular}
\end{table*}
\endgroup

\end{document}